\documentclass[aps,prb,shownopacs,twocolumn,superscriptaddress]{revtex4-1}
\usepackage{bm,color,amsmath,amssymb,mathrsfs,latexsym,graphicx,psfrag}

\newcommand{\bra}[1]{\left\langle#1\right|}
\newcommand{\ket}[1]{\left|#1\right\rangle}

\newcommand{\beq}{\begin{equation}}
\newcommand{\eneq}{\end{equation}}

\begin{document}

\title{Inversion Symmetric Topological Insulators }

\author{Taylor L. Hughes}
\affiliation{Department of Physics, University of Illinois, 1110 West Green St, Urbana IL 61801} 
\author{Emil Prodan}
\affiliation{Department of Physics, Yeshiva University, New York, NY 10016}
\author{B. Andrei Bernevig}
\affiliation{ Department of Physics, Princeton University, Princeton, NJ 08544} 

\begin{abstract} We analyze translationally-invariant insulators with inversion symmetry that fall outside the current established classification of topological insulators. These insulators exhibit no edge or surface modes in the energy spectrum and hence they are not edge metals when the  Fermi level is in the bulk gap. However, they do exhibit \emph{protected}  modes in the entanglement spectrum localized on the cut between two entangled regions. Their entanglement entropy \emph{cannot} be made to vanish adiabatically, and hence the insulators can be called topological.  There is a  direct connection between the  inversion eigenvalues of the Hamiltonian band structure and the mid-gap states in the entanglement spectrum. The classification of protected entanglement levels is given by an integer $n\in Z$, which is the difference between the negative inversion eigenvalues at inversion symmetric points in the Brillouin zone, taken in sets of two.  When the  Hamiltonian describes  a Chern insulator or a non-trivial time-reversal invariant topological insulator, the entirety of the entanglement spectrum exhibits \emph{spectral flow}.  If the Chern number is zero for the former, or time-reversal is broken in the latter, the entanglement spectrum does \emph{not} have spectral flow, but, depending on the inversion eigenvalues, can still exhibit protected mid-gap bands similar to impurity bands in normal semiconductors. Although spectral flow is broken (implying the absence of real edge or surface modes in the original Hamiltonian), the mid-gap entanglement bands cannot be adiabatically removed, and the insulator is `topological.'  We analyze the linear response of these insulators and provide proofs and examples of when the inversion eigenvalues determine a non-trivial charge polarization, a quantum Hall effect, an anisotropic 3D quantum Hall effect, or a magneto-electric polarization. In one-dimension, we establish a link between the product of the inversion eigenvalues of all occupied bands at all inversion symmetric points and charge polarization. In two dimensions, we prove a link between the product of the inversion eigenvalues and the parity of the Chern number of the occupied bands. In three dimensions, we find a topological constraint on the product of the inversion eigenvalues thereby showing that some $3$D materials are protected topological metals, we show the link between the inversion eigenvalues and the $3$D Quantum Hall Effect, and analyze the magneto-electric polarization ($\theta$ vacuum) in the absence of time-reversal symmetry.
\end{abstract}

\date{\today}


\maketitle

One of the most active fields of research  in recent years  has been the study of non-trivial topological states of matter. The paradigm example of such a state is the  Quantum Hall Effect, with its Integer (IQHE) and Fractional (FQHE) versions. More recently, examples of topological phases that do not require external magnetic fields have been proposed, the first being Haldane's Chern Insulator model.\cite{Haldane1988} Although this state has not been experimentally realized, a time-reversal invariant (TRI) version has been proposed and discovered.\cite{Kane2005A,kane2005B,bernevig2006a,bernevig2006c,koenig2007,fu2007a,hsieh2008}

 Recent work in the theory of topological insulators\cite{Kane2005A,kane2005B,bernevig2006a} showed that an important consideration is not only which symmetries the state breaks, but which symmetries must be preserved to ensure the stability of the state. A periodic table classifying the topological insulators and superconductors has been created. The table organizes the possible topological states according to their space-time dimension and the symmetries that must remain protected: time-reversal, charge conjugation, and/or chiral symmetries.\cite{qi2008B,schnyder2008,kitaev2008} The most interesting entries in this table, from a practical standpoint, are the 2 and 3D TRI topological insulators which have been already found in nature.\cite{bernevig2006c,koenig2007,fu2007a,hsieh2008} These are insulating states classified by a $Z_2$ invariant that requires an unbroken time-reversal symmetry to be stable. There are several different methods to calculate the $Z_2$ invariant,\cite{kane2005B,fu2006,moore2007,roy2009a,fu2007a,qi2008B,Prodan:2009oh,Prodan:2009mi} and a non-trivial value for this quantity implies the existence of an odd number of gapless Dirac fermion boundary states as well as a non-zero magneto-electric polarizibility in 3D.\cite{qi2008B,essin2009}

The current classification of the topological insulators covers only the time-reversal, charge conjugation, or chiral symmetries and does not exhaust the number of all possible topological insulators. In principle, for every discrete symmetry, there must exist topological insulating phases with distinct physical properties, and a topological number that classifies these phases and distinguishes them from the ``trivial" ones.  So far, in our discussion we have used the term ``topological" cavalierly so before proceeding, we should ask: what makes an insulator topological? We start by first defining a trivial insulator: this is the insulator that, upon slowly turning off the hopping elements and the hybridization between orbitals on different sites, flows adiabatically into the atomic limit. In most of the existent literature on non-interacting topological insulators, it is implicitly assumed that non-trivial topology implies the presence of gapless edge states in the energy spectrum of a system with boundaries. However, it is well known from the literature on  topological phases that such systems can theoretically exist without exhibiting gapless edge modes.\cite{Freedman2004}  Hence, the edge modes cannot be the only diagnostic of a topological phase and, consequently, the energy spectrum alone, with or without boundaries, is insufficient to determine the full topological character of a state of matter. In the bulk of an insulator, it is a known fact that the topological structure is encoded in the eigenstates rather than in the energy spectrum. As such, one can expect that entanglement  - which only depends on  the eigenstates - can provide additional information about the topological nature of the system. However we know that topological entanglement entropy (or the sub-leading part of the entanglement entropy),\cite{Hamma2004,kitaev2006,levin2006} the preferred quantity used to characterize topologically ordered phases, does not provide a unique classification, and, moreover, vanishes for any non-interacting topological insulator, be it time-reversal breaking Chern Insulators or TRI topological insulators. 
However, as we will see, careful studies of the full entanglement spectrum \cite{li2008} helps in characterizing these states.\cite{ryu2006,li2008, BrayAli2009,Flammia2009,Thomale2010A,Thomale2010B,Pollmann2010, turner2009, fidkowski2010, Kargarian2010,prodan2010,Sterdyniak2010}

The \emph{total} entanglement entropy can be continuously deformed to zero for trivial insulators, since the atomic limit to which every trivial insulator can be adiabatically continued (by the above definition) is completely local and has flat featureless bands.  We could therefore suggest that a \emph{non-trivial} topological state in a non-interacting translationally invariant insulator should be defined as having an entanglement entropy which cannot be adiabatically tuned to zero. However, even this definition cannot be entirely correct, as the entanglement entropy strongly depends on the nature of the cut made in the system. For a single particle entanglement spectrum with eigenvalues $\{\xi_a\}$ the entanglement entropy is determined via
\begin{equation}
S_{ent}=-\sum_{a}\left(\xi_a\log \xi_a +(1-\xi_a)\log (1-\xi_a)\right).
\end{equation}
For IQH states on the sphere, the many-body wavefunction is a single Slater determinant of occupied Landau orbitals, and hence an orbital cut  \cite{li2008} would result in zero entanglement entropy since all orbitals are fully occupied or unoccupied leading to a set of $\{\xi_a\}$ which are all $0$'s or $1$'s and do not contribute to $S_{ent}.$ Similarly, for a translationally-invariant Chern or TRI topological insulator on a lattice, a momentum space-cut would always give zero entanglement entropy since the Hamiltonian is diagonal in this basis. A spatial cut, however, would show mid-gap bands in the entanglement spectrum both the IQHE and in the topological insulator case(\emph{i.e.} a set of eigenvalues spanning the `gap' between $0$ and $1$) similar to the ones in the real energy spectrum.\cite{ryu2006,haldane2009, BrayAli2009,turner2009, fidkowski2010, prodan2010} These mid-gap states give large contributions to the  entanglement entropy. In fact, for such states, the entanglement entropy for the spatial cut cannot be made to vanish by any adiabatic changes in the Hamiltonian. We hence propose that a translationally invariant insulator can be classified as topological if it cannot be adiabatically connected to a state with zero entanglement entropy for \emph{at least one kind of cut of the system}. Explicitly, an insulator should be characterized as topological if it has protected mid-gap states in the single-particle entanglement spectrum that cannot be pushed to eigenvalues $0$ or $1$  by any adiabatic changes of the Hamiltonian. 

In the current paper we analyze the physics of insulators with inversion symmetry based on the above definition. While some of the inversion-symmetric insulators exhibit protected edge modes in the energy spectrum with boundaries (\emph{e.g.} a Chern insulator with inversion symmetry), most do not. However, they can still be topological because their entanglement spectrum for a spatial cut exhibits protected mid-gap bands of states. This was first pointed out for 3D strong topological insulators with inversion symmetry and soft time-reversal breaking in Ref. \onlinecite{turner2009} . Although it was indicated in Ref. \onlinecite{turner2009} that the entanglement spectrum cannot distinguish between a TRI and inversion invariant topological insulator, and one with TRI slightly broken (compared to the bulk gap), we show that it can distinguish these states. In Section \ref{sec:entanglement} we explicitly show that inversion symmetric topological insulators have two types of entanglement spectrum, both with protected mid-gap states. The characteristic which distinguishes the two types of entanglement spectra (and the two cases from Ref. \onlinecite{turner2009}) is the presence or absence of spectral flow.  For non-trivial TRI or  Chern insulators the entanglement spectrum exhibits spectral flow, very much like their energy spectrum. Heuristically this means that the filled and empty bulk states are connected via an interpolating set of states localized on the partition between the two entangled regions.
Spectral flow in the energy spectrum implies spectral flow in the entanglement spectrum. If time-reversal is broken for TRI topological insulators, or for T-breaking insulators with vanishing Chern number, we show that  such spectral flow is interrupted in both the energy and entanglement spectra, and the occupied bands are disconnected from the unoccupied ones.  One could then assume that, in systems without a continuous spectral connection between the bulk entanglement bands, one could push all the mid-gap entanglement bands to entanglement eigenvalues $0,1$ and hence to a trivial insulator with vanishing entropy on every cut. We find this not to be the case for  inversion symmetric insulators  whose set of inversion eigenvalues changes between two inversion symmetric points in the Brillouin zone. In this case,  while most entanglement eigenvalues can be made $0$ or $1,$ there is a set of protected mid-gap states that form bands which give the insulator nonzero entanglement entropy (even when the spectral flow has been destroyed). For the case of time-reversal and inversion invariant topological insulators, this was shown in Ref. \onlinecite{turner2009} by breaking time-reversal softly. We find a formula relating the number of protected mid-gap bands in the entanglement spectrum to the inversion eigenvalues of the system at inversion symmetric points. At inversion symmetric $k$, and for cuts which separate the system into two equal halves, the entanglement spectrum has protected entanglement edge or surface modes at \emph{exactly} $\xi=1/2.$ This means that there is no finite-size level repulsion (splitting) between these modes which is a common feature for real boundary or interface states in the energy spectrum.  Even if the original system has other topological invariants, such as the Chern number, or  the TRI $Z_2$ invariant which are all trivial, the number of entanglement edge modes can be nonzero.  

In Section \ref{sec:response} we analyze the physical response of a subset of these inversion symmetric insulators, and show important implications for the charge polarization, the parity of the Chern number, the 3D quantum Hall effect,  and the topological magneto-electric polarizability. Namely, given the set of inversion eigenvalues for the occupied bands at all inversion symmetric points in the Brillouin zone, we provide explicit, compact formulas and complementary derivations for determining the physical responses which only depend on the inversion eigenvalues. Additionally,  as an alternative perspective, we also show that some of the inversion topological invariants are equivalent to the wavefunction monodromy, which, in principle is an experimentally measurable quantity. In particular, we show the following: in one dimension, the product of inversion eigenvalues over all occupied bands and over all inversion symmetric points is related to the quantized charge polarization. In two dimensions, the product over the inversion eigenvalues determines the parity of the Chern number of the occupied bands. In three dimensions, we show several things: first, we prove a topological restriction for the product of inversion eigenvalues of any insulator: it must always equal $+1$. As such, some systems are protected metals, which cannot be made insulating with weak scattering. Second, we show that, depending on the product of  inversion eigenvalues in different inversion symmetric planes we will have 3D quantum Hall effects on different planes in the sample. We then show that, depending on the inversion eigenvalues, several inversion symmetric systems can exhibit a quantized magneto-electric polarizability, even though the Hamiltonian my not be adiabatically continuable to a time-reversal invariant topological insulator.

 Finally in Section \ref{sec:examples} we end the paper with several examples of such insulators and with numerical results. While most of the examples we chose have a topological response connected with an inversion topological invariant, we  stress that the presence of mid-gap states in the entanglement spectrum is not intrinsically related to the presence of a non-trivial topological response. For example, two identical copies of a strong-topological insulator with inversion and time-reversal symmetry has a trivial $Z_2$ invariant and thus  a trivial response. However, this system will exhibit protected mid-gap entanglement states. So, while some inversion invariant insulators have protected topological responses, some do not. The situation is even more complicated: in several cases, we prove that inversion eigenvalues by themselves cannot uniquely determine the response. We can show that a nontrivial quantum spin Hall state and two copies of a Chern insulator each with Chern number unity have identical inversion eigenvalues but obviously represent very different states of matter. The question of relating \emph{ all} the inversion eigenvalues to a response function remains, in the cases where possible, still unsolved.

\section{Preliminaries}\label{sec:prelim}

Let us start our discussion with some observations for simple  two- and four-band model Hamiltonians, which will be referenced throughout the paper. These properties of the models that we discuss are only dependent on a generic inversion symmetry and not other symmetries present ``accidentally" in these simple models.

\begin{figure}
  \includegraphics[width=7cm]{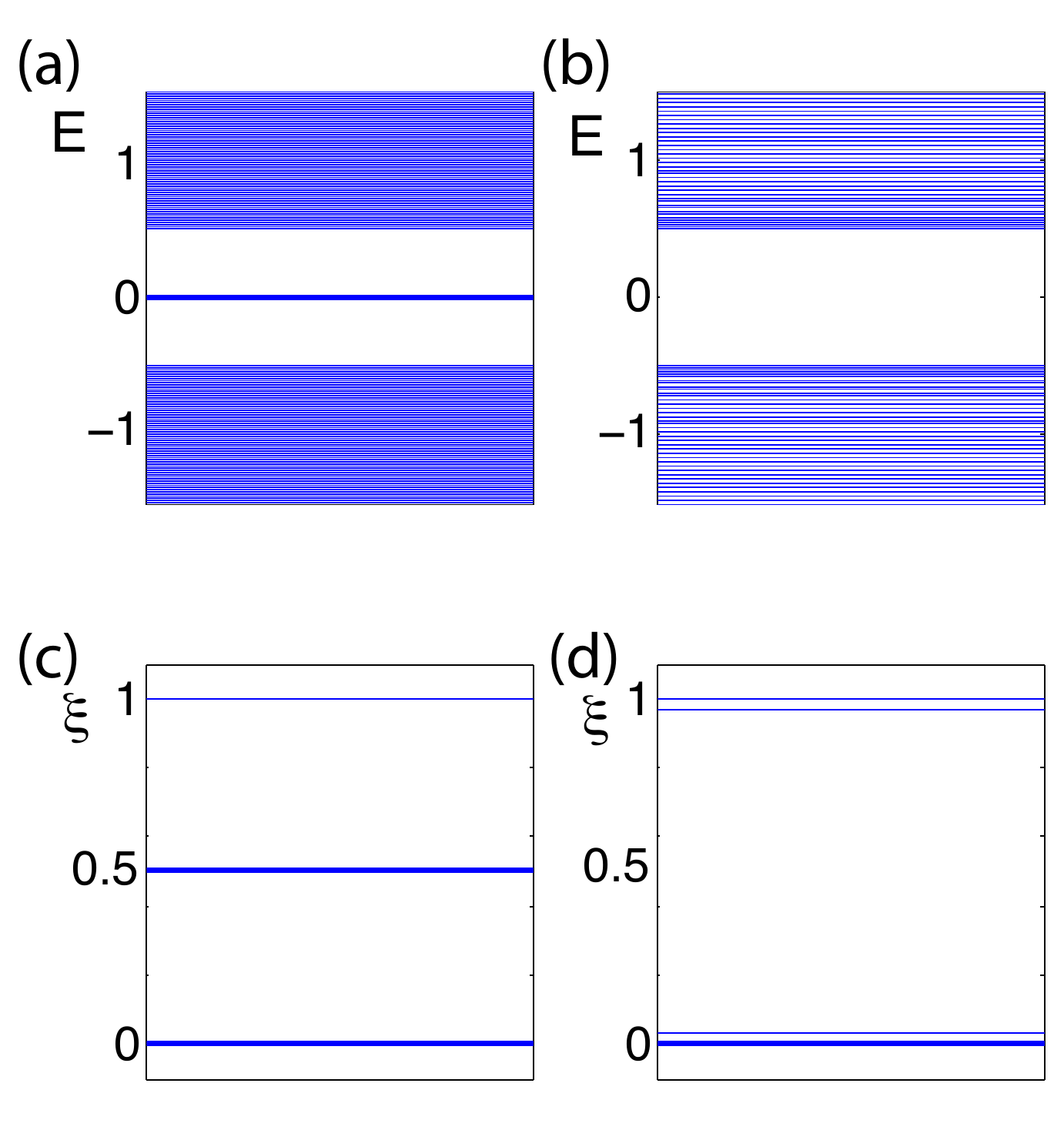}\\
  \caption{Energy Spectra for the simple 1D 2-band model with open boundary conditions for (a)$\alpha=0\;m=-1$ (non-trivial)(b)$\alpha=0\; m=1$ (trivial) Entanglement spectra for the two cases are shown in (c) and (d) respectively for a half-filled Fermi sea ground state with periodic boundary conditions.  }
 \label{fig:simple1d2band}
\end{figure}

{\it 2 band model.} The two band model is
\begin{equation}\label{H2}
\begin{array}{c}
H_2=\sum\limits_{x}\left[c_{x}^\dagger \frac{\alpha+\hat{\sigma}_3-i\hat{\sigma}_1}2c_{x+1}+h.c. +c_{x}^\dagger \left(1+m \right)\hat{\sigma}_3 c_{x}\right],
\end{array}\nonumber
\end{equation}
where $\alpha,m$ are two parameters and $\hat{\sigma}_a$ are Pauli's matrices. The model is symmetric under the inversion operation $c_x$$ \rightarrow$$ \hat{\sigma}_3 c_{-x}$. The Bloch representation of $H_2$ takes the simple form:
\begin{eqnarray}
\hat{H}_2(k)=\alpha \cos k+ \sin(k)\hat{\sigma}_1+(1+m-\cos k)\hat{\sigma}_3,\label{eq:simple1d2band}
\end{eqnarray} 
and the inversion is implemented by the operator ${\cal P}=\hat{\sigma}_3$: 
\begin{equation}
\hat{\sigma}_3 \hat{H}(k) \hat{\sigma}_3= \hat{H}(-k).
\end{equation} 
$H_2$ is gapped, except when $m$=$-2$ or 0.  The energy spectrum of the model with open boundary conditions is presented in Fig.~\ref{fig:simple1d2band}(a-b), for $\alpha$=0 and $m$$=$$\mp 1$, respectively. Throughout this paper, $\hat{P}_k$ will denote the projector onto the occupied bands at momentum $k$, which is a $K$$\times$$K$ matrix ($K$= total number of  bands), whose entries depend on $k$. 

The two special points $k_{\mbox{\tiny{inv}}}$=$0,\pi$ where $\hat{H}_2(k)$ is mapped onto itself by inversion will play a special role in the following discussion. We are going to examine the (non-zero) eigenvalues $\zeta(0)$ and $\zeta(\pi)$ of $\hat{P}_{k=0}{\cal P}\hat{P}_{k=0}$ and $\hat{P}_{k=\pi}{\cal P}\hat{P}_{k=\pi}$, respectively, for the 3 different insulating phases of $H_2$. What we find is the following: \smallskip

1. $\zeta (0)=\zeta (\pi)=-1$, for $m>0$.\smallskip

2. $\zeta (0)=\zeta (\pi)=+1$, for $m<-2$.\smallskip

3. $\zeta (0)=-\zeta (\pi)=+1$, $-2<m<0$.\smallskip

\noindent We can  form a $Z_2$ topological invariant:
\begin{equation}\label{eq:ChiP}
\chi_{{\cal P}}=\prod_{k_{\mbox{\tiny{inv}}},i\in\;occ.} \zeta_{i} (k_{\mbox{\tiny{inv}}}),
\end{equation}
which is topologically stable since one cannot change its value without closing the gap of the Hamiltonian. The expression of $\chi_{\cal P}$ is similar to invariants formed for time-reversal and inversion invariant topological insulators in 2D and 3D.\cite{fu2007a} 

For the model of Eq.~\ref{H2}, $\chi_{{\cal P}}$ takes the values $\chi_{\cal P}$=$+1$ for the insulating phases with $m$$\notin$$[-2,0]$, and $\chi_{\cal P}$=$-1$ for $m$$\in$$[-2,0]$. A direct calculation indicates that the phase with $\chi_{\cal P}$=$-1$ displays a single end mode at each end of a system with open boundaries, while the phases with $\chi_{{\cal{P}}}=+1$ do not display any end modes. This simple model indicates that, indeed, systems with inversion symmetry do posses non-trivial topological phases that, apparently, can be classified by the $\chi_{\cal P}$ invariant.  As we shall see in the following sections, $\chi_{{\cal{P}}}$ can be linked to a physical response of the system but it cannot completely classify the topological phases of a system with inversion symmetry, even in 1D.

{\it 4 band model.} It is instructive to  repeat a similar analysis on a 4-band inversion symmetric model. For this we use the following Hamiltonian, written directly in the Bloch representation:
\begin{equation}
\begin{array}{c}
\hat{H}_4(k)=\sin(k) \hat{\Gamma}_1 +\sin(k) \hat{\Gamma}_2 \medskip \\ 
+(1-m-\cos k)\hat{\Gamma}_0 +\delta \hat{\Gamma}_{24}+\epsilon \cos(k) (1+\hat{\Gamma}_0)
\end{array}\label{eq:simple1d4band}
\end{equation}
where $\hat{\Gamma}_1=\sigma^{z}\otimes\tau^x,\;\hat{\Gamma}_2=1\otimes\tau^y,\;\hat{\Gamma}_0=1\otimes\tau^z,$ and $\hat{\Gamma}_{24}=\sigma^{x}\otimes\tau^z.$  The Pauli's matrices $\tau^a,\sigma^a$ act in the orbital and spin spaces, respectively.  $\hat{H}_4(k)$ is symmetric under inversion, which is implemented by ${\cal P}=\hat{\Gamma}_{0}$, and is gapped except for a few  values of the parameters $\delta,\epsilon,$ and $m.$  Note that this system also has an accidental time-reversal symmetry with $T=(i\sigma^y\otimes 1)K$, but this can be broken without affecting the stability of the topological state or removing the mid-gap modes in the entanglement spectrum. The two lower energy bands  are assumed occupied, and in this case $\hat{P}_0{\cal P}\hat{P}_0$ and $\hat{P}_\pi{\cal P}\hat{P}_\pi$ are  4$\times$4 matrices, each displaying two non-zero eigenvalues $\zeta_i(0)$ and $\zeta_i(\pi)$, $i$=$1,2$. 
\begin{figure*}
  \includegraphics[width=14cm]{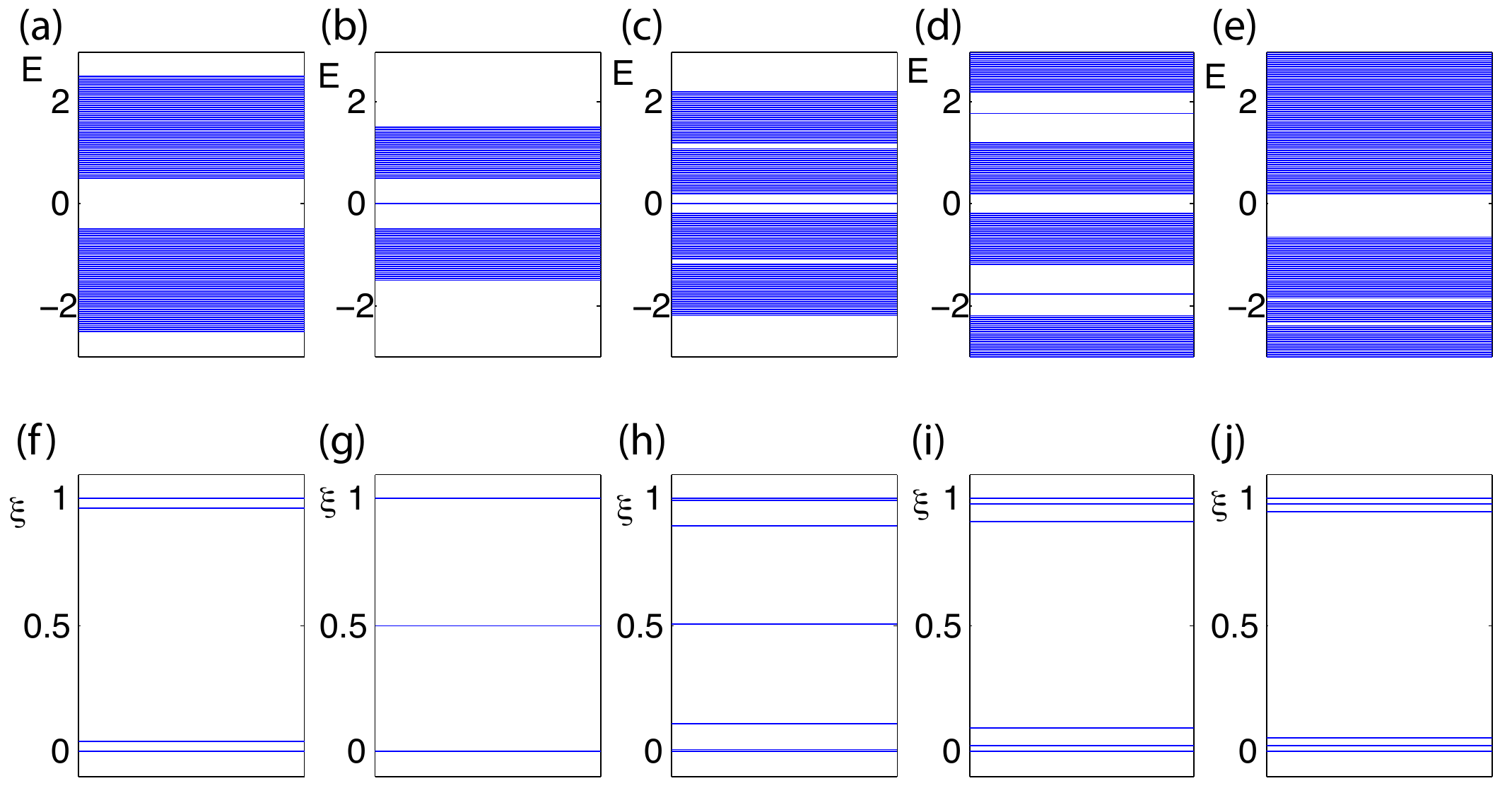}\\
  \caption{(a,b,c,d,e) Energy spectra for $H_4$ in cases 1,2,3,4,5 respectively with open boundary conditions. (f,g,h,i,j) Entanglement spectra for $H_4$ in cases 1,2,3,4,5 respectively with periodic boundary conditions. }
 \label{fig:4bandEnergyEspec}
\end{figure*}

We are going to present the inversion eigenvalues for the insulating phases of the model. There are six such phases (we discuss only five of them) and their energy spectra with open boundary conditions are shown in Fig.~\ref{fig:4bandEnergyEspec}(a-e). Choosing representative values for the parameters, we find:\smallskip

\noindent Case 1) $b=\delta=\epsilon=0$ and $M<0$:
\begin{equation}
\begin{array}{c}
\zeta_1(0)=\zeta_2(0)=+1 \smallskip \\
\zeta_1(\pi)=\zeta_2(\pi)=+1.
\end{array}\label{eq:4bandCase1}
\end{equation}
and consequently $\chi_{\cal P}=+1$.\smallskip  

\noindent Case 2) $b=\delta=\epsilon=0$ and $M=0.5$:
\begin{equation}
\begin{array}{c}
\zeta_1(0)=\zeta_2(0)=-1 \smallskip \\
\zeta_1(\pi)=\zeta_2(\pi)=+1.
\end{array}
\end{equation}
and consequently $\chi_{\cal P}=+1$. \smallskip

\noindent Case 3) $b=1$, $\delta=0.7$, $\epsilon=0$ and $M=<0$:
\begin{equation}
\begin{array}{c}
\zeta_1(0)=-1, \ \zeta_2(0)=+1 \medskip \\
\zeta_1(\pi)=\zeta_2(\pi)=+1.
\end{array}
\end{equation}
and consequently $\chi_{\cal P}=-1$.\smallskip
 
\noindent Case 4) $b=1$,$\delta=1.7$, $\epsilon=0$ and $M=0.5$:
\begin{equation}
\begin{array}{c}
\zeta_1(0)=-1, \ \zeta_2(0)=+1 \medskip \\
\zeta_1(\pi)=+1,\ \zeta_2(\pi)=-1.
\end{array}
\end{equation}
and consequently $\chi_{\cal P}=+1$.\smallskip
 
\noindent Case 5) $b=1$, $\delta=1.7$, $\epsilon=0.7$ and $M=0.5$:
\begin{equation}
\begin{array}{c}
\zeta_1(0)=+1,\ \zeta_2(0)=-1 \medskip \\
\zeta_1(\pi)=+1,\ \zeta_2(\pi)=-1.
\end{array}\label{eq:4bandCase5}
\end{equation}
and consequently $\chi_{\cal P}=+1$. \smallskip
 
The 4-band model reveals a far richer internal  structure. Case 1) can be identified with a trivial topological phase and, based on the value of $\chi_{\cal P}$ and on the presence of end modes seen in Fig.~\ref{fig:4bandEnergyEspec}(c), one will be inclined to classify case 3) as a non-trivial topological insulator. But one will have clear difficulties with labeling  cases 2), 4) and 5). This is a clear indicative that $\chi_{\cal P}$ alone is not enough for a full classification and that additional topological invariants are needed for a complete picture. 

It is instructive to consider the atomic limit of the model. By the atomic limit we mean the limit of the adiabatic process in which the hopping terms between {\it different} sites are tuned to zero. Since the bands are dispersion-less and completely local (disentangled) in this  limit, it makes sense to talk about the parity of an entire band (or orbital), since its inversion eigenvalues at $k=0,\pi$ are identical. For a model with 2 occupied bands, the atomic limit can lead to the following cases, depending of how the occupied atomic orbitals behave under inversion: two occupied bands of parity $+$ (labeled $++$), two occupied bands of parity $-$ (labeled $--$), and one band of parity $+$ and one band of parity $-$ (labeled $+-$).  These three options give the complete classification of the trivial inversion symmetric insulators with 2 occupied bands in 1D. Now, a direct calculation will show that Cases 1), 4) and 5) can be connected to their atomic limits without closing the insulating gap and that Case 1) can be identified with $++$ trivial insulator, while both Cases 4) and 5) can be identified with the $+-$ trivial insulator. Note that Cases $4$ and $5$ can be adiabatically connected to each other without closing the bulk insulating gap. The $--$ trivial insulator occurs in our 4-band model if we take the large $m$ limit. 

Based on the above discussion and on the absence/presence of the end modes in Fig.~\ref{fig:4bandEnergyEspec}, we can consider the Cases 1, 4, 5 as completely trivial and Case 3 as non-trivial, but Case 2 is still uncharacterized. It cannot be continued to the trivial atomic limit without closing the bulk gap, it displays end modes, yet  $\chi_{\cal P}=+1$. To distinguish this phase we must carefully consider the parity eigenvalues. We see that when the $k_{\mbox{\tiny{inv}}}$ points are considered separately, $\chi_{{\cal{P}}}(k_{\mbox{\tiny{inv}}})\equiv\prod_{i\in occ.}\zeta_{i}(k_{\mbox{\tiny{inv}}})=1.$ Thus, at each $k_{\mbox{\tiny{inv}}}$ there are an even number of bands with negative parity eigenvalues. In this situation, when for each $k_{\mbox{\tiny{inv}}}$ the local product over the eigenvalues of the occupied bands is trivial $(+1)$ we can define a second invariant
\begin{equation}
\begin{array}{c}
\chi^{(2)}_{{\cal{P}}}\equiv \prod_{k_{\mbox{\tiny{inv}}},i\in occ./2}\zeta_{i}(k_{\mbox{\tiny{inv}}}),
\end{array}
\end{equation} 
where the product over bands is defined to be the product of \emph{half} of the bands with negative parity eigenvalues at each $k_{\mbox{\tiny{inv}}}$. Note, that we do not require there to be an even number of filled bands, just an even number of negative parity eigenvalues. Out of the five cases discussed above, the $\chi^{(2)}_{{\cal{P}}}$ invariant can only be defined for cases 1 and 2 for which $\chi^{(2)}_{\cal P}$ is trivial/non-trivial, respectively. As we shall see, the $\chi^{(2)}_{\cal P}$ invariant is relevant and important for inversion symmetric insulators in 3D.  

The invariant $\chi$=$|n_1$$-$$n_2|$, where $n_1$ and $n_2$ are the number of negative parity eigenvalues at $k$=0 and $\pi$, respectively, is also useful for classifying the inversion symmetric insulators. This invariant generically indicates how many times the insulating gap must close when one takes the atomic limit. As we shall see, $\chi$ also gives the number of robust mid-gap modes in the entanglement spectrum localized on a single cut-boundary. For our 4-band model, $\chi$=0 for Cases 1), 4) and 5); $\chi$=1 for Case 3) and $\chi$=2 for Case 2). 

To conclude, the 2- and 4-band explicit models show that the systems with inversion symmetry can display topologically distinct phases, {\it i.e.} they cannot be continuously deformed into one another without closing the insulating gap. These phases can be characterized by the different values of $\chi$, $\chi_{P}$ or $\chi^{(2)}_{P}$ (when they can be consistently defined). In the remainder of the manuscript, we will try to elucidate the physical signatures of these topological phases and to give the physical meaning of the invariants mentioned above. For this, we will investigate the properties of the entanglement spectrum, which we demonstrate to contain robust edge modes, and the physical responses of the inversion symmetric insulators.


\section{Entanglement Spectrum of Topological Insulators With Inversion Symmetry}\label{sec:entanglement}

In this section we discuss the bi-partite single-particle entanglement spectra for inversion symmetric topological insulators.   Previous work on entanglement spectra in translationally invariant topological insulators was carried out in Refs.~\onlinecite{ryu2006,haldane2009,fidkowski2010,turner2009}, where it was shown that the primary contributions to entanglement arise from states localized near the spatial cut between regions A and B.  Additionally the entanglement spectrum of  disordered Chern insulators has been investigated in Ref.~\onlinecite{prodan2010}. The first indication that the presence of inversion symmetry is important for the structure of the entanglement spectrum was presented in Ref.~\onlinecite{turner2009}. Here it was shown that, while the physical edge spectrum of a time-reversal and inversion invariant topological insulator is gapped in the presence of an added Zeeman field (which does not close the bulk gap), the entanglement spectrum still contains a gapless mode.  The authors of that work link the existence of mid-gap states for each cut  in the entanglement spectrum with the existence of a $\theta=\pi$ vacuum characteristic of a TRI non-trivial topological insulator.\cite{qi2008B} Although, as we mentioned before (and will discuss more in Sec. \ref{sec:response}), there is not always a direct and unique connection between the physical response and protected states in the entanglement spectrum, this was an essential indication that inversion symmetry could support topological states and that the properties of the entanglement spectra were closely connected with inversion symmetry.

We start this section by  detailing how to obtain the entanglement spectrum for noninteracting insulators. We then look at the entanglement spectrum of topological insulators and show that there are two  fundamental properties which may be present (i) protected mid-gap states at entanglement eigenvalue $\xi=1/2$ and (ii) spectral flow in the entanglement spectrum.  In the presence of inversion symmetry, there can be mid-gap states in the entanglement spectrum, and these may or may not be connected to the entanglement bulk band edges via a spectral flow pattern.  Hence there are two distinct types of non-trivial entanglement spectra.  One example that we will see is  TRI  topological insulator parent states which are inversion symmetric but which may have time-reversal slightly broken. The time-reversal invariant case has both protected mid-gap modes and spectral flow, while the $T$-broken case only has protected mid-gap modes. 

For now we focus solely on insulators with a generic inversion symmetry and show its consequences on the entanglement spectrum. First, we show that if the system is cut exactly in half, then there can be mid-gap states in the entanglement spectrum  located exactly at a value of  $1/2$. This is equivalent to the statement that the mid-gap eigenvalues of the flat band Hamiltonian, for cuts exactly in half, exhibit no finite-size level repulsion.  We then give an expression for the number of $1/2$ eigenvalues in the entanglement spectrum as a function of the numbers of negative inversion eigenvalues at inversion symmetric points. Ref \onlinecite{turner2009} also points out the existence of multiple exact midgap states in the entanglement spectrum but does not relate them to the difference of the inversion eigenvalues \emph{between} inversion symmetric points.

 In everything presented  below, it is very important to clarify that by a spatial cut in a translationally invariant system we mean a cut \emph{between} primitive unit cells. This point is important when we consider systems with partially broken translation symmetry \emph{e.g.} the dimerized models in Sec. \ref{sec:examples}. The physics is independent of the choice of the unit cell. For example, in a multi-orbital system with a one-site unit cell, the cut should \emph{not} be made through the orbitals on the same site. 

\subsection{Obtaining the Entanglement Spectrum of an Insulator}

All of the models we study are free fermion Hamiltonians. To find the single-particle entanglement spectrum we use Peschel's method.\cite{Peschel2004} We begin by assuming a quadratic Hamiltonian of an insulator with $\alpha = 1... K$ quantum states per site which is translationally invariant:
\beq
H= \sum_k c_{\alpha,k}^\dagger H_{\alpha \beta} (k) c_{\beta, k}
\eneq and the canonical transformation $U$ that diagonalizes it:
\beq
U^\dagger H U = \mbox{Diag}(E_n).
\eneq $U$ is the matrix of eigenvectors $u^n(k)$ of energy $E_n(k)$:
\beq
U(k) = (u^1(k), u^2(k).... u^K(k))
\eneq where each $u^n(k)$ is a $K$-component vector. In general, we will use $k$ to denote the wave-vector of components $k_x$, $k_y$, etc.. 

The relationship between the normal mode operators  $\gamma_{\beta k}$ and the electron creation operators is
\beq
c_{\alpha  k} = U_{\alpha \beta}(k)  \gamma_{\beta k} = u^n_\alpha(k)  \gamma_{n k}
\eneq

To calculate the single-particle entanglement spectrum we simply need the correlation function:
\begin{equation}
C_{ij}^{\alpha \beta} = \langle c_{i \alpha}^\dagger c_{j \beta} \rangle,
\end{equation}
where $c_{i \alpha}^\dagger$ creates an electron in state $\alpha$ at site $i$. The expectation value is taken in the ground state. We can view this correlator as a matrix $\hat{C}_{ij}$, with entries that depend on $i$ and $j$. We have:
\begin{eqnarray}
C_{ij}^{\alpha \beta}&=& \sum_{k_1,k_2} e^{i k_1 i- i k_2 j} \langle c_{\alpha k_1}^\dagger c_{\beta k_2} \rangle\nonumber\\ & =& \sum_k e^{i k(i-j)} \sum_{n \in occ.} u^{n}_\alpha(k)^* u^n_\beta(k) \nonumber\\ &=&  \sum_k e^{i k(i-j)} P_k^{\alpha \beta},
\end{eqnarray}
or more compactly
\begin{equation}\label{CVP}
\hat{C}_{ij}= \sum_k e^{i k(i-j)} \hat{P}_k.
\end{equation}
 We want to make a translationally invariant cut along the $y$-direction so that $k_y$ is still a good quantum number ($k_y$ is a short-hand notation for all the momenta parallel to the cut, so we are implicitly also treating systems in $2,3$ dimensions). Thus we have:
\beq
\hat{C}_{ij}(k_y)    = \frac{1}{L} \sum_{k_x} e^{i k_x (i-j)} \hat{P}_{k_x,k_y},
\eneq  
 where $L$ is the total number of sites along the cut. Following Peschel, for the entanglement spectrum, we restrict  $i,j$ to be in region A, which is an explicit cut in position space. There are several physical choices for cuts, but for topological insulators we will show that a spatial cut can distinguish between topological and trivial insulators. 

As an aside, note that for an insulator the spectrally flattened Hamiltonian matrix where the states above/below the gap are flattened to energies $+1/2$ and $-1/2$ respectively is given by:
\beq
\begin{array}{c}
\hat{H}_{\text{flat}}(k_x,k_y) = \frac{1}{2} - \hat{P}_{k_x,k_y}.
\end{array}
\eneq
The two above expressions of the  correlation function and of the flat band Hamiltonian explicitly show that the entanglement spectrum, i.e. the eigenvalues of the restricted $C_{ij}^{\alpha \beta}$ are identical to the energy levels of the flat band Hamiltonian with open boundaries in region A shifted by a constant. As such, if the flat band Hamiltonian is topological (i.e. has protected edge states), then immediately we know the entanglement spectrum will have states localized on the cut. This agrees with the results of Ref.~\onlinecite{fidkowski2010}. The interesting thing is that the flattened Hamiltonian can have mid-gap states even when the un-flattened one does not. 
\subsection{Properties of the Entanglement Spectrum}

We would like to first get an intuitive idea of how the entanglement spectrum of an insulator should look. The one-body correlation function over the \emph{ full system} (not only over region A) is a projector. It is the real space representation of the projector onto the occupied bands, and as such only has the eigenvalues $0$ and $1$. This is shown explicitly in Appendix \ref{app:Gprojector}. When we make a cut, the eigenvalues of the one-body correlator deviate from $0,1$ but most of them only deviate slightly.  However, in the topologically non-trivial case, we must get entanglement ``edge" -modes similar to the edge states, but localized on the entanglement cut, because we are really diagonalizing the spectrum of the open boundary flat band Hamiltonian of a topological insulator.

From now on we choose to cut the system \emph{exactly} in half. A cut exactly in half will enable us to  show the existence of exact degeneracies rather than exponential degeneracies only arising in the thermodynamic limit. Our choice of cut does not matter in the thermodynamic limit, where it \emph{cannot} physically matter whether we make a cut exactly in the middle or away from the middle. It is however the case that if we cut the system in two identical halves, we can prove things exactly, otherwise we can just give arguments.
 
The one-body matrix, computed in the basis $c^\dagger_{i,\alpha}|0\rangle$, takes the block form 
 \beq
 C= \left(\begin{array}{cc} C_L & C_{LR} \\C_{RL} & C_R\end{array}\right)
\eneq where $C_L$ is the matrix of the left-half  (the one we diagonalize for the entanglement spectrum) $(C_L)_{ij} =C_{ij}^{\alpha \beta}$, $i,j \in A$;  $C_R$ is the matrix of the right-half $(C_R)_{ij} =C_{ij}^{\alpha \beta}$, $i,j \in B$;  $C_{LR}$ is the left-right coupling, $(C_{LR})_{ij} =C_{ij}^{\alpha \beta}$, $i \in A$, $j \in B$; with $C_{RL} = C_{LR}^\dagger$. Since $\hat{C}_{i,j}=\hat{C}_{i+n,j+n}$, the following extra property is true if the cut is exactly symmetric (in which case a proper translation of $A$ gives $B$):
\beq
C_R=C_L.
\eneq We also find that $C_{LR}=C^{\dagger}_{LR}.$
Moreover, the projector property $C^2=C$:
\beq
\left(\begin{array}{cc} C_L & C_{LR} \\C_{LR}^\dagger& C_L\end{array}\right)\left(\begin{array}{cc} C_L & C_{LR} \\C_{RL} & C_L\end{array}\right) = \left(\begin{array}{cc} C_L & C_{LR} \\ C_{LR}^\dagger & C_L\end{array}\right)
\eneq 
gives the following additional identities:
\begin{equation}
\begin{array}{l}
C_L(1- C_L) = C_{LR}^\dagger C_{LR},\medskip \\ 
C_{LR} C_{LR}^\dagger  = C_{LR}^\dagger C_{LR} \medskip \\
C_L C_{LR}+ C_{LR} C_L = C_{LR}.
\end{array}
\end{equation}
Using the last equation, if $\psi$ is an eigenstate of the entanglement spectrum matrix $C_L$ with eigenvalue (probability) $p$:
\beq
C_L \psi = p \psi
\eneq
then $C_{LR} \psi$ is also an eigenstate with eigenvalue $1-p$:
\begin{equation}
\begin{array}{c}
C_L C_{LR} \psi=  C_{LR} \psi -  C_{LR} C_L \psi \medskip \\
= C_{LR} \psi - p C_{LR} \psi = (1-p) C_{LR} \psi.
\end{array}
\end{equation} 
If $p=1/2$, $\psi$ and $C_{LR} \psi$ have the same 1/2 entanglement probability, but as we shall see, this does not automatically mean that the $p=1/2$ entanglement probability is doubly degenerate because  $\psi$ and $C_{LR} \psi$ are not linearly independent, in general.


\subsubsection{Properties of Entanglement spectrum with time-reversal symmetry}

The entanglement spectrum maintains the symmetries of the original Hamiltonian. For example, for time-reversal symmetry of the original Hamiltonian $T\hat{H}(k) T^{-1} = \hat{H}(-k)$ (equivalently, $T \hat{P}_k T^{-1} = \hat{P}_{-k}$):
\begin{eqnarray}
&T\hat{C}_{ij}(k_y) T^{-1} = \sum_{k_x} T e^{i k_x (i-j)} \hat{P}_{k_x, k_y} T^{-1} =  \nonumber \\ &= \sum_{k_x}  e^{- i k_x (i-j)} T \hat{P}_{k_x, k_y} T^{-1} =  \nonumber \\ &= \sum_{k_x}  e^{- i k_x (i-j)}  \hat{P}_{-k_x, -k_y}   =  \nonumber \\ &=   \sum_{k_x}  e^{ i k_x (i-j)}  \hat{P}_{k_x, -k_y} = \hat{C}_{ij}(-k_y)
\end{eqnarray}
\noindent 
so we see that the correlator also has  time-reversal symmetry, and for spin $1/2$ particles for which $T^2$$=$$-1$, the entanglement levels come in pairs at $k$ and $-k$. Thus, there are entanglement Kramers' doublets at time-reversal invariant points where $k \equiv -k \mod G$ where $G$ is a reciprocal lattice vector.

\subsection{Inversion Symmetric Topological Insulators}

In this section we give explicit arguments that the entanglement spectrum of an insulator with inversion symmetry (and without any other symmetry) can have mid-gap states pinned at \emph{exactly} $1/2$, \emph{without level repulsion} when cut exactly in half. An integer number of such modes is robust without splitting, so the classification of the entanglement spectra of insulators with inversion symmetry is given by an integer $Z$ (compare with the $Z_2$ case where an even number of modes would be unstable).  As an example, in 1D, if a bulk insulator has a number $n_1$ of filled bands with negative inversion eigenvalues at $k=0$ and a number $n_2$ at $k=\pi$, we give explicit arguments that the entanglement spectrum for a system with periodic boundary conditions (when the system is cut exactly in half) will have $2|n_1- n_2|$ protected mid-gap modes at exactly $1/2$. In more than 1D, there will be conserved  momenta parallel to the cut (say $k_y$), for an insulator cut in the $x$ direction. When cut  exactly in half, there will be $2|n_1-n_2|$ zero modes situated at the $K^1_y = - K^1_y \mod G_y$ for which, in the periodic bulk (before the cut), there were $n_1$ negative inversion eigenvalues at $(k_x, k_y)=( 0, K_y^1)$ and $n_2$ negative inversion eigenvalues at $(k_x, k_y)=( \pi, K_y^1)$.  We illustrate this explicitly with several examples in Sec. \ref{sec:examples}.

\subsubsection{Properties of Entanglement Spectrum With Inversion Symmetry}

With inversion symmetry:
\beq
{\cal P} \hat{H}(k) {\cal P}^{-1} = \hat{H}(-k), \ \  {\cal P}^{2} =1; \ \ {\cal P} = {\cal P}^{-1}
\eneq we can define a unitary matrix $B_{ij}(k)$ which connects the bands at $k$ and $-k$:
\beq
\ket{u_i(-k)} = B^\ast_{ij}(k) {\cal P} \ket{u_j (k) } 
\eneq  where the indices $i,j$ run over the occupied bands $1,..., N$. In fact, by performing simple band crossings between the $N$ bands below the gap (which does not influence the physics in the gap which depends only on the ground state), we can make the bands non-degenerate in which case we can use $B^{\ast}_{ij}(k)  = e^{i\phi(k)} \delta_{ij}$, but we do not need to choose this gauge here. Since $\hat{P}_k=\eta(E_F-\hat{H}(k))$, where $\eta(x)$ is the Heaviside function, we have: 
\beq
{\cal P} \hat{P}_k {\cal P}  = \hat{P}_{-k},
\eneq which can be used to show:
\beq
{\cal P} \hat{C}_{ij} {\cal P} = \sum_k e^{ik(i-j)} \hat{P}_{-k} = \hat{C}_{ji}  = \hat{C}_{ij}^\dagger .
\eneq


We now want to  relate the appearance of these $1/2$ eigenvalues with the inversion eigenvalues of the occupied bands.  
We first consider the one-dimensional case where we will be able to infer the behavior of the insulator just from $k_x=0, \pi.$ In principle only two sites in the $x$ direction should be enough to reveal the physics. Of course, with just two sites, our cut has to be made right in the middle of the two-site problem, \emph{i.e.} we are computing the entanglement spectrum of one site vs. the other site. This seems a bit problematic at first because if we are looking for the properties of the \emph{energy} spectrum in a topological insulator phase the wavefunctions of the states localized on each end will overlap and the degeneracy of these low-energy end states will be lifted because of the small size. Crucially, we show that the flat-band Hamiltonian does not have such finite-size eigenvalue repulsion between the edge modes \emph{even when these modes rest on top of each other on the same site}. That is, even if we bring the ends close to each other \emph{e.g.} on the same site (which is the meaning of the one-site entanglement spectrum), it is still true that the end modes do not exhibit level repulsion and are degenerate. This statement is true in higher dimensions where the end states become propagating edge and surface states. We prove this statement for several particular cases, which indicate it is true in the thermodynamic limit. We first show that for one occupied band (we do not particularize to a specific model), there are two mid-gap modes at exactly $1/2$ if the inversion eigenvalue  at $k=0, \pi$ are opposite. Then we repeat this procedure for a chain of four sites cut in half. We then show that for two occupied bands (we do not particularize to a specific model), there are two mid-gap modes at exactly $1/2$ if there is one  inversion eigenvalues  at $k=0, \pi$ opposite (while the other two are the same), whereas there are four mid-gap modes at exactly $1/2$ if both inversion eigenvalues at $k=0$ are opposite from the ones at $k=\pi$ (i.e. at one momentum both are negative and at the other momentum both are positive). We again do this for a chain of two sites cut in half, then for a chain of four sites cut in half. The main conclusion to be drawn from this is that in the flat-band Hamiltonian (entanglement spectrum), these mid-gap modes \emph{do not} experience eigenvalue repulsion. It is physically clear  that, although our proofs are only for two and four site flat band Hamiltonians (cut in half for the entanglement spectrum), level repulsion will \emph{not} set in  for larger systems: level repulsion between edge modes gets \emph{weaker} as the distance between them is increased. Finally, at the end, we look at the general case of $N$ occupied bands for the two-site problem and prove that the number of $1/2$ modes in the entanglement spectrum is $2|n_1-n_2|$  where $n_1, n_2$ are the number of negative eigenvalues at $k=0, \pi$. As there is no level repulsion when all modes are spatially on top of each other, we do not expect level repulsion when the number of sites is increased to the thermodynamic limit. We check this numerically for several examples with larger system sizes (\emph{e.g.} 100 sites).  Our exercise shows that time-reversal invariant insulators with inversion symmetry (or even the case with $T$ slightly broken) are not the only inversion symmetric topological insulators with protected entanglement mid-gap states. These are but one of a whole series of inversion symmetric insulators with mid-gap entanglement modes.

\subsubsection{One Occupied Band, Two-Site Problem}

First we look at a generic case with one occupied band, two-sites, and periodic boundary conditions.  In this case, $k$-space contains only the points $k$=$0, \pi$. The wavefunction of the occupied band is $\ket{\psi_1(k)}$:

\beq
\hat{H}(k) \psi_1(k) = \varepsilon(k) \psi_1(k)
\eneq
with inversion eigenvalues:
\beq
{\cal P} \ket{\psi_1(0) } = \zeta(0) \ket{\psi_1(0)}, \  {\cal P} \ket{\psi_1(\pi) } = \zeta (\pi) \ket{\psi_1(\pi)}.
\eneq  Since ${\cal P}$ is a unitary operator which squares to unity, we have: ${\cal P}^\dagger = {\cal P}(= {\cal P}^{-1})$ and by taking scalar products in the above we have:
\beq
(\zeta(0) - \zeta(\pi)) \langle \psi_1(0) |\psi_1(\pi) \rangle = 0
\eneq Hence if $\zeta(0) = - \zeta(\pi)$ (the eigenvalues can never be zero due to $\det {\cal P}=1$) we have  $ \langle \psi_1(0) |\psi_1(\pi) \rangle = 0$.  Notice that the Hamiltonian $\hat{H}(k)$ does not impose any restrictions on the wavefunctions at different momenta $k$,  \emph{i.e.}  at $k$=0 and $k$=$\pi$ we are effectively diagonalizing independent Hamiltonians. What allows us to relate wavefunctions at $k=0,\pi$ is that they are both eigenstates of the same matrix ${\cal P}$ (it is important to recall that ${\cal P}$ is $k$-independent). 


For the two-site problem ($i=1,2$):
\beq
\begin{array}{l}
\hat{C}_L =\hat{C}_{11}= \frac{1}{L} \sum_k \hat{P}_{k}
= \frac{1}{2}(\hat{P}_0 + \hat{P}_\pi).
\end{array}
\eneq
The eigenstates of the original Hamiltonian have opposite inversion eigenvalues then per the above:
\beq
\hat{P}_0\ket{  \psi_1(\pi)} = \hat{P}_\pi\ket{ \psi_1(0)} =0
\eneq which means that $\psi_1(0)$ and $\psi_1(\pi)$, the original Hamiltonian eigenstates,  are also the eigenstates of the entanglement spectrum, with two eigenvalues at $1/2$:
\beq
\begin{array}{l}
C_L \ket{  \psi_1(0)}  = \frac{1}{2} \ket{  \psi_1(0)} ; \ C_L \ket{  \psi_1(\pi)}  = \frac{1}{2} \ket{  \psi_1(\pi)}.
\end{array}
\eneq
We see that the original Hamiltonian can change, leading to a change of $\psi_1(k)$, but as long as the inversion eigenvalues remain fixed and opposite to each other, and as long as we can take the flat-band limit (both of which mean no gap closing), the  eigenvalues of the entanglement spectrum will be fixed at $1/2.$ It does not matter what the actual explicit model for $\hat{H}(k)$ is.  If the inversion eigenvalues at $k=0, \pi$ are not opposite, there is no reason why   $ \langle \psi_1(0) |\psi_1(\pi) \rangle = 0$, and the $1/2$ modes might not exist or will not be protected.

\subsubsection{One Occupied Band, Four-Site Problem}\label{sec:1band4site}

With one occupied band with opposite inversion eigenvalues, the two-site Hamiltonian has exact $1/2$ modes in the  entanglement spectrum. This is the first indication that the modes are stable and experience zero level repulsion. We now show that the generic four-site Hamiltonian, with  one occupied band also has exact $1/2$ modes without level repulsion. This strongly suggests that these modes are stable in the thermodynamic limit, as long as the entanglement spectrum  is computed for a system cut exactly in half.  But, in the thermodynamic limit we know there can be no physical difference between a cut in half and any other cut except for exponentially suppressed finite-size level splittings. Thus there will be asymptotic zero modes in the thermodynamic limit regardless of the cut.

For the four-site problem,  $k$-space contains four momenta $k_j =  \frac{j \pi}{2}$ , $j=0,1,2,3$.  Call the occupied eigenstate of the Hamiltonian, as above $\ket{\psi_1(k)}$, $\hat{H}(k) \psi_1(k) = \varepsilon(k) \psi_1(k)$. For a system cut in half, the entanglement spectrum is given by diagonalizing the matrix $C_{ij}= C_{i-j}$ with $i,j =1,2$:
\beq
C_L= \left( {\begin{array}{cc}
 \hat{C}_{11} & \hat{C}_{12}  \\
 \hat{C}_{12}^\dagger & \hat{C}_{22}   \\
 \end{array} } \right)
\eneq
We have:
\beq
 \hat{C}_{11}=\hat{C}_{22} = \frac{1}{4} (\hat{P}_0 + \hat{P}_\pi + \hat{P}_{\frac{\pi}{2}} +\hat{P}_{\frac{3 \pi}{2}})\equiv \hat{C}_0
\eneq
\beq
\hat{C}_{12} = \frac{1}{4} (\hat{P}_0 - \hat{P}_\pi  +i \hat{P}_{\frac{\pi}{2}} - i \hat{P}_{\frac{3 \pi}{2}})\equiv \hat{C}_1.
\eneq  
The eigenstate of $C_L$ corresponding to the eigenvalue $\xi$ takes the form $(\psi_A, \psi_B)$, which satisfy the equation:
\beq
\hat{C}_0 \psi_A+ \hat{C}_1 \psi_B = \xi \psi_A,\;\;\; \hat{C}_1^\dagger \psi_A + \hat{C}_0 \psi_B = \xi \psi_B \label{oneoccbandfoursiteproblem}
\eneq 
Due to the presence of inversion symmetry, irrespective of the inversion eigenvalues, we showed before that ${\cal{P}} \hat{C}_{ij} {\cal{P}} = \hat{C}_{ji} = \hat{C}_{ij}^\dagger$, which renders the second equation of Eq.~\ref{oneoccbandfoursiteproblem}:
\beq
\hat{C}_1 {\cal{P}} \psi_A + \hat{C}_0 {\cal{P}} \psi_B = \xi {\cal{P}} \psi_B.
\eneq 
We see that this is consistent with the first equation of Eq.~\ref{oneoccbandfoursiteproblem} if  $\psi_B = m {\cal{P}} \psi_A$, with $m^2=1$. The eigenvalue equation to  solve is then:
\beq
(\hat{C}_0 + m \hat{C}_1 {\cal{P}}) \psi_A = \xi \psi_A
\eneq Also because of inversion symmetry, we have that:
\beq
\hat{P}_{\frac{3 \pi}{2}} = {\cal{P}} \hat{P}_{\frac{\pi}{2}} {\cal{P}}.
\eneq For the one-band problem, we know that:
\beq
\hat{P}_0 {\cal{P}} = \zeta (0) \hat{P}_0;\;\;\;\;\; \hat{P}_\pi {\cal{P}} = \zeta (\pi) \hat{P}_\pi
\eneq where $\zeta(0),\zeta (\pi)$ are the inversion eigenvalues at $k=0, \pi$ of the occupied band $\psi_1(k)$.  Hence to find the entanglement spectrum we need to diagonalize the following operator:
\begin{eqnarray}
\hat{F}&=&\frac{1}{4}[(1+ m \zeta(0)) \hat{P}_0 + (1- m \zeta (\pi)) \hat{P}_\pi \nonumber\\
&+& \hat{P}_{\frac{\pi}{2}} {\cal{P}}({\cal{P}} + i m) + {\cal{P}} \hat{P}_{\frac{\pi}{2}} ({\cal{P}}- im)]. \label{oneoccbandfoursiteproblemdiagonalization}
\end{eqnarray}

For the half-mode, we pick an ansatz:
\beq
\psi_A= a \psi_1(0) + b \psi_1(\pi),
\eneq which we show can diagonalize $\hat{F}$ for an appropriate choice of $a, b$:
\beq
\begin{array}{l}
a= -(1+ i m \zeta (\pi)) \langle \psi_1 ({\frac{\pi}{2}}) |\psi_1(\pi)\rangle \medskip \\
b=  (1+ i m\zeta(0)) \langle \psi_1 ({\frac{\pi}{2}}) |\psi_1(0)\rangle .
\end{array}
\eneq 
This choice of $a$ and $b$ makes  $(\hat{C}_0 + m \hat{C}_1 {\cal{P}})\psi_A$ independent of both  $\psi_1 (\frac{\pi}{2}) $, and ${\cal{P}} \psi_1 (\frac{\pi}{2})$ in general  (i.e. $[ \hat{P}_{\frac{\pi}{2}} {\cal{P}}({\cal{P}} + i m) + {\cal{P}}  \hat{P}_{\frac{\pi}{2}} ({\cal{P}}- im)]  \psi_A=0$), as it should in order for our ansatz to be an eigenstate. With this choice of $a, b$ we find (by taking $\frac{1}{4}[(1+ m \zeta(0)) \hat{P}_0 + (1- m \zeta(\pi)) \hat{P}_\pi] \psi_A$) that in general the entanglement spectrum eigenvalue is dependent on $\langle \psi_1(0) |\psi_1(\pi)\rangle$, and hence the mode is not fixed at $1/2$. However, if the inversion eigenvalues at $k=0, \pi$ are opposite $\zeta(\pi) = - \zeta(0)$ then  $\langle \psi_1(0) |\psi_1(\pi)\rangle=0$ and we find the eigenvalue of the entanglement spectrum of our ansatz to be:
\beq
\xi = \frac{1}{4} (1+ m \zeta(0))
\eneq 
 Recall that have the liberty to choose the values of $m=\pm1,$ which is equivalent to saying  $m=\pm\zeta(0)$. If we pick $m= \zeta(0)$, then our ansatz gives an eigenstate with entanglement eigenvalue equal to exactly  $1/2$. The other choice leads to $\xi=0$, so is of no interest to us.  The eigenstate at $1/2$ is $(\psi_A, \epsilon_0 {\cal{P}} \psi_A)$, where:
\beq
\psi_A =  i \langle \psi_1 ({\frac{\pi}{2}}) |\psi_1(\pi)\rangle  \psi_1(0)    + \langle \psi_1 ({\frac{\pi}{2}}) |\psi_1(0)\rangle   \psi_1(\pi) 
\eneq 
 In the pathological case when $\langle \psi_1(\frac{\pi}{2} ) |\psi_1(\pi)\rangle= \langle \psi_1(\frac{\pi}{2} ) |\psi_1(0)\rangle=0$, both $\psi_1(0), \psi_1(\pi)$ are $1/2$ modes, but in general only their combination $\psi_A$ gives a robust $1/2$ mode. 

A second stable $1/2$ eigenvalue can be found by picking $m=\zeta(\pi) = - \zeta(0)$, but using a different ansatz: 
\begin{equation}
\begin{array}{l}
\psi'_A= a' \psi_1(\frac{\pi}{2}) + b' {\cal P} \psi_1(\frac{\pi}{2}).
\end{array}
\end{equation} 
With this choice for $m$, the $\hat{F}$ matrix to be diagonalized is:
\beq
\frac{1}{4}[\hat{P}_{\frac{\pi}{2}} {\cal{P}}({\cal{P}} - i\zeta(0) ) + {\cal{P}}  \hat{P}_{\frac{\pi}{2}} ({\cal{P}}+ i\zeta(0))]. 
\eneq
After straightforward calculations, we find the exact $1/2$ mode to be:
\beq
\begin{array}{l}
\psi'_A=  \psi_1(\frac{\pi}{2}) + i\zeta(0) {\cal{P}} \psi_1(\frac{\pi}{2}).
\end{array}
\eneq 
Thus, for a completely generic Hamiltonian and its eigenstates  we have shown that for one occupied band, if  the inversion eigenvalues at $0,\pi$ in the bulk are the opposite of each other, there are two exact midgap states at $1/2$ in the entanglement spectrum. These mid-gap states have linearly independent eigenstates:  $(\psi_A, \zeta(0) {\cal{P}} \psi_A),$  $(\psi'_A, -\zeta(0) {\cal{P}} \psi'_A)$ and $\psi_A, \psi'_A$ as above.  The standard intuition about interface or boundary states is that the  eigenvalues repel less as the  length of the system is increased, and  the entanglement spectrum for the case analyzed here will have two exact $1/2$ modes in the thermodynamic limit, when cut exactly in half. Numerical simulations on specific models agree. In the thermodynamic limit, there can be no physical difference between a half cut and a cut away from  half, and the levels will asymptote to $1/2$ in the infinite size limit if the cut is not exactly in half. 

\subsubsection{Two Occupied bands, Two-Site Problem}

On our way to the most general case we now analyze the two-site, two occupied band problem. This  follows in the same fashion as the previous example, except that we now have an extra complication. Namely, there are more options for the sets of  occupied-band inversion eigenvalues. 

We have the two occupied bands of the original Hamiltonian:
\beq
\hat{H}(k) \psi_1(k) = \epsilon_1(k) \psi_1(k);  \hat{H}(k) \psi_2(k) = \epsilon_2(k) \psi_2(k)
\eneq For $k=0, \pi$  we generically have:
\beq
\langle \psi_1(0) |\psi_2(0)\rangle =\langle \psi_1(\pi) |\psi_2(\pi)\rangle =0.
\eneq
We denote the inversion eigenvalues for $\psi_1(0)$, $\psi_2(0)$, $\psi_1(\pi)$, $\psi_2(\pi)$ by $\zeta_1 (0),\zeta_2 (0),\zeta_1(\pi),\zeta_2(\pi) $ respectively. The order of the occupied inversion eigenvalues at each inversion symmetric momentum can be changed without affecting the topological structure so we assume that all negative inversion eigenvalues are listed first.  We now calculate the number of midgap $1/2$ eigenvalues in the entanglement spectrum. If the number of negative inversion eigenvalues at $k=0$ is the same as the number of negative inversion eigenvalues at $\pi$ (which means the number of positive inversion eigenvalues is also the same), it is easy to prove that in general there are no protected $1/2$ modes because the entanglement eigenvalues depend on the overlap of bands at the two inversion symmetric momenta (cf. Eq \ref{eq:Eoverlap}). If the number of negative eigenvalues at the inversion symmetric points is different, then we distinguish two cases:

\textbf{Case 1:}
The number of negative eigenvalues at $k=0$ differs from the number of negative eigenvalues at $k=\pi$ by $\pm 2$ (\emph{i.e.} they are both different):
\beq
\zeta_1(0) \zeta_1(\pi) =\zeta_1(0) \zeta_2(\pi) = \zeta_2(0) \zeta_1(\pi) =\zeta_2(0) \zeta_2(\pi) =-1
 \eneq implies:
 \begin{eqnarray}
&\langle \psi_1(0) |\psi_1(\pi)\rangle =\langle \psi_1(0) |\psi_2(\pi)\rangle =0 \nonumber \\ &\langle \psi_2(0) |\psi_1(\pi)\rangle =\langle \psi_2(0) |\psi_2(\pi)\rangle =0.
\end{eqnarray}
 
 The one-site entanglement spectrum obtained by cutting the system in half is obtained by diagonalizing the operator $C_L = (\hat{P}_0 + \hat{P}_\pi)/2$ where $\hat{P}_0= \sum_{i=1}^2 \ket{\psi_i(0)} \bra{\psi_i(0)}$,  $\hat{P}_\pi= \sum_{i=1}^2 \ket{\psi_i(\pi)} \bra{\psi_i(\pi)}$. Due to their inversion eigenvalues, eigenstates at $\pi$ have zero eigenvalue under the projector at $0$ (and vice-versa) but unit eigenvalue under the projector at $\pi.$ We see that the modes at $1/2$ in the entanglement spectrum are given by  exactly the occupied eigenstates of the original Hamiltonian  $\psi_1(0)$, $\psi_2(0)$, $\psi_1(\pi)$, $\psi_2(\pi)$. There are exactly $4$ of them, twice the difference  between negative and positive eigenvalues at the two inversion symmetric points. 

\textbf{Case 2}
The number of negative eigenvalues at $k=0$ differs from the number of negative eigenvalues at $k=\pi$ by $\pm 1$: this implies that at one $k$ point, both inversion eigenvalues are identical. Without loss of generality, let this point be $k=\pi$ and let the eigenvalue products be:
\begin{eqnarray}
& \zeta_1(0) \zeta_1(\pi) =\zeta_1(0) \zeta_2(\pi) =-1 \nonumber \\ & \zeta_2(0) \zeta_1(\pi) =\zeta_2(0) \zeta_2(\pi) =1
 \end{eqnarray}  which renders the following inner products to be zero:
 \beq
 \langle \psi_1(0)\vert \psi_2(0)\rangle=\langle \psi_1(0) |\psi_1(\pi)\rangle =\langle \psi_1(0) |\psi_2(\pi)\rangle =0.\nonumber
 \eneq   Consider the eigenvalue problem:
 \beq
 \frac{1}{2} (\hat{P}_0 + \hat{P}_\pi) \psi_A = \alpha \psi_A.
 \eneq  We expand the state $\psi_A$ into the (non-orthogonal) set of eigenstates  $\psi_1(0)$, $\psi_2(0)$, $\psi_1(\pi)$, $\psi_2(\pi)$:
 \beq
\ket{ \psi_A} = a_1 \ket{\psi_1(0)}+ a_2 \ket{ \psi_2(0)} + b_1 \ket{\psi_1(\pi)} + b_2 \ket{\psi_2(\pi)} 
 \eneq As $\psi_1(0)$ is orthogonal with all the other eigenstates at both $k=0,\pi$ since it has a different inversion eigenvalue, it is then obvious to see that the first $1/2$ mode solution is $(a_1, a_2, b_1, b_2) = (1,0,0,0)$. To find another $1/2$ mode, we must expand in the three remaining eigenstates: $ \ket{ \psi_A} =a_2 \ket{ \psi_2(0)} + b_1 \ket{\psi_1(\pi)} + b_2 \ket{\psi_2(\pi)}  $. There is a slight complication with this expansion since nothing guarantees that the states  $\ket{ \psi_2(0)} , \ket{\psi_1(\pi)}, \ket{\psi_2(\pi)} $ are orthogonal: in fact, in the generic case, they are not. Moreover, it is not clear that they are even linearly independent. We will assume that the states are linearly independent. This is a perfectly valid procedure since if the   $\ket{ \psi_2(0)} , \ket{\psi_1(\pi)}, \ket{\psi_2(\pi)} $ are not independent, we will simply get a non-trivial nullspace. However, the non-zero eigenvalues are still good eigenvalues of the entanglement matrix. The matrix to diagonalize is:
 \beq
 \left( {\begin{array}{ccc}
 \frac{1}{2} &  \langle \psi_2(0) |\psi_1(\pi) \rangle  &  \langle \psi_2(0) |\psi_2(\pi) \rangle  \\
  \langle \psi_2(0) |\psi_1(\pi) \rangle^\ast & \frac{1}{2} & 0\\
  \langle \psi_2(0) |\psi_2(\pi) \rangle^\ast &0 & \frac{1}{2}
 \end{array} } \right)\label{eq:Eoverlap}
\eneq with an obvious $1/2$ eigenvalue  for the state $(a_2, b_1, b_2) = (0,- \langle \psi_2(0) |\psi_2(\pi) \rangle, \langle \psi_2(0) |\psi_1(\pi) \rangle  )$ ($a_1=0$). In the nongeneric case when $\langle \psi_2(0) |\psi_2(\pi) \rangle = \langle \psi_2(0) |\psi_1(\pi) \rangle =0$, $\psi_2(0)$ is the other $1/2$ eigenvalue. 
We have hence proved the existence of  two  exact $1/2$ eigenvalues for the two site problem, cut in half, when the difference between the number of negative inversion eigenvalues at $0,\pi$ is $\pm 1$.

As in the one-band case, this argument can be extended analytically to a system with four sites as shown in Appendix \ref{app:foursite} indicating that  the conclusions hold for chains longer than two-sites.

\subsubsection{$N$ Occupied bands, Two-Site Problem}
We now show that the two-site  problem with $n_1$ negative inversion eigenvalues at $k=0$ and $n_2$ negative inversion eigenvalues at $k=\pi$  with a total number $N$ of occupied bands contains $2|n_1- n_2|$ zero modes in the entanglement spectrum when a real-space cut is made on a system with periodic boundary conditions (\emph{i.e.} there are two cuts). The simplest case, which should be obvious from our previous examples, is that  all the  $N$ inversion eigenvalues at $k=0$ are  identical and are the opposite of the $N$ eigenvalues at $k=\pi$. In this case, the projector at  one of the inversion symmetric $k$'s annihilates all the eigenstates at the other inversion symmetric $k$, and the $2N$ occupied eigenstates of the original two-site Hamiltonian are  also the eigenstates of the entanglement spectrum at fixed eigenvalue $1/2$. Due to their orthogonality, they are linearly independent. From here it is clear that our formula is  correct for this case.

Now we will prove the more general formula. Let $n_1$ and $n_2$ be the number of eigenvectors  for the $-1$ eigenvalue of $\hat{P}_0 {\cal P} \hat{P}_0$ and $\hat{P}_\pi {\cal P}\hat{P}_\pi$, respectively, and assume $n_1 > n_2$. Recall that $K$ is the number of orbitals per site, so ${\cal P}$ is a $K$$\times$$K$ matrix acting on ${\bm C}^K$. ${\cal P}$ has $\pm 1$ eigenvalues and we denote the invariant subspaces corresponding to the positive/negative eigenvalue by ${\cal H}_\pm$ (${\cal H}_- + {\cal H}_+={\bm C}^K$).  

Now the subspaces $\hat{P}_0 {\bm C}^K$ and $\hat{P}_\pi {\bm C}^K$ are  invariant   under the inversion operation ${\cal P}$, and $\hat{P}_0 {\bm C}^K \cap {\cal H}_-$ is precisely the subspace spanned by the $n_1$ eigenvectors of $\hat{P}_0 {\cal P} \hat{P}_0$ corresponding to its negative eigenvalue. Similarly, $\hat{P}_\pi {\bm C}^K \cap {\cal H}_-$ is precisely the subspace spanned by the $n_2$ eigenvectors of $\hat{P}_\pi {\cal P}\hat{P}_\pi$ corresponding to its negative eigenvalue. Since $\dim [\hat{P}_0 {\bm C}^K \cap {\cal H}_-]=n_1$ and $\dim [\hat{P}_\pi {\bm C}^K \cap {\cal H}_-]=n_2$, with $n_1>n_2$, we can always find $n_1-n_2$ vectors $\Psi_n$ in $\hat{P}_0 {\bm C}^K \cap {\cal H}_-$ that are orthogonal to any vector in $\hat{P}_\pi {\bm C}^K \cap {\cal H}_-$. Since these vectors are in ${\cal H}_-$, they are also orthogonal to any vector in $\hat{P}_\pi {\bm C}^K \cap {\cal H}_+$. In other words, $\Psi_n$'s are $n_1-n_2$ vectors in $\hat{P}_0 {\bm C}^K$ perpendicular to all the vectors in $\hat{P}_\pi {\bm C}^K$. Consequently:
\begin{equation}
C_L\Psi_n = \frac{1}{2}(\hat{P}_0+\hat{P}_\pi)\Psi_n = \frac{1}{2}\Psi_n,
\end{equation}
for all $n_1-n_2$ vectors $\Psi_n$. Following the same arguments, we can find $n_1-n_2$ vectors in $\hat{P}_\pi {\bm C}^K \cap {\cal H}_+$ that are orthogonal to $\hat{P}_0 {\bm C}^K$, and consequently another set of $n_1-n_2$ eigenvectors with eigenvalue $\frac{1}{2}$. In total, there are $2(n_1-n_2)$ robust modes of $C_L$ at $1/2$. For a more explicit proof see Appendix \ref{app:manybands}.

\subsubsection{Extension to higher dimensions}
The extension to higher dimensions is just a matter of reinserting the extra momenta which are conserved in the presence of the cut. As an example let us consider 2D with an entanglement cut parallel to the y-axis so that $k_y$ is a conserved quantum number.  The exact mid-gap modes in the entanglement spectrum will exist only at inversion symmetric points in the momentum parallel to the cut, i.e. $k_y=0, \pi$. However, since  bands become continuous when $k_y$ is finely discretized,  in the thermodynamic limit the existence of $1/2$ modes at these two discrete k-points implies the existence of mid-gap bands. The number of bands  will be equal to the number of $1/2$ modes. These bands typically disperse away from $1/2$ but do not have to connect with the ``bulk" entanglement bands at entanglement eigenvalues close to $0,1$  (in the special cases of Chern insulators and time-reversal symmetric non-trivial insulators, they do connect and have spectral flow, as pointed out previously). However, since the modes at $1/2$ are robust upon changes in the original Hamiltonian that do not close the band gap, the mid-gap bands cannot be entirely pushed to the entanglement bulk band edges  at $0$ or $1.$ This means the system is a non-trivial insulator (\emph{i.e.} $S_{ent}$ cannot be made to vanish) if the number of negative (or positive) inversion eigenvalues is different between inversion symmetric points.    We would like to know at which inversion symmetric $k_y$  the exact $1/2$ modes will appear occur. The answer is simple: there will be exactly $2|n_1- n_2|$  $1/2$ modes  at  $k_y=0$  when the number of negative inversion eigenvalues at $(k_x, k_y)= (0,0) $  differs by $|n_1 - n_2|$ from the number of negative inversion eigenvalues at  $(k_x, k_y)= (\pi, 0) .$ Additionally  there will be exactly $2|n_{1}'- n_{2}'|$  $1/2$ modes  at  $k_y=\pi$  when the number of negative inversion eigenvalues at $(k_x, k_y)= (0,\pi) $  differs by $|n_{1}' - n_{2}'|$ from the number of negative inversion eigenvalues at  $(k_x, k_y)= (\pi, \pi) .$ The generalization to higher dimensions is the trivial extension of this. 

\subsection{Spectral Flow in the Entanglement Spectrum}\label{subsec:flow}
As mentioned earlier in this section the two important features of the entanglement spectra of inversion symmetric insulators are protected mid-gap modes, and spectral flow. What we mean by spectral flow is a continuous connection between the valence and conduction bulk entanglement bands through the entanglement edge states (an example is seen in Fig. \ref{fig:ChernInsulator}h,i ).  For a TRI topological insulator in 2D and 3D, or for a Chern insulator in 2D, the entanglement spectrum mirrors the energy spectrum of the open-boundary Hamiltonian. In fact, we have already shown an explicit map between the entanglement spectrum and the energy spectrum of the open boundary spectrally flattened Hamiltonian. \cite{fidkowski2010} This implies that if there is spectral flow between the conduction and valence bands in the energy spectrum then such a flow exists in the entanglement spectrum. In fact, this is the only case where there is true spectral flow in the entanglement spectrum. Out of the entire set of inversion invariant topological insulators only a small subset have spectral flow. Instead most non-trivial systems simply exhibit protected mid-gap states (or bands) but these do not continuously interpolate between the bulk entanglement bands. 

There is a nice example which illustrates this dichotomy. Let us consider the 3D strong topological insulator with both inversion and time-reversal symmetries. If we preserve ${\cal{P}}$ but break T,  spectral flow generically disappears from the energy spectrum because gaps are opened in the surface state spectrum. The degeneracies which existed in the entanglement spectrum at the time-reversal invariant momenta when T is preserved, are almost all broken, with the exception of the protected degeneracy for states at $\xi=1/2.$  This degeneracy splitting breaks the spectral flow in the entanglement spectrum, and opens gaps at the TR-invariant momenta as shown for a specific model in Fig.~\ref{fig:QSHInsulator}h,i.  As such, the entanglement spectrum is capable of distinguishing the subtle difference between topological insulators with T and ${\cal{P}}$ symmetry from topological insulators with only ${\cal{P}}$ symmetry.

\section{The linear response}\label{sec:response}
To date, some of the most spectacular features of topological insulators are their responses to external fields. The Chern insulators exhibit a quantized Hall effect and the 3D TRI topological insulators exhibit a topological magneto-electric effect. The topological invariants which distinguish these states from trivial insulators are directly connected with the corresponding response coefficient. In fact, a whole ladder of topological responses was uncovered in Ref. \onlinecite{qi2008B}. With this precedent one would hope that the inversion invariant topological insulators would also exhibit some type of defining physical response. However, this turns out to be true in only a limited set of the inversion invariant topological insulators. The situation is quite varied (remember we only have generic access to the information held in the inversion eigenvalues): some insulators have unique well defined topological responses, some systems can exhibit one of several allowed topological responses, and for others it is unclear if there is any topological response at all. We will see examples of all three cases in Sec. \ref{sec:examples}. In this section though we focus on the first case where insulators do exhibit a unique response which is the most interesting physical case. We discuss responses  in 1,2, and 3 dimensions and then we briefly mention how the general pattern might be extended to higher dimensions to make contact with Refs. \onlinecite{qi2008B,kitaev2008} in Appendix \ref{app:higherD}.
\subsection{1D inversion symmetric insulators}

The following discussion applies to a generic one dimensional $K$-band insulator with $N$ occupied bands, and with a generic inversion symmetry \emph{i.e.} 
\beq
{\cal P} \hat{H}(k) {\cal P}^{-1} = \hat{H}(-k),
\eneq 
where the inversion matrix ${\cal P}$ is unitary and squares to the identity:
\beq
{\cal P}^\dagger {\cal P}= 1, \ \ \ {\cal P}{\cal P} =1.
\eneq 
As explicitly shown in Appendix \ref{app:PolOdd}, the charge polarization $P_1$ of a 1D  insulator behaves as:
\beq\label{Find1}
P_1 \rightarrow -P_1+je
\eneq
under inversion, where $j$ is a gauge-dependent integer. This shows that the polarization of 1D systems with inversion symmetry can take only two values, 0 and $e/2$, modulo a gauge-dependent integer multiple of $e.$\cite{zak1989} We will prove that if the $\chi_{{\cal P}}$ invariant, defined as the product of all inversion eigenvalues of the occupied bands, takes the value 1, then $P_1=0$, and if $\chi_{{\cal P}}=-1$, then $P_1=e/2$. More precisely, we will establish that:
\begin{equation}
 P_1 = \frac{e}{2\pi i}\mbox{Log}\left [ \prod_{i=1}^N \zeta_i(0)\zeta_i(\pi)\right ].\label{Find2}
\end{equation}
 The integer ambiguity of the logarithm is identical to the integer ambiguity of the polarization.  

For this, we define the $k$-dependent $N$$\times$$N$ unitary matrix $\hat{B}(k)$
\begin{equation}
B_{ij}(k)=\langle u_{i,-k}\vert {\cal P} \vert u_{j,k}\rangle,\label{eq:appBdef}
\end{equation}
where the indices $i$ and $j$ run only over the occupied bands. The inversion eigenvalues $\zeta_i(0)$ and $\zeta_i(\pi)$ coincide with the eigenvalues of the matrix $\hat{B}(k)$, when evaluated at the special inversion $k$-points $k$=0 and $\pi$. It is then obvious that the determinant of $\hat{B}(k)$ at these $k_{\mbox{\tiny{inv}}}$ points is the product of the inversion eigenvalues at that inversion invariant point:
\begin{equation}\label{Det}
\det [\hat{B}(k_{\mbox{\tiny{inv}}})] = \prod_{i=1}^N \zeta_i(k_{\mbox{\tiny{inv}}}).
\end{equation}

We now turn to the calculation of the polarization
\begin{equation}
P_1=\frac{e}{2\pi} \int_{-\pi}^\pi dk A(k),
\end{equation}
where $A(k)$ is the adiabatic connection:
\begin{equation}
A(k)=-i\sum_{i\in {\rm{occ}}}\langle u_{i,k}\vert\nabla_{k}\vert u_{i,k}\rangle .
\end{equation}
We will use the following important relation, which is proven in Appendix \ref{app:PolZ2}:
\begin{eqnarray}
&A(-k) = -A(k)+ i \mbox{Tr}[\hat{B}(k)  \nabla_k \hat{B}^\dagger(k) ].   
\end{eqnarray}
The last term can be written in the equivalent form: 
\begin{equation}
\mbox{Tr}[\hat{B}(k) \nabla_k \hat{B}^\dagger(k)]  = - \nabla_k \mbox{Log}\big [\det[\hat{B}(k)]\big].
\end{equation}
We can now proceed as follows:
\begin{equation}
\begin{array}{l}
P_1=\frac{e}{2\pi} \int_{0}^\pi dk [A(k)+A(-k)]   \medskip \\
= \frac{e}{2\pi i} \int_{0}^\pi dk \ \nabla_k \mbox{Log}\big [\det [\hat{B}(k)]\big],
\end{array}
\end{equation}
with the final answer:
\begin{equation}
P_1=\frac{e}{2\pi i}\left[\mbox{Log}\big [\det [\hat{B}(\pi)]\big]-\mbox{Log}\big [\det [\hat{B}(0)]\big]\right ].
\end{equation}
This, together with Eq.~\ref{Det} and the fact that the determinants can take only the values $\pm1$, so that $\det [\hat{B}]=1/\det [\hat{B}]$, prove the statement of Eq.~\ref{Find2}.

We mention that similar arguments were used in Ref. \onlinecite{qi2008B} to classify 1D particle-hole symmetric insulators via a $Z_2$ invariant. In fact, the $Z_2$ invariant found there is exactly the value of the charge polarization modulo an integer. For a 1D model with both inversion and particle-hole symmetry, such as the 1D lattice Dirac model, the invariants coincide. In the non-trivial phase, the 1D Dirac model exhibits mid-gap energy modes bound to the ends of an open chain. The requirement of particle-hole symmetry restricts these modes to lie at zero energy if there are an odd number of them. An even number on each end is not stable and the degeneracy can be lifted, which is another manifestation of the $Z_2$ nature. The minimal case is one mode on each end and with particle-hole symmetry at half filling one mode is filled and one is empty. This leads to an  excess charge of $+e/2$ on the side with the filled state and $-e/2$ on the empty side. If we break particle-hole symmetry but keep inversion symmetry then both modes can be empty or filled, but they are empty or filled together because inversion symmetry connects the two modes. This means that the excess charge  is either $+e/2$ on \emph{both} ends or $-e/2$ on \emph{both} ends. However, because of the gauge-variance of the polarization this is equivalent to the polarization in the particle-hole symmetric case. Thus, both insulators have the same topological electric response. One can form a complimentary argument by using the effective response action for particle-hole symmetric insulators given in Ref. \onlinecite{goldstone1981,qi2008B}:
\begin{equation}
S_{eff}=\frac{1}{2}\int dx dt P_1 \epsilon^{\mu\nu}F_{\mu\nu}
\end{equation}\noindent where $F_{\mu\nu}$ is the field-strength tensor of the externally applied electro-magnetic field. The argument for the quantization of $P_1$ is as follows. In the partition function the phase due to this term is $e^{iS_{eff}}$ and under particle-hole symmetry $P_1\to -P_1.$ Thus if our system is to be particle-hole symmetric we must have $e^{2iS_{eff}}=1.$ For constant $P_1$ the integral gives $2\pi n$ for integer $n$ and we have $e^{4\pi i n P_1}=1$ and thus $P_1=0,1/2\;\; {\textrm{mod}}\; Z$ in dimensionless units.  Since $P_1\to -P_1$ under inversion symmetry the \emph{same} argument holds and $P_1$ is quantized there as well. A similar argument for magneto-electric polarizability of 3D insulators with inversion symmetry was given in Ref. \onlinecite{turner2009}.
As an aside, we recall that in 1D we also had an invariant $\chi^{(2)}_{{\cal{P}}}$ which helped classify the 4-band insulator example. We do not know of any response related to this invariant in 1D inversion invariant insulators.

Additionally,  Eq.~\ref{Find2} can be derived using an alternative approach based on a monodromy argument.  The effect of a magnetic flux $\Phi$ through a large one dimensional ring can always be gauged away by a transformation: $\Psi$$\rightarrow$$e^{-i\Phi}\Psi$, which is equivalent to a translation in $k$-space by  $\Phi$. The evolution of the states in response to an adiabatically slowly varying magnetic flux can be understood from the evolution of the Bloch states in response to an adiabatic translation of $k$-space. \cite{simon1983} The monodromy $\hat{U}(k,k_0)$ describes the evolution of the occupied Bloch states when the $k$ of the Bloch Hamiltonian $\hat{H}(k)$ is adiabatically varied, and it is the unique solution of the equation:
\begin{equation}\label{AdTransp}
\begin{array}{c}
\frac{d}{d k} \hat{U}(k,k_0) =i [\hat{P}_k,\partial_k \hat{P}_k] \hat{U}(k,k_0),
\end{array}
\end{equation}
with the initial condition $\hat{U}(k_0,k_0)$=$\hat{P}_{k_0}$, assuming that we start the evolution from $k_0$.  

\begin{figure}
  \includegraphics[width=3cm]{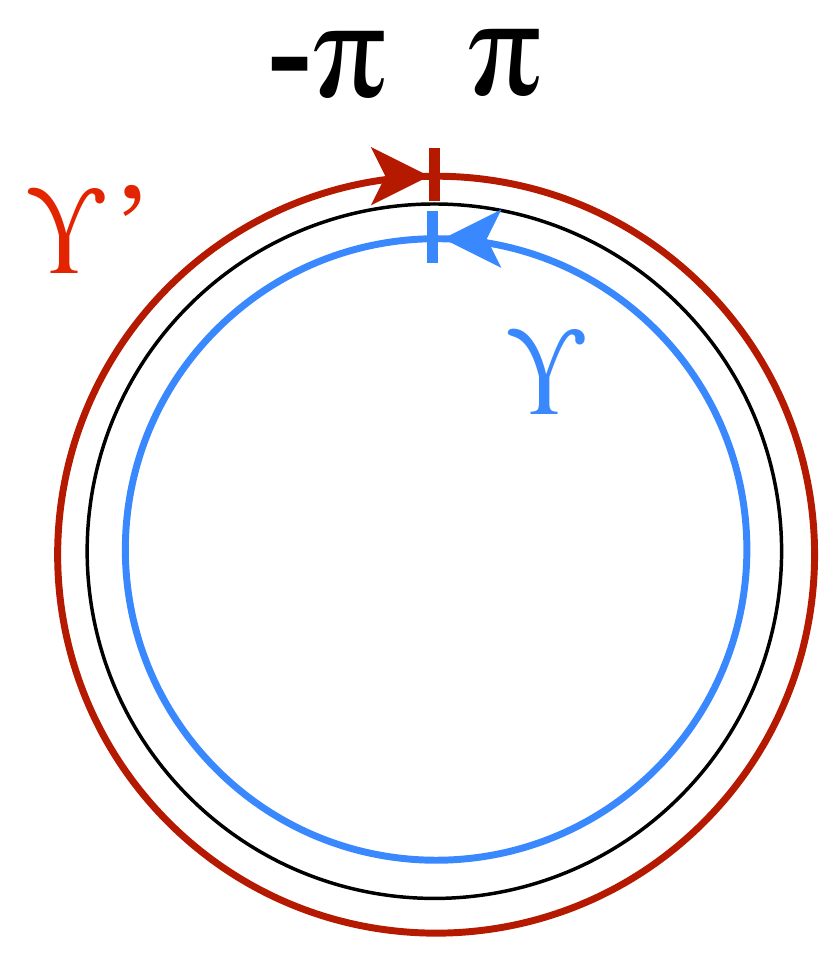}\\
  \caption{The adiabatic transport is carried over $\gamma_{1,2}$.}
 \label{Paths}
\end{figure}

The monodromy $\hat{U}(k_0,k_0)$ maps the space of occupied Bloch states $\hat{P}_{k_0}{\bm C}^K$ at $k_0$ into the space of occupied Bloch states $\hat{P}_k{\bm C}^K$ at $k$. Since $k=\pm \pi$ are the same, $\hat{U}(\pi,-\pi)$ takes the space $\hat{P}_{-\pi}{\bm C}^K$ into itself, and is a unitary operator that we call $\hat{U}_{\gamma}$, where $\gamma$ is one of the paths shown in Fig.~\ref{Paths}. $U_\gamma$ gives the change occurring after a full quantum of magnetic flux has been pumped through the system. If bases $\{\psi_i(k)\}$ were pre-chosen for all $\hat{P}_k{\bm C}^K$ spaces, the matrix $U_{ij}(k)=\langle \psi_i(k)|\hat{U}(k,k_0)|\psi_j(k_0)\rangle$ satisfies the parallel transport equation
\begin{equation}
\begin{array}{c}
\frac{d}{dk}\hat{U}(k)=\hat{A}(k) \hat{U}(k),
\end{array}
\end{equation}
where $\hat{A}(k)$ is the full non-abelian adiabatic connection discussed by Wilczek and Zee in Ref.~\onlinecite{wilczek1984}.

There is a direct relation between the determinant of the monodromy and the line integral of the abelian connection (the trace of the full connection):
\begin{equation}\label{MonPol}
\begin{array}{l}
\det[\hat{U}(k_f,k_i)]=\exp\left ({\int\limits_{k_i}^{k_f} \mbox{Tr}[\hat{A}(k)]dk} \right ).
\end{array}
\end{equation}
 Indeed, working with predefined basis sets for $\hat{P}_k{\bm C}^K$ and breaking the interval $k_f,k_i$ in small subintervals $k_f$, $k_n$, $\ldots$, $k_i$, we have, up to second order corrections:
\begin{equation}
\begin{array}{l}
\hat{U}(k_f,k_i)=\hat{U}(k_f,k_n)\hat{U}(k_n,k_{n-1}) \ldots \hat{U}(k_1,k_i) \medskip \\
=\big(I+(k_f-k_n)\hat{A}(k_n)\big) \ldots \big(I+(k_1-k_i)\hat{A}(k_i)\big).
\end{array}
\end{equation}
Taking the determinant on both sides and using some elementary identities, we obtain
\begin{equation}
\begin{array}{l}
\det[\hat{U}(k_f,k_i)]=\big(1+(k_f-k_n)\mbox{Tr}[\hat{A}(k_n)]\big) \medskip \\
\ \ \ \ \ldots \big(1+(k_1-k_i)\mbox{Tr}[\hat{A}(k_i)]\big) \medskip \\
=e^{(k_f-k_n)\mbox{Tr}[\hat{A}(k_n)]} \ldots e^{(k_1-k_i)\mbox{Tr}[\hat{A}(k_i)]}, 
\end{array}
\end{equation}
from which the identity of Eq.~\ref{MonPol} follows. The identity is valid in higher dimensions too.

Assuming $k_0$=$-\pi$, a conjugation of Eq.~\ref{AdTransp} with the inversion operator ${\cal P}$ gives:
\begin{equation}
\begin{array}{c}
\partial_k \{{\cal P}\hat{U}(k,-\pi){\cal P}^{-1}\} =  i[\hat{P}_{-k},\partial_k \hat{P}_{-k}] {\cal P}\hat{U}(k,-\pi){\cal P}^{-1},\nonumber
\end{array}
\end{equation}
with the initial condition ${\cal P}\hat{U}(-\pi,-\pi){\cal P}^{-1}$=$\hat{P}(\pi)$. This is just the equation for $\hat{U}(-k,\pi)$, which shows that ${\cal P}\hat{U}(k,-\pi){\cal P}^{-1}$ coincides with $\hat{U}(-k,\pi)$. Equivalently we can think that ${\cal P}$ sends $\gamma$ into  $\gamma'$ in Fig.~\ref{Paths}. Now obviously $\hat{U}_{\gamma} \hat{U}_{\gamma'}$ equals the identity, therefore:
\begin{equation}
\det [ \hat{U}_{\gamma} {\cal P}  \hat{U}_{\gamma} {\cal P}^{-1} ]=1.
\end{equation}
Using the elementary properties of the determinant, we conclude that $\det [ \hat{U}_{\gamma} ]^2$=1, hence $\det [ \hat{U}_{\gamma} ]$ can take only two values:
\begin{equation}\label{Find3}
\det [ \hat{U}_{\gamma} ]=\pm 1.
\end{equation}

The following calculation will show that the cases $\det [ \hat{U}_{\gamma} ]$=$\pm 1$ correspond to $P_1$=0 and $P_1$=$\frac{e}{2}$ (mod $Z$), respectively. Indeed:
\begin{equation}\label{Monodromy}
\begin{array}{c}
\det [\hat{U}_{\gamma} ] = \det [ \hat{U}(\pi,0) \hat{U}(0,-\pi) ] \medskip \\
= \det [ \hat{U}(\pi,0) {\cal P}\hat{U}(0,\pi) {\cal P}^{-1}  ] \medskip \\
= \det [ \hat{U}(\pi,0) \hat{P}_0 {\cal P}\hat{P}_0\hat{U}(0,\pi) \hat{P}_\pi {\cal P}^{-1}\hat{P}_\pi  ].
\end{array}
\end{equation}
In the last line we have inserted the projectors $\hat{P}_{0,\pi}$ to see explicitly the spaces on which ${\cal P}$ is acting. Using again the elementary properties of the determinant and the fact that $\hat{U}(\pi,0)\hat{U}(0,\pi)$=1, we obtain:
\begin{equation}
\det [ \hat{U}_{\gamma} ]=\det [\hat{P}_0 {\cal P} \hat{P}_0 ] \det [ \hat{P}_\pi {\cal P}^{-1}  \hat{P}_\pi ]
\end{equation}
or
\beq\label{Find4}
\begin{array}{c}
\det [\hat{U}_{\gamma} ]=\prod\limits_{i=1}^N\zeta_{i}(0)\zeta_{i}(\pi).
\end{array}
\eneq\noindent 
 and thus
\beq
P_1=\frac{e}{2\pi i}{\textrm{Log}}\det[\hat{U}_{\gamma}]
\eneq\noindent (cf.  Eq.~\ref{MonPol}).

For a topological insulator, no matter what definition one uses, it is always the case that, when the hopping terms between the neighbors are adiabatically turned off, that is, when one takes the atomic limit, the insulating gap of the system closes at some point in the process. One can investigate this issue using the inversion eigenvalues directly, as we have focused on, but here we  see how the physical response enters into the picture. Note that the inversion eigenvalues can be easily computed for simple models, but may not always be available, especially for complex materials. $P_1$ or $U_\gamma$ are physically measurable, so they can provide physical signatures of the non-trivial state. In the atomic limit, the bands have no dispersion, so in this limit $\hat{U}_{\gamma}$ is just the identity matrix. If $\mbox{Det}[\hat{U}_{\gamma}]$=$-1$, it is obvious that $U_\gamma$ cannot be smoothly connected to the identity and the insulator is topological. However, we also must note that if $\mbox{Det}[\hat{U}_{\gamma}]$=$1$, it is \emph{not} necessary that the insulator is trivial.

We can  refine the investigation by asking when can $\hat{U}_\gamma$ be smoothly connected to the identity?  For this, let us look again at Eq.~\ref{Monodromy}:
\begin{equation}\label{Node1}
\begin{array}{c}
\hat{U}_{\gamma} = \hat{U}(\pi,0) \hat{P}_0{\cal P}\hat{P}_0\hat{U}(0,\pi) \hat{P}_\pi{\cal P}\hat{P}_\pi.
\end{array}
\end{equation}
The first term, $\hat{U}(\pi,0) \hat{P}_0{\cal P}\hat{P}_0\hat{U}(0,\pi)$, is just $\hat{P}_0{\cal P}\hat{P}_0$ parallel transported from $k$=$0$ to $k$=$\pi$. The parallel transport does not alter the eigenvalues of $\hat{P}(0){\cal P}\hat{P}(0)$, which are pinned at $\pm 1$ (recall ${\cal P}^2$=1). The eigenvalues of $\hat{P}(\pi){\cal P}\hat{P}(\pi)$ are also pinned at $\pm 1$. So what we have in Eq.~\ref{Node1} is a product of two operators with eigenvalues pinned at $\pm 1$ and because of that the eigenvalues cannot change  under any smooth deformation of the Hamiltonian that keeps the insulating gap open and preserves inversion symmetry. Now, if $\hat{U}_{\gamma}$ can be deformed into the identity, then $\hat{U}(\pi,0) \hat{P}_0{\cal P}\hat{P}_0\hat{U}(0,\pi)$ can be turned into the inverse of $\hat{P}_\pi{\cal P}\hat{P}_\pi$ and this requires that $\hat{P}_0{\cal P}\hat{P}_0$ and $\hat{P}_\pi{\cal P}\hat{P}_\pi$ have identical eigenvalues, counting the degeneracy too. The conclusion is: if the set of inversion eigenvalues of 
\begin{equation}\label{Operators}
\hat{P}_0{\cal P}\hat{P}_0 \ \mbox{and} \ \hat{P}_\pi{\cal P}\hat{P}_\pi
\end{equation}
are not identical, then $\hat{U}_{\gamma}$ cannot be connected to the identity. Since the eigenvalues are restricted to just $\pm 1$ values, then the following integer:
\begin{equation}
\begin{array}{c}
\chi = \frac{1}{2}\left |\mbox{Tr}\left \{ \hat{P}_0{\cal P}\hat{P}_0 - \hat{P}_\pi{\cal P}\hat{P}_\pi \right \}\right |,
\end{array}
\end{equation}
tells how many eigenvalues are different for the two matrices in Eq.~\ref{Operators}. To reach the atomic limit, we need to flip  $\chi$  inversion eigenvalues, and that will require a minimum of $\chi$  gap closings. Unfortunately, we were not able to find an expression of the topological invariant $\chi$  solely in terms of the monodromy $\hat{U}_\gamma$, but we know there are precisely $\chi$ topological obstructions when trying to connect the monodromy to the identity. The topological invariant $\chi$ also gives the number of robust edge modes seen in the entanglement spectrum on a single cut as shown in Sec. \ref{sec:entanglement}. 

\subsection{2D inversion symmetric insulators}

The physical response of 2D inversion symmetric insulators is much richer than that of 1D. Based solely on the inversion eigenvalues, one can define several invariants, the first of which is the isotropic extension of $\chi_{P}$ to 2D \emph{i.e.}
\begin{equation}
\chi_{\cal{P}}= \prod_{k_{\mbox{\tiny{inv}}}; i\in occ.}\zeta_i (k_{\mbox{\tiny{inv}}})
\end{equation}
where $k_{\mbox{\tiny{inv}}}$ runs over all four inversion invariant $k$-points. We show that 
\begin{equation}\label{ChernParity}
\chi_{\cal{P}}=(-1)^{C_1},
\end{equation}\noindent where $C_1$ is the first Chern number of the ground state. Thus if $\chi_{\cal{P}}$ is negative the system \emph{must} be in a quantum Hall state, and if it is positive it is in a  state with an even Chern number which can be zero. Thus, only if it is negative are we sure it is in a topological insulator state. 

We will prove the statement of Eq.~\ref{ChernParity} in two ways, first using a band crossing argument and then a monodromy argument. Let us begin by assuming we are in a trivial insulator state in an atomic limit with N occupied bands and that we have inversion symmetry.  We can reach any non-trivial topological insulator state from this limit through a series of Hamiltonian deformations that will lead us through band crossings. Our assumption of an atomic limit implies that $\chi_{\cal{P}}=+1$ initially. If we want to generate $\chi_{\cal{P}}=-1$ we need to have an odd number of band crossings between bands with \emph{opposite} inversion eigenvalues. Assume we have an odd number of such crossings. This implies that there must be an odd number of crossings at the inversion invariant points. This is true since crossings occurring at non-invariant $k$ are accompanied by a crossing at $-k$ always giving an even number of eigenvalue switches which will not affect the value of $\chi_{\cal{P}}.$  Thus we only need to consider the odd number of crossings occurring at the invariant momenta. The generic Hamiltonian of each crossing between opposite inversion eigenvalue states near an inversion invariant momentum is $H=p_1\sigma^1+p_2\sigma_2+m_{eff}\sigma^3$ where $(p_1,p_2)$ is the momentum away from the inversion invariant point, $m_{eff}$ is a term parameterizing the distance to the band crossing, and $\sigma^3$ is the inversion operator projected onto the two crossing bands. Exactly at the inversion invariant momentum the Hamiltonian reduces to $m_{eff}\sigma^3$ as it must in order to commute with ${\cal P}$ there. As the crossing occurs $m_{eff}$ switches sign and an inversion eigenvalue of the occupied band is changed. We note that this Hamiltonian is a 2D massive Dirac Hamiltonian which is switching the sign of the mass. At such a crossing, the Chern number changes by $\pm 1$ and thus we see that going through an odd number of crossings changes the parity of the Chern number. Thus if $\chi_{\cal{P}}=-1$ the parity of the Chern number is odd since the parity is even in the initial atomic limit \emph{i.e.}  $C_1=0$. The inversion eigenvalues thus give us a rough way to characterize the quantum Hall effect in an insulator. 

\begin{figure}
  \includegraphics[width=4cm]{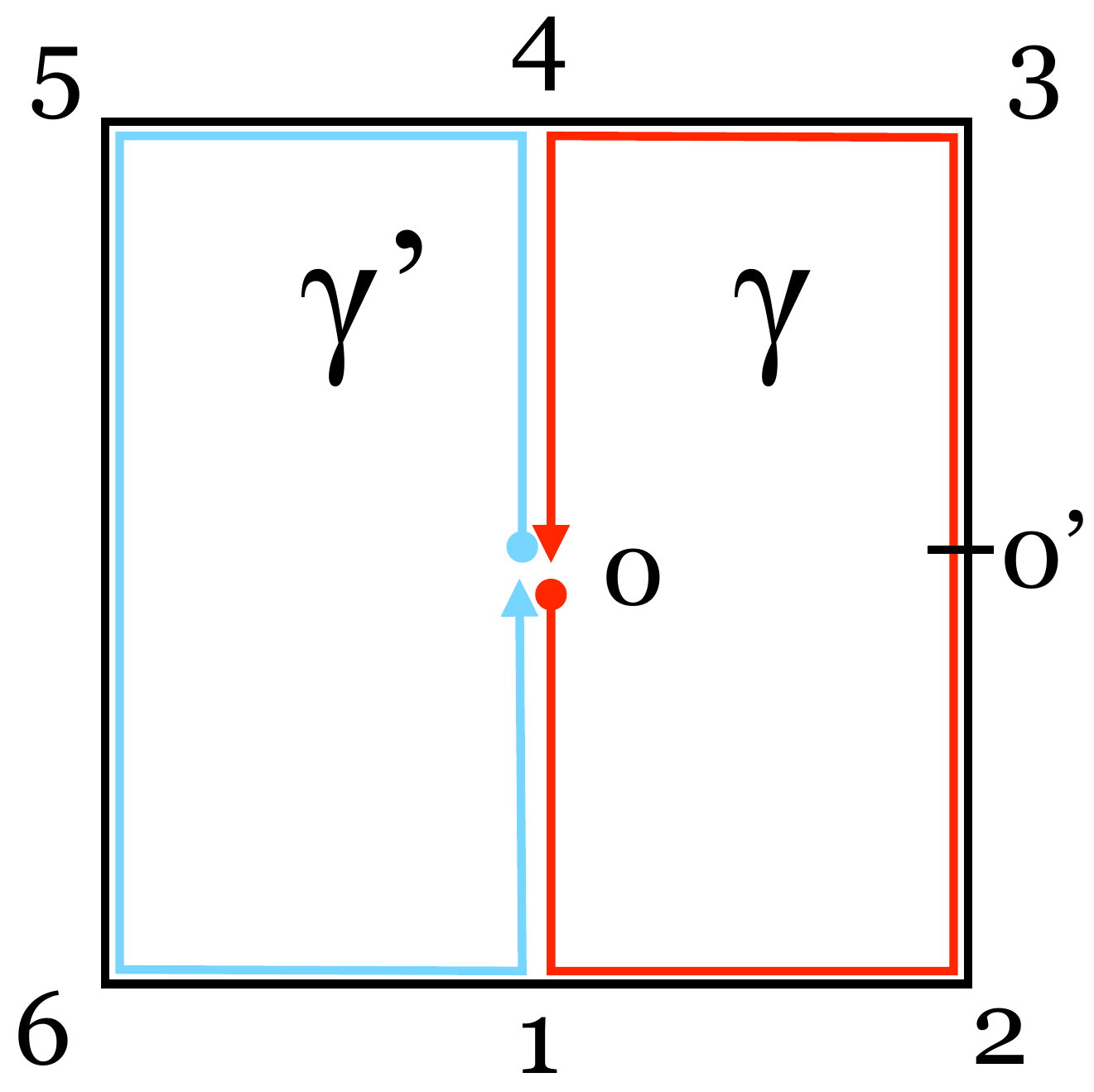}\\
  \caption{The two paths $\gamma$ and $\gamma$ used in the monodromy argument in 2D.}
 \label{2DPaths2}
\end{figure}

The corresponding monodromy argument proceeds as follows. We consider the monodromy corresponding to the path 01234560 in Fig.~\ref{2DPaths2}, starting from the middle of the Brillouin zone and continuing on its rim. This path can be also be viewed  as the composition $\gamma$+$\gamma'$ of the paths $\gamma$=012340 and $\gamma'$=045610, and the monodromy corresponding to $\gamma$+$\gamma'$ can be written as a product of partial monodromies:
\begin{equation}\label{Monodromy1}
\begin{array}{c}
\hat{U}_{\gamma+\gamma'}=\hat{U}_{01}\hat{U}_{16}\hat{U}_{65}\hat{U}_{54}\hat{U}_{43}\hat{U}_{32}\hat{U}_{21}\hat{U}_{10}.
\end{array}
\end{equation}
Since the products $\hat{U}_{10}\hat{U}_{01}$, $\hat{U}_{16}\hat{U}_{54}$, $\hat{U}_{65}\hat{U}_{32}$ and $\hat{U}_{43}\hat{U}_{21}$ are all equal to the identity, taking the determinant of Eq.~\ref{Monodromy1} leads us to conclude that $\det [\hat{U}_{\gamma+\gamma'}]$=1. This is not surprising and is related to the fact that $\det [\hat{U}_{\gamma+\gamma'}]=e^{2i \pi  C_1}$ (see Eq.~\ref{MonPol}) and the first Chern number is an integer.

The next observation is that inversion sends $\gamma$ into $\gamma'$ and consequently:
\begin{equation}
\begin{array}{c}
\det [\hat{U}_{\gamma+\gamma'}]=\det [ \hat{U}_{\gamma}{\cal P}\hat{U}{\gamma}{\cal P} ]=\det [\hat{U}_{\gamma}]^2.
\end{array}
\end{equation}
The conclusion is that the determinant of $\hat{U}_\gamma$ can only take  the values $\pm 1.$ Since the path $\gamma$ encircles half of the Brillouin zone, and the adiabatic curvature is symmetric when $k
\rightarrow -k,$ it is also true that $\det [\hat{U}_{\gamma}]= e^{i\pi C_1}$ (see Eq. \ref{MonPol}).

We now take a closer look at $\hat{U}_\gamma$. Since  inversion sends 10 into 40 and 20' into 30', we have:
\begin{equation}\label{Monodromy1A}
\begin{array}{c}
\hat{U}_{\gamma}=\hat{U}_{04}\hat{U}_{43}\hat{U}_{30'}\hat{U}_{0'2}\hat{U}_{21}\hat{U}_{10} \medskip \\
={\cal P}\hat{U}_{01}{\cal P} \hat{U}_{12}{\cal P} \hat{U}_{20'} {\cal P} \hat{U}_{0'2} \hat{U}_{21} \hat{U}_{10} 
\end{array}
\end{equation} 
Inserting the appropriate projectors to specify explicitly on which spaces are the ${\cal P}$ operators acting, taking the determinant and using its elementary properties together with the fact that $\hat{U}_{ij}\hat{U}_{ji}$ equals the identity, we obtain:
\begin{equation}
\begin{array}{c}
(-1)^{C_1}=\det [\hat{U}_{\gamma}]= \det [ \hat{P}_{0,0}{\cal P} \hat{P}_{0,0} ] \det [ \hat{P}_{0,\pi}{\cal P} \hat{P}_{0,\pi} ] \medskip \\
\times [\det\{ \hat{P}_{\pi,\pi}{\cal P} \hat{P}_{\pi,\pi} ] \det [ \hat{P}_{\pi,0}{\cal P} \hat{P}_{\pi,0}],  
\end{array}
\end{equation}
which completes monodromy argument for Eq.~\ref{ChernParity}.

We will briefly mention two other interesting inversion invariants one can define in 2D. The first is an anisotropic invariant defined by taking the product of the inversion eigenvalues at only two invariant momenta in the 2D Brillouin zone. We provide an example in Sec. \ref{sec:examples} with $\chi_{\cal{P}}=+1$ but where this anisotropic invariant is non-trivial.  This invariant has interesting implications for the charge polarization, but there are some subtleties that we will illustrate. To begin, assume we have an inversion invariant, insulator Hamiltonian $\hat{H}(k_x,k_y)$ with N occupied bands such that $\prod_{i=1}^{N}\zeta_{i}(0,0)\zeta_{i}(0,\pi)=-1.$ This implies that the 1D Hamiltonian $\hat{H}(0,k_y)$ has a charge polarization $P_1=e/2$ since this restricted Hamiltonian is inversion invariant. Now we want to know if this non-trivial anisotropic invariant is enough to specify the full polarization of the entire 2D system. The answer is no. To see why, we slightly deform $\hat{H}(0,k_y)$ away from the $k_y$ axis. The 1D Hamiltonian $\hat{H}(\delta k_x,k_y)$ is \emph{not} generically invariant under inversion symmetry because it gets mapped onto the Hamiltonian $\hat{H}(-\delta k_x, k_y)$ and thus the polarization does not have to remain quantized with value $e/2.$\cite{zak1989} In fact, by the time we have deformed all the way to the Hamiltonian $\hat{H}(\pi,k_y)$ the polarization, which must again be quantized since this Hamiltonian does have inversion, can be completely different. So while we can think of $\hat{H}(k_x,k_y)$ as a gapped interpolation between $\hat{H}(0,k_y)$ and $\hat{H}(\pi,k_y)$, inversion symmetry, and thus the value of the polarization, is not preserved along the interpolation. Intuitively this makes sense because it is exactly when the polarization changes its quantized value that the system has an odd Chern number.  An odd Chern number is allowed because a quantum Hall effect is not forbidden by the requirement of inversion symmetry. Thus, if we have 2D inversion symmetry the anisotropic invariant \emph{does not} determine the 2D charge polarization. 

As an aside, since we clearly know why inversion symmetry fails, we immediately know how to fix the problem. We fix it by requiring a \emph{reflection} symmetry (\emph{i.e.} parity symmetry) about an axis, instead of inversion. Without loss of generality we can have a reflection symmetry ${\cal{M}}$ such that ${\cal{M}}\hat{H}(k_x,k_y){\cal{M}}^{-1}=\hat{H}(k_x,-k_y).$ We will see that the reflection eigenvalues will specify the polarization in the y-direction (for this choice of reflection symmetry). First, we see that the inversion invariant momenta are also reflection invariant momenta. Thus $[\hat{H}(k_{inv}),{\cal{M}}]=0,$ and we can label the states at these points by their reflection eigenvalues which we will also call $\zeta_i (k_{inv}).$ Now let us assume the same setup as the previous paragraph with  $\prod_{i=1}^{N}\zeta_{i}(0,0)\zeta_{i}(0,\pi)=-1.$ The point is that now when we adiabatically deform away from the $k_y$ axis the 1D Hamiltonian $\hat{H}(\delta k_x,k_y)$ is invariant under reflection. Additionally, the y-component of the polarization is quantized as long as reflection is a good symmetry. Thus, $\hat{H}(k_x,k_y)$ is a gapped interpolation along which reflection symmetry is preserved and thus the y-component of the polarization is fixed and quantized to $e/2$ for each 1D Hamiltonian. This argument immediately implies that 
$\prod_{i=1}^{N}\zeta_{i}(\pi,0)\zeta_{i}(\pi,\pi)=-1.$ Thus the parity of the Chern number is always \emph{even} when reflection symmetry is required. In fact, it always vanishes because the quantum Hall effect is incompatible with reflection symmetry. Notice that the reflection eigenvalues do not uniquely specify the polarization in the x-direction since the two eigenvalue conditions can be satisfied by choosing $\prod_{i=1}^{N}\zeta_{i}(0,0)\zeta_{i}(\pi,0)= \prod_{i=1}^{N}\zeta_{i}(0,\pi)\zeta_{i}(\pi,\pi)=\pm 1.$ Thus, the anisotropic invariants in the x-direction can take either value.

The other invariant one can define is the isotropic extension of $\chi^{(2)}_{\cal{P}}$ to 2D. If $\chi_{\cal{P}}=+1$ and the product over the inversion eigenvalues at \emph{every} invariant momentum is separately trivial then $\chi^{(2)}_{\cal{P}}$ is well-defined because there are an even number of negative inversion eigenvalues at each invariant momentum and by definition $\chi^{(2)}_{\cal{P}}$ is the product over half of those negative eigenvalues. This is the $Z_2$ topological invariant which indicates a quantum spin Hall effect\cite{fu2007a} when an insulator has both inversion and time-reversal symmetry. Unfortunately this topological invariant does not yield a unique topological response. We can understand this in several ways. By explicit construction take two decoupled copies of the QAHE state each with $C_1=1$ to give an IQHE with $C_1=2$ or take two decoupled copies of QAHE one with $C_1=1$ and the other with $C_1=-1. $ The first system breaks time-reversal and still gives an IQHE while the second preserves time reversal and will not give a quantum Hall effect. These systems both have $\chi_{\cal{P}}=+1$ but $\chi^{(2)}_{\cal{P}}=-1.$ We can immediately see why this invariant does yield a unique response in the presence of an additional time-reversal symmetry because this requires the total Chern number to vanish which only leaves the possibility of a quantum spin Hall state.  Thus it is time-reversal symmetry which restricts the allowed physical response. We can understand this by a simple symmetry argument as well. The quantum spin Hall effect is even under both inversion symmetry $(x,y)\rightarrow (-x,-y)$ and parity $(x,y)\rightarrow (x,-y)$ We see that in even space dimensions inversion symmetry is just a rotation. The quantum Hall effect is even under inversion but odd under parity. Thus having a quantum Hall effect is compatible with inversion symmetry but not parity (and not time-reversal). This is why the \emph{inversion} classification does not distinguish between the doubled quantum Hall state and a quantum spin Hall state. This type of ambiguity exists in every dimension where one can define a Chern number invariant. However, as we saw with the polarization, if we consider the eigenvalues of a reflection symmetric Hamiltonian we can eliminate the possibility of a non-zero Chern number thus leaving a quantum spin Hall state as the alternative.

\subsection{3D Inversion Symmetric Insulators}\label{sec:3dins}

We show that the isotropic extension of $\chi_{\cal P}$ in 3D
\begin{equation}
\chi_{\cal{P}}=\prod_{k_{\mbox{\tiny{inv}}}; i\in occ.}\zeta_i (k_{\mbox{\tiny{inv}}}),
\end{equation}
where the product runs over all 8 inversion symmetric $k$-points, is \emph{always} trivial. We can first see this by using a band crossing argument. Start from the atomic limit in which bands have identical inversion eigenvalues at all inversion symmetric points. For $\chi_{\cal{P}}$ to be non-trivial (equal to $-1$)  there has to be an odd number of band crossings between bands of opposite inversion eigenvalues at the inversion invariant momenta when beginning from this trivial limit. Without loss of generality let us consider one crossing between two bands with opposite inversion eigenvalues at all inversion symmetric points  in the Brillouin zone. The crossing can happen at either an inversion symmetric point (by tuning one parameter) or at a non-inversion symmetric point $k$ in the Brillouin zone. In the latter case, there are actually two crossings at $k$ and $-k$ because of inversion symmetry.  To initially close the gap between the two bands one needs a \emph{quadratic} touching in at least one of the directions, otherwise we would be creating a  nonzero Chern number Fermi surface (after gap closing) out of a zero Chern number surface(before the closing, in the trivial limit).  If the crossing starts at an inversion symmetric point, the quadratic touching and the gap reopening at the inversion symmetric point will switch the inversion eigenvalues of the bands at that point and make $\chi_{\cal{P}}= -1$. However, the system will no longer be an insulator: it will have two crossings somewhere else in the Brillouin zone. That is, although the gap has opened at the inversion symmetric point,  the gapless points have been moved away. The quadratic touching effectively splits into multiple 3D Dirac points.  In the generic situation, there are two gapless points in the Brillouin zone, with relative position fixed by inversion, and in order to gap the system, we need to annihilate the two Dirac points. Note that a 3D Dirac point is locally stable even if inversion is not preserved. On a 2D surface surrounding a 3D Dirac point ($\hat{H}_{local}(k)=k_i A_{ij} \sigma_j$, ($i,j=1,2,3$)), the Chern number is $C={\textrm{sgn}}(\det(A_{ij})).$ The degeneracy points are stable unless two Dirac points of opposite Chern numbers annihilate. Inversion symmetry forces the points at $k, -k$ to have opposite Chern numbers.  So, by inversion, annihilation of only two Dirac points in the whole Brillouin zone can only happen at another inversion symmetric point, in which case inversion eigenvalues are  switched again to give $\chi_{\cal{P}} = 1$ for an insulator. If the gap first closes at a non-symmetric point $k$, we find the same end result: generically two Dirac points are created close to $k$ and two close to $-k$. They can annihilate in pairs, always switching an \emph{even} (possibly zero in this case, since the $4$ dirac points can annihilate two by two at non-inversion symmetric points in the BZ) number of inversion eigenvalues.

Another similar way of understanding the 3D band crossings is the following: since we are considering a two-band crossing in 3D there are three varying parameters and we cannot find a gapped phase with $\chi_{\cal{P}}=-1,$ though it is possible to find a gapless phase. If we have an even number of band crossings then we can open a gap but this means that two inversion eigenvalues are switched leaving $\chi_{\cal{P}}=+1.$ There is also a deeper reason why this cannot be done. Let us look at the simplest case of a single two-band  crossing at the $\Gamma$-point. The effective Hamiltonian can be put in the form $H_{eff}=p_1\sigma^1+p_2\sigma^2+p_3\sigma^3$ which is a (chiral) Weyl-fermion Hamiltonian. From the Nielsen-Ninomiya no-go theorem this type of Hamiltonian cannot arise without a partner fermion with the opposite chirality.\cite{NIELSEN1981} Thus all two-band crossings must generically occur in pairs and there cannot be an odd number of negative parity eigenvalues in an insulating system. Conversely, if no states at the inversion invariant momenta cross the Fermi-level and $\chi_{\cal{P}}=-1$ we immediately know that the system contains a gapless point(s) somewhere in the Brillouin zone.

We now provide an alternate proof that the product of all the inversion eigenvalues of all the occupied bands must be positive in 3D. This will prove useful when considering possible 3D QHE states. Assume a gapped insulator Hamiltonian $\hat{H}(k_x, k_y, k_z)$ which has $N$ occupied bands and is invariant under inversion symmetry.  We can take the plane $k_z=0$, and think of $\hat{H}(k_x, k_y, 0)$ as a $2D$ inversion symmetric Hamiltonian. We have already proved  that the inversion eigenvalues of a 2D inversion symmetric Hamiltonian are related to the Chern number \emph{i. e.}:
\begin{eqnarray}
&\prod_{i=1}^{N} &\zeta_{i}(0,0,0) \zeta_{i}(\pi,0,0) \zeta_{i}(0,\pi,0) \zeta_{i}(\pi,\pi,0)\nonumber\\ &=& (-1)^{C_{1}(\hat{H}(k_x, k_y, 0))}.
\end{eqnarray}\noindent The same thing is true for the inversion symmetric 2D Hamiltonian $\hat{H}(k_x,k_y,\pi)$:
\begin{eqnarray}
&\prod_{i=1}^{N}&\zeta_{i}(0,0,\pi) \zeta_{i}(\pi,0,\pi) \zeta_{i}(0,\pi,\pi) \zeta_{i}(\pi,\pi,\pi)\nonumber\\ &= &(-1)^{C_{1}(\hat{H}(k_x, k_y, \pi))}.
\end{eqnarray}\noindent  Hence
\begin{eqnarray}
&\prod_{i=1}^{N}&\zeta_{i}(0,0,0) \zeta_{i}(\pi,0,0) \zeta_{i}(0,\pi,0) \zeta_{i}(\pi,\pi,0)\nonumber\\ & &\cdot\; \zeta_{i}(0,0,\pi) \zeta_{i}(\pi,0,\pi) \zeta_{i}(0,\pi,\pi) \zeta_{i}(\pi,\pi,\pi)\nonumber\\  &=&(-1)^{C_{1}(\hat{H}(k_x, k_y, 0)) + C_{1}(\hat{H}(k_x, k_y, \pi))}.
\end{eqnarray}\noindent Since $\hat{H}(k_x, k_y, k_z)$ is gapped due to our assumption of an insulator, we can think of it as an adiabatic interpolation between $\hat{H}(k_x,k_y,0)$ and $\hat{H}(k_x,k_y,\pi)$  by varying the parameter $k_z.$ Since this interpolation preserves the $U(1)$ charge \emph{conservation} symmetry \emph{i.e.} there is no superconductivity, it means that the Chern number cannot change from $k_z=0$ to $k_z=\pi.$ Thus,
\beq
C_{1}(\hat{H}(k_x, k_y, 0)) = C_{1}(\hat{H}(k_x, k_y, \pi)) = C_{1}.
\eneq
Hence
\beq
\prod_{k_{\mbox{\tiny{inv}}}; i\in occ.}\zeta_i (k_{\mbox{\tiny{inv}}}) =(-1)^{2 C_1}= 1.
\eneq

\subsubsection{Anisotropic Invariants and the 3D Quantum Hall Effect}
We have seen that the isotropic invariant in 3D, which is constructed by multiplying the inversion eigenvalues at all invariant momenta, is always trivial for an insulator. However we can form anisotropic invariants by considering the products of inversion eigenvalues over planes or lines of the Brillouin zone which are mapped onto themselves under inversion.

First consider a plane in the 3D Brillouin zone which is mapped onto itself under inversion. To be explicit, take the plane  $k_z =\pi$. If the product of inversion eigenvalues of all the occupied bands in that plane is
\beq
\prod_{i=1}^{N}\zeta_i(0,0,\pi) \zeta_i(\pi,0,\pi) \zeta_i(0,\pi,\pi) \zeta_i(\pi,\pi,\pi) = -1
\eneq we have a $3D$ QHE\cite{halperin1987} with 3D Hall conductance
\beq
\sigma_{xy}= \text{odd integer} \times \frac{2 \pi}{c}
\eneq where $c$ is lattice constant in the $z$ direction. From our above proof we know that the product of  eigenvalues in the $k_z=0$ plane must  also be $-1$. The proof of the 3D QHE is simple: $\hat{H}(k_x, k_y, k_z)$ can be thought of as an adiabatic continuation  of $\hat{H}(k_x, k_y, \pi)$ (since we assumed it to be an insulator)  As such, each $k_z$ plane $\hat{H}(k_x, k_y, k_z=\text{constant})$ has an odd integer QHE. Multiplying by the momentum, we get the above 3D Hall conductance. It is important to note that this argument depends crucially on the fact that the Chern number remains unchanged when performing adiabatic deformations as long as charge \emph{conservation} symmetry is preserved (\emph{i.e.} we don't allow superconducting perturbations).

In general the $3$D quantum Hall effect directions can be inferred from the eigenvalue formulas. The general formula for the 3D Hall conductance in terms of the inversion eigenvalues is 
\begin{eqnarray}
\sigma_{\alpha \beta \perp \gamma }& =&{\textbf{G}}_{\gamma}( 2 n+1/2\nonumber \\&-& \frac{1}{2} \prod_{i \in \text{occ.}}\prod_{ k_{inv} \in \text{plane} \perp G_\gamma}  \zeta_i(k_{inv})).
\end{eqnarray}
This expression above gives the 3D Hall conductance as a product of the inversion eigenvalues in a plane $\alpha \beta$ perpendicular to the $\gamma (=x,y,z)$ direction. ${\textbf{G}}_\gamma$ is the reciprocal lattice vector in the $\gamma$ direction and we have left out the units of $e^2/h$. If the product over all the bands of the inversion symmetric eigenvalues in the $\alpha \beta$ plane  is $-1$ we can see that the Hall conductance is an odd integer in that plane and hence cannot be zero. If the product is $+1$ then the 3D Hall conductance is even and can be zero.

Finally, just as in 2D we can consider the product of inversion eigenvalues along a single inversion invariant line in the 3D Brillouin zone. Again inversion symmetry does not allow us to determine the polarization but a reflection symmetry about a \emph{plane} allows us to specify the polarization perpendicular to that reflection plane. The argument is basically the same as in 2D so we did not include it here. The result, however, is explicitly illustrated with the 3D dimer model shown in Sec. \ref{sec:examples}.

\subsubsection{Topological Metal State}

As a corollary to the above, when the product over the inversion eigenvalues of several bands at all points in the Brillouin zone is negative:
\begin{equation}
\chi_{{\cal{P}}}=\prod_{k_{\mbox{\tiny{inv}}}; i\in occ.}\zeta_i (k_{\mbox{\tiny{inv}}}) =-1,\nonumber
\end{equation}
the system is a metal protected from opening a gap from infinitesimal perturbations. 
We can gain some intuition about why this metal state exists by looking at the effective Hamiltonian near a band crossing between states with opposite inversion eigenvalues at an inversion invariant momentum.  In the most generic case, in which we do not have any extra point-group symmetries the effective Bloch Hamiltonian expanded for small $k$ near $k_{inv}$ :
\beq
\hat{H}(k) =( M +A_{ij} k_i k_j)\sigma_z + k_i B_{i \alpha} \sigma_{\alpha} 
\eneq where $i,j =1,2,3$ while $\alpha = 1,2$. Notice that since the bands have opposite parity, the mixing elements between them have to be odd in $k.$ The Hamiltonian reduces to $M\sigma^z$ at $k_{inv}$ and thus the sign of $M$ dictates the occupied inversion eigenvalue. When $M$ is tuned thru zero, we have a phase transition with an eigenvalue switch between $+ $ and $-.$ For a given $A_{ij}$ matrix, notice that on at least one side of the switch, we must have a gapless phase: if $M>0$ and $Det(A_{ij}) <0$ we have a metallic phase, or if $M<0$ and $Det(A_{ij})>0$ we have gapless Dirac points away from $k_{inv}.$ There are of course two Dirac points, which can then annihilate at another $k_{inv}$ point with $\pm$ eigenvalues and can re-open the gap and give $\chi_{{\cal{P}}}=+1.$  Because of the different eigenvalues under inversion, we only need to tune one parameter to get a band crossing in this case. 

\subsubsection{Magneto-electric Polarization and Inversion Symmetry}

Although the first 3D isotropic invariant we mentioned is always trivial, Ref. \onlinecite{turner2009}  argues that inversion invariant insulators in $3D$, which come from strong topological insulators with softly broken time-reversal invariance, can support an isotropic topological magneto-electric response (\emph{i.e.} a $\theta= \pi$ vacuum).  Their  argument uses the transformation properties of the effective response action which we recount here. The topological response action in $3D$ for an insulator coupled to an electromagnetic field is
\begin{equation}
S_{eff}\left[A_{\mu}\right]=\frac{e^2}{2\pi h}\int d^4 x \theta {\textbf{E}}\cdot{\textbf{B}}
\end{equation}  where ${\textbf{E}},{\textbf{B}}$ are external applied electric and magnetic fields, and $\theta $ is an intrinsic quantity which is proportional to the magneto-electric polarizability.\cite{qi2008B} For translationally invariant systems the magneto-electric polarizability for time-reversal invariant insulators is
\begin{equation}
P_3=\frac{\theta}{2\pi}=\frac{1}{32\pi^3}\int d^3 k \epsilon^{ijk}{\rm{Tr}}\left[\hat{A}_{i}\hat{F}_{ij}-\frac{2i}{3}\hat{A}_{i}\hat{A}_{j}\hat{A}_{k}\right]
\end{equation}\noindent where $\hat{A}_{i}(k)$ is the non-abelian adiabatic connection, and $\hat{F}_{ij}(k)$ is the non-abelian adiabatic curvature.
Under time-reversal symmetry $P_3 \rightarrow -P_3$ and thus, for time-reversal invariant insulators $P_3=0,1/2$ (in units of $2 \pi$). $P_3$ is not a gauge invariant quantity and only defined modulo an integer.\cite{qi2008B} Note that under time-reversal ${\textbf{B}}$ \emph{does not change} in the effective action since it is an externally applied field. Only intrinsic quantities such as $P_3$ get acted upon with time-reversal. Using these two values of $P_3$ we can physically define a non-trivial time-reversal invariant  topological insulator as one with $P_3=1/2.$ It has been shown that this definition is equivalent to the band structure definition of strong topological insulators.\cite{
fu2007b,fu2007a,moore2007,roy2009a,qi2008B,wang2009} In the presence of time-reversal and inversion symmetries there is an elegant topological invariant one can define
\begin{eqnarray}
(-1)^{2P_3}=\prod_{i\in occ./2}\prod_{k_\alpha\in \{k_{inv}\}}\zeta_{i}(k_{\alpha})\label{eq:kaneInvariant}
\end{eqnarray}\noindent which is the product of the inversion eigenvalues at every invariant momenta in the 3D Brillouin zone for half the occupied bands. Since we have time-reversal, half the bands just means one band out of each Kramers' pair. We provide a new  physical proof of this equation in Appendix \ref{app:kaneFuFormula}.

The additional insight of Ref. \onlinecite{turner2009} is that $P_3$ is also odd under inversion symmetry.  This means that the values of $P_3$ are still quantized to be $0,1/2$ even when the system does not have time-reversal symmetry but only inversion. We find that this argument holds for topological responses in all even spacetime dimensions (see Appendix \ref{app:higherD}). Thus, there are inversion symmetric topological insulators in $3D$ where $P_3=1/2$,  analogous to the case in $1D$ where  $P_1=e/2.$ This argument is an indicator that inversion symmetric insulators can support non-trivial topological states with interesting response properties.  Now we can ask the question, is Eq. \ref{eq:kaneInvariant} still valid when only inversion symmetry is preserved?
 If we only have inversion symmetry and no time-reversal symmetry, then this formula continues to apply if we can adiabatically connect the system to the T and ${\cal{P}}$ invariant limit without breaking inversion symmetry. However this is not the only case,  and we will show exactly when the inversion eigenvalues in 3D indicate a non-trivial magneto-electric polarizability protected by inversion symmetry in the following section. Our arguments use many of the results discovered in the previous sections.

\subsubsection{Magneto-electric Polarizability for Inversion Invariant Insulators}
We begin by reintroducing the unitary matrix
\begin{equation}
B_{ij}(k)=\bra{u_{i,-k}}{\cal{P}}\ket{u_{j,k}}
\end{equation}\noindent where $u_{i,k}$ is a Bloch function with $i,j$ labeling which occupied band and $k$ is the Bloch momentum.  An important property of the matrix $B$ is
\beq
B(-k) = B^\dagger(k)
\eneq \noindent which is true because ${\cal{P}}$ is unitary and squares to the identity matrix. This means that at the inversion symmetric points, the matrix is real and symmetric.
For inversion symmetric insulators, we prove in Appendix \ref{app:windNumb} that the non-Abelian adiabatic connection satisfies:
\beq \hat{A}_{i}(-k) = - B \hat{A}_{i}(k) B^\dagger + i B(k) \nabla_i B^\dagger(k)
\eneq which implies that the adiabatic curvature is gauge covariant via:
\begin{eqnarray}
& \hat{F}_{ij}(-k) = B(k) \hat{F}_{ij}(k) B^\dagger(k).
\end{eqnarray}  From this the magneto-electric polarizability is easy, but tedious (see Appendix \ref{app:windNumb}) , to obtain:
\begin{eqnarray}
2P_3&=&- \frac{1}{24 \pi^2} \int d^3 k \epsilon^{ijk} Tr[(B(k) \partial_i B^\dagger(k))\nonumber\\ & &\cdot (B(k) \partial_j B^\dagger(k))  (B(k) \partial_k B^\dagger(k))  ].\label{eq:P3winding}
\end{eqnarray} This proves that $P_3$ is either integer or half-integer depending on whether the RHS (which is an integer winding number) is even or odd. Since $P_3$ is itself defined mod $1$ it means that $P_3$ defines a $Z_2$ classification that indicates a non-trivial topological response.

With time-reversal symmetry, states pair up in Kramers' doublets and are degenerate at time-reversal invariant points (same as inversion symmetric points). If we break all accidental degeneracies in a time-reversal invariant system we are still left with an \emph{even} number of occupied bands - they are degenerate in pairs at time-reversal symmetric points. The $B(k)$ matrix then, at most, decomposes into diagonal blocks of decoupled  $U(2)$ matrices. The winding number can be non-trivial in this case because the homotopy group $\pi_3(U(2)) = Z$ (the $Z_2$ nature appears because $P_3$ is defined as the winding number mod 2).   Once we lose time-reversal symmetry the bands do not have to be degenerate at the $k_{inv}.$ In fact,  once time reversal is softly broken, the bands at $k_{inv}$ experience eigenvalue repulsion: before TR was broken, the inversion eigenvalues of the Kramers' doublets had to be identical.  As such, it seems that once time-reversal is broken we can completely separate the occupied bands and isolate them  from each other at all points in the Brillouin zone.  This would imply  that the matrix $B$ could be reduced to an $N\times N$ diagonal matrix with $U(1)$ phases on the diagonal. In other words,  $B(k)$ will map from 3D momentum space into $U(1)^{\otimes N}.$ Since $\pi_{3}(U(1)^{\otimes N})=0$ this implies that the winding number in Eq. \ref{eq:P3winding} always vanishes (note that by making this statement we are implicitly assuming that all of the mappings are smooth which we will come back to later). This trivial reasoning would seem to imply that we cannot get a topological insulator with inversion symmetry.  Fortunately, this line of reasoning fails because of the non-trivial global constraint that the product of all the inversion eigenvalues must be positive.  Before we deal with the effects of the constraint we draw several conclusions about a Hamiltonian with only one occupied band.  The non-Abelian form of the winding number implies that with only one occupied band $P_3\equiv 0.$  Thus, no matter what the inversion eigenvalue content, there is no contribution to $P_3$ from a system with a single occupied band. This also holds true for a band that can be completely separated and untangled from all the other bands at all  points in the Brillouin zone.  Such isolated bands do not contribute to a non-trivial $P_3.$ This is in contrast to the statement at the end of Ref. \onlinecite{wan2010} which seems to state that a non-trivial $P_3$ only requires an odd number of pairs of negative inversion eigenvalues. As a counter-example, a single occupied band can have a single pair of negative parity eigenvalues (which occur at different $k_{inv}$) and it has vanishing $P_3.$ We give such a Hamiltonian in Eq. \ref{eq:3dQHEHam}. We will see that  the important thing to consider is pairs of negative parity eigenvalues at the \emph{same} $k_{inv}.$

The global constraint on the inversion eigenvalues is crucial for our discussion. We first give an explicit example to gain intuition about its importance. Let us assume that $N=2$ and that the occupied bands have  $\zeta_{1}(0,0,0)=\zeta_{2}(0,0,0)=-1$ and all other inversion eigenvalues at all other $k_{inv}$ are $+1.$ The naive reasoning from above implies that by perturbing this Hamiltonian while preserving inversion symmetry we should be able to separate these bands so that they are isolated with no intermingling degeneracies. However, if we could make the two bands non-degenerate everywhere we would contradict the constraint $\chi_{\cal{P}}=+1.$ This is easy to see because if we could separate the two bands we could consider a different insulating ground state with only one of the previous two bands occupied and then the product of the inversion eigenvalues of the single occupied band would be negative.  But we have proved that this is not possible, so we cannot make the bands fully non-degenerate over the entire Brillouin zone.
Alternatively, if we could separate the bands, we would have a single band that has eigenvalues $-+++$ in the $k_z=0$ plane  and $++++$ on the $k_z=\pi$ plane. If we consider the 3D Hamiltonian as a gapped interpolation between these two planes then we have adiabatically connected 2D Hamiltonians with odd and even Chern number respectively. This cannot happen and is thus another contradiction.
Hence, we can be sure that the two occupied bands are degenerate at at least two points in the Brillouin zone (by inversion symmetry). These two points are exactly enough to transfer an even Chern number from one plane to the other which fixes the Chern number issue. Due to this topological obstruction, the $\hat{B}(k)$ map cannot be reduced to a map from $T^3$ into $U(1)\otimes U(1)$, but instead it is a true map from $T^3$ into $U(2)$ and therefore its winding can take non-trivial values. Additionally, the winding number for each $U(2)$ block is additive due to the trace in the winding number.

The conclusion at this step is that, in general, the $B(k)$ matrix can be decomposed into one and two dimensional diagonal blocks. The one dimensional blocks have no contribution to the winding number. The two dimensional blocks come from paired bands that cannot be disentangled. Following the previous arguments, one can see that this happens only for the following inversion eigenvalues patterns at the eight $k_{\mbox{\tiny{inv}}}$ (here we restrict the projector to just these bands):\smallskip
 
\noindent${^-_-} {^+_+} {^+_+} {^+_+}  {^+_+} {^+_+} {^+_+} {^+_+}, \ {^-_-} {^-_-} {^-_-} {^+_+}  {^+_+} {^+_+} {^+_+} {^+_+}, \ {^-_-} {^-_-} {^-_-} {^-_-}  {^-_-} {^+_+} {^+_+} {^+_+}, \ {^-_-} {^-_-} {^-_-} {^-_-}  {^-_-} {^-_-} {^-_-} {^+_+} $
\smallskip

\noindent Note that the pattern $(^+_-)$  at the \emph{same} $k_{\mbox{\tiny{inv}}}$ is excluded in all these cases. The reason is because, if the pattern $(^+_-)$ shows up at a $k_{\mbox{\tiny{inv}}}$, then it will necessarily show up at another $k_{\mbox{\tiny{inv}}}$ in order to make $\chi_{\cal P}$=1. In this case, a simple exercise will show that the inversion eigenvalues can always be separated in two groups with the product of inversion eigenvalues in each group equal to one. Thus, there is no topological obstruction and the bands can be untangled. The same thing happens for all the other possible inversion eigenvalues that are not listed above. We will discuss more about this below and we now return to the calculation of the winding number.

We first discuss the winding number in the continuum (sphere) and then focus on the lattice (torus). We know from Ref. \onlinecite{kitaev2008} that when considering the isotropic topological invariants such as $P_3$ it is sufficient to think of momentum space as a sphere $S^3$ instead of the Brillouin zone torus $T^3.$ Allowing for the full torus structure gives a rich set of anisotropic states which we will consider later, but for now we assume that the momentum space topology is spherical. This effectively reduces the number of invariant momenta we need to consider to just two: the origin and the `point at infinity.' Equivalently we could think of the torus with unrestricted inversion eigenvalues at $k_{inv}=(0,0,0)$ but with the inversion eigenvalues at all the other $k_{inv}$ constrained to be equal band by band. 
Now consider a Hamiltonian with $N$ occupied bands (again $N$ does not have to be even, a crucial difference with the time-reversal case). Since we are in 3D we know that the product of all the inversion eigenvalues must be $+1.$  This means that there can only be an even number of inversion eigenvalues that are different between $k=0$ and $k=\infty.$  For example, the number of negative eigenvalues at $k= \infty$ must have the same parity (i.e. even or odd) as the number of negative eigenvalues at $k=0.$  It is clear that, by exchanging parity eigenvalues between bands, either at $k=0, \infty$, we can split the bands into two classes: (i) bands with $\chi_{P}=+1$ and (ii) pairs of bands with $\chi_{P}=-1$ for each band. We see case (i) when the eigenvalues match at $k=0$ and $\infty$ and case (ii) when they are opposite.  For case (i) the bands can be isolated from each other, but in case (ii) they must generically be in tangled pairs where the $\chi_{P}=+1$ for the pair. Thus we can understand both cases by considering just $2$ occupied bands. For example, a trivial case is that of  two bands with no negative inversion eigenvalues. This is a realization of case (i) where the inversion eigenvalues of each band separately multiply to $+1.$ The global constraint does not prevent us from isolating all the occupied bands and  thus, with all eigenvalues positive, $P_3=0.$  All realizations of case (i) are trivial for the same reason. 

The first interesting case is that of two bands with a single pair of negative inversion eigenvalues at the same point (say $k=0$) but positive inversion eigenvalues \emph{at} the other point. We consider that case now. The important consequence of the global constraint, as we just saw, is that the two bands with the negative eigenvalues can never be completely separated from each other - the (generically) two degeneracy points between them cannot be lifted or annihilated. $B(k)$ restricted to these two-bands  is a $U(2)$ matrix and generically takes the form:
\beq
B(k) = e^{i \phi(k) } (f(k) I + i g_a(k) \sigma_a)
\eneq where
\beq
(f(k))^2 + g_a(k) g_a(k) =1.
\eneq\noindent There is a global $\pm$ sign ambiguity in the choice of $f(k)$ but once the sign is chosen at one point, smoothness  assures the signs at the other points. This ambiguity does not have implications for the final result.  If we substitute this form into Eq. \ref{eq:P3winding} all of the dependence on $\phi (k)$ (\emph{i.e.} the $U(1)$ part) drops out as long as $e^{i\phi(k)}$ is smooth (see Appendix \ref{app:U1ind}. Since all loops in $S^3$ are contractible we can gauge transform $B(k)$ to remove the k-dependent phase so the assumption of smoothness is not an issue.
What remains is the winding of the $SU(2)$ part which must be an integer. Now, with only the $SU(2)$ part we know that due to $B(k) = B^\dagger (-k)$
\beq
f(k) = f(-k), \;\;\; g_a(k) = -g_a(-k).
\eneq If we think of $S^3= R^3\cup \{\infty\}$
then it is easy to consider the general form of $B.$ The function $f(k): S^3\to R $ and its derivative vanishes at $k=0$ (and $k=\infty$). This is a Morse function for the sphere and we can expand it around the origin to find:
\begin{eqnarray}
f(k) &=&N(k)( M+k_a C_{ab} k_b+\ldots)
\end{eqnarray} where $C_{ab}$ is a $3\times 3$ constant matrix with three non-zero eigenvalues of the same sign\cite{Milnorbook}, and  $N(k)$ is a normalization factor which is even in $k$ and constrains the matrix $B$ to have unit determinant.

Without loss of generality we choose the case when the eigenvalues of $C_{ab}$ are all positive (another reason the eigenvalues have to have the same sign is to fix the boundary condition at $k=\infty$ so that it is independent of the path taken to get there). This choice determines which sign of $M$ will lead to the non-trivial phase.   Similarly we can expand the function
\begin{equation}
 g_a(k) = N(k) (D_{ab} k_b +\ldots)
\end{equation}\noindent for a $3\times 3$ constant matrix $D_{ab}$ with no restriction on the eigenvalues.
Generically,  $D_{ab}$ will have non-zero determinant (\emph{i.e.} it will have rank 3). In cases where the determinant of $D_{ab}$ is tuned to zero, we have to look use a higher order Taylor-expansion in \emph{both}  $f(k)$ and $g_a(k)$ - maintaining even terms in $f(k)$ and odd terms in $g_a(k).$ As long as the boundary conditions are fixed, which requires us to keep terms in $f(k)$ with higher order than $g_a(k)$, this will not change the result. Without loss of generality we take the case $\det D\neq 0,$ and by rescaling and rotating we transform to the momentum space basis $(k_1,k_2,k_3)$ with $C_{ab}=D_{ab}=\delta_{ab}.$ For this choice we have:
\beq
N(k) = \frac{1}{\sqrt{(M+k^2)^2 + k^2}}
\eneq\noindent with $k^2=k_{1}^2+k_{2}^2+k_{3}^3.$
 For this $B(k)$ we have
\beq
 B(0) = \text{sgn}(M) I_{2\times 2}, \;\;\;\;\;\;\;\;\;  B(\infty) = I_{2\times 2}.
\eneq
 By explicit calculation, we find that:
 \beq
 P_3 = \frac{1}{\pi} \int_0^\infty \frac{(M-k^2)}{((M+k^2)^2 + k^2)^2}  k^2 dk = \frac{\text{sign}(M) -1 }{4}.
\eneq We notice that when there is an eigenvalue switch when passing from zero to infinity, we have $P_3 =1/2,$ whereas when there is no eigenvalue switch, we have $P_3 = 0$. Although more terms can be kept in the expansion around the origin this does not influence the result of the winding number, as long as the eigenvalues of $B(0)$ and $B(\infty)$ are not changed.

We have seen from this simple argument that for two bands which cannot be separated, the contribution to $P_3$ depends only on the change in inversion eigenvalues. If there is only a single pair of bands which is in case (ii) \emph{i.e.} cannot be untangled from its partner,  then the other $N-2$ occupied bands are not constrained and may  each be isolated away from all other bands. Each of these isolated bands does not contribute to $P_3$ and thus the non-trivial value of $P_3$ only comes from the two tangled bands. 
To finish the proof we must consider the case when there are more than one set of tangled occupied bands. If, for example, there are four bands which have negative eigenvalues at $k=0$ (for simplicity we fix all the eigenvalues at $k=\infty$ to be positive) we can generically isolate the four bands into two pairs. The two pairs can be separated from each other and all other bands, but the bands making up a single pair must share degeneracies from the arguments above. Once we have decoupled the inversion eigenvalues of  the two pairs, we can remove all the accidental degeneracies and isolate the pairs from each other because a combined  pair of bands with negative inversion eigenvalues by itself does not contradict the global constraint. Since the pairs can be isolated, they contribute independently to $P_3.$ Each pair will contribute a half-integer giving $P_3= n = 0\; {\textrm{mod}}\; Z.$ We see there is an even odd effect so that an odd number of pairs of bands with negative inversion eigenvalues is non-trivial while an even number of pairs is trivial. 

To complete the picture we will discuss how these arguments carry over to the lattice case when momentum space is a torus $T^3.$ We now have eight invariant momenta to consider which can lead to many more different combinations of inversion eigenvalues.  We will not enumerate all the possible phases but instead construct the necessary general principles to classify such states.  We again consider a set of N occupied bands which does not have to be an even number. In general the only restriction is that the product of all the inversion eigenvalues of all the occupied bands is equal to one. We can generically perform band crossings only between the occupied bands to split the bands into three possible classes:  (i) a set of $n_{+}$  bands with positive inversion eigenvalues at all $k_{inv}$ , (ii)  a set of  $n^{e}_{-}$  bands with an even number of negative eigenvalues on each band, and a set of $n^{o}_{-}$  bands where the product of eigenvalues on each band is $-1$ and the product \emph{cannot} be made equal to $+1$ via band crossings among the other bands in $n^{o}_{-}.$ From the arguments above, the sets of  $n_{+}$ and $n^{e}_{-}$ bands contribute nothing to $P_3,$ because they can be isolated one-by-one from all of the occupied bands.  Note that the $n^{e}_{-}$ can contribute to non-trivial 3D QHE states in the same manner shown above. The number of bands in the third class $n^{o}_{-}$ must be an even number to satisfy the global constraint.  We will now show that value of $P_3 = (1/2)n^{o}_{-}\;{\textrm{mod}}\; Z.$ We call this process the band decoupling process and we give explicit examples of this band decoupling picture for lattice models shown in Appendix \ref{app:invExamples}.

We know that the only bands which can contribute to $P_3$ are those in $n^{o}_{-}.$ We can consider these bands  as a set of $n^{o}_{-}/2$ decoupled pairs. Each one of them contributes a $U(2)$ block to the $B(k)$ matrix. We show in Appendix \ref{app:smoothPhase} that once the band decoupling process is finished then the $U(1)$ phase in each of the  $n^{o}_{-}/2$ $U(2)$ blocks is smooth and can thus be completely eliminated from consideration. Thus each $U(2)$ block can be contracted to $SU(2).$ Since this implies that we are really considering maps from $T^3\to SU(2)$, which have the same dimension, we can connect the winding number of each $SU(2)$ block of $B(k)$ to the degree of the map. This argument follows along the same lines shown in Ref. \onlinecite{wang2009} so we will not include all the details.  To calculate the degree of the map from $T^3$ to an $SU(2)$ block of $B(k)$ we can choose any point in $SU(2).$ For example,  we could choose $-I_{2 \times 2}.$    Since $B(k)= B^\dagger(-k),$ any $k$ which is not inversion symmetric contributes to the degree of the map twice,  \emph{i.e.} if $-I_{2 \times 2}$ occurs at $k$ it will occur at  $-k$, thereby leaving $P_3$ unchanged. The only contributions, therefore,  come from the inversion symmetric points, and hence the winding number counts the number of inversion symmetric points that have $-I_{2 \times 2}$ as their inversion eigenvalues.
This implies that we can simply apply the Kane-Fu formula\cite{fu2007a} to the bands in $n^{o}_{-}$ to determine the value of $P_3.$

 \section{Simple Example Models}\label{sec:examples}
In this section we provide a set of explicit models that illustrate the majority of the technical details discussed in the previous sections. For each model we list the interesting phases, inversion eigenvalue structure, and describe what the entanglement spectra should look like. Additionally we provide figures showing the entanglement spectra for each model which confirms our analytic formulae. The models we chose are similar to ones used in many contexts especially in the field of topological insulators. Combined with the models introduced in Sec. \ref{sec:prelim} these examples provide valuable intuition about inversion invariant topological insulators and the similarities and differences between the states protected by inversion symmetry and those protected by other discrete symmetries \emph{e.g.} charge-conjugation, or time-reversal. 
\subsection{1D Models}
We have already introduced two illustrative 1D models for which we will analyze the entanglement spectrum. Additionally we will introduce a 1D model of a dimerized chain.\cite{su1979}

\subsubsection{Simple two band model}
\begin{figure}
  \includegraphics[width=7cm]{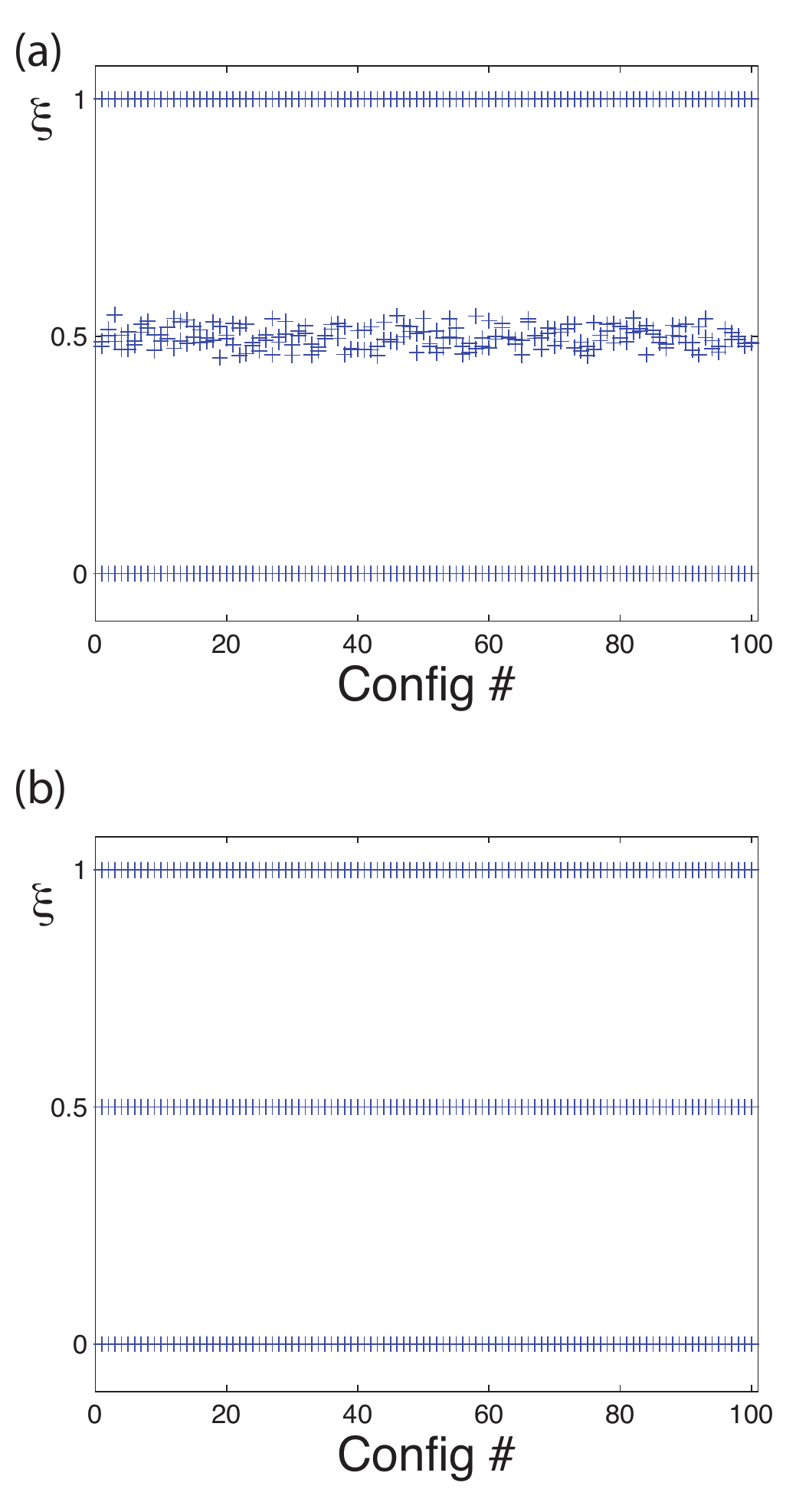}\\
  \caption{Entanglement spectrum for the two-band 1D model with (a) random site disorder (b) random site disorder which is inversion symmetric around the center of the 1D chain. We show the entanglement spectra for many different random disorder configurations.}
 \label{fig:TwoBandDisorder}
\end{figure}

Here we focus on the Hamiltonian given in Eq. \ref{eq:simple1d2band} and reproduced here
\begin{eqnarray}
\hat{H}_2(k)=\alpha \cos k+ \sin(k)\hat{\sigma}_1+(1+m-\cos k)\hat{\sigma}_3.\nonumber
\end{eqnarray} 
 This model has one occupied band, and in 1D there are two inversion symmetric momenta $k=0,\pi.$  There is a phase transition in this model between two insulating phases as a function of the parameter $m$ and we have previously characterized the two phases of this model by examining the inversion eigenvalues of the occupied band. For $m>0$ the occupied band has two negative inversion eigenvalues. This implies that this system can be adiabatically continued to an atomic limit where the occupied atomic orbital has a negative inversion eigenvalue. In the atomic limit the entanglement entropy is identically zero and so we expect on physical grounds that there should be no stable entanglement eigenvalues at $1/2.$ Using our inversion criterion we see that this is the case since the inversion eigenvalues do not change between $0$ and $\pi.$ This same characterization applies for $m<-2$ where the two occupied bands have positive inversion eigenvalues. The final case is for $-2<m<0$ where $\zeta (0)=-\zeta (\pi)=+.$ This cannot be adiabatically connected to an atomic limit. Here the inversion criterion implies we should see a pair of entanglement modes at $1/2$ which is shown in Fig. \ref{fig:simple1d2band}c . Thus the entanglement spectrum is a good indicator for a non-trivial topological insulator.  The trivial case is shown in Fig. \ref{fig:simple1d2band}d which has no $1/2$ mode. We can compare these results to the values of the topological invariant $\chi$ which takes on values  $\chi=0$ when $m<-2$ or $m>0$ and $\chi=1$ when $-2<m<0.$ The number of required $1/2$ modes in the entanglement spectrum is $0,$ $0$ and $2$ respectively which is exactly what we found numerically. 
 
 To illustrate the protection due to inversion symmetry we also consider $H_2$ with an onsite disorder term added
 \begin{equation}
 H=H_2+\sum_{i}W_i c^{\dagger}_{i}c_{i}
 \end{equation}\noindent where the $W_i$ are randomly chosen from a uniform distribution $[-W/2,W/2].$ If it is purely random, uncorrelated disorder the mid-gap entanglement modes are no longer protected as shown in Fig. \ref{fig:TwoBandDisorder}a. Next we mirror the disorder around the center of the chain to make a system with inversion symmetric disorder. Although unphysical, this helps illustrate the fact that only inversion symmetry is required for the protected mid-gap entanglement states as we show in Fig. \ref{fig:TwoBandDisorder}b where the cut is along the remaining inversion center.  
\subsubsection{Simple four band model}
The Hamiltonian for the simple four band model was given in Eq. \ref{eq:simple1d4band} and the set of inversion eigenvalues for the different phases were listed in Eqs. \ref{eq:4bandCase1}-\ref{eq:4bandCase5}. For convenience we reproduce the Hamiltonian here
\begin{equation}
\begin{array}{c}
\hat{H}_4(k)=\sin(k) \hat{\Gamma}_1 +\sin(k) \hat{\Gamma}_2 \medskip \\ 
+(1-m-\cos k)\hat{\Gamma}_0 +\delta \hat{\Gamma}_{24}+\epsilon \cos(k) (1+\hat{\Gamma}_0).
\end{array}\nonumber
\end{equation}
 From the inversion criterion we see that Case 2) and Case 3) should have entanglement modes at $1/2$. In fact, Case 2) should have a pair of modes localized on each cut, one for each occupied band that flips the sign of the inversion eigenvalue. The entanglement spectra for the five cases are shown in Fig. \ref{fig:4bandEnergyEspec}f-h and they agree with the analytic prediction. For cases 1-5 the invariant $\chi=0,2,1,0,0$ yielding $0,4,2,0,0$ entanglement modes at $1/2$ which is what we found numerically. 
\subsubsection{Dimerized Chain}
\begin{figure}
  \includegraphics[width=7cm]{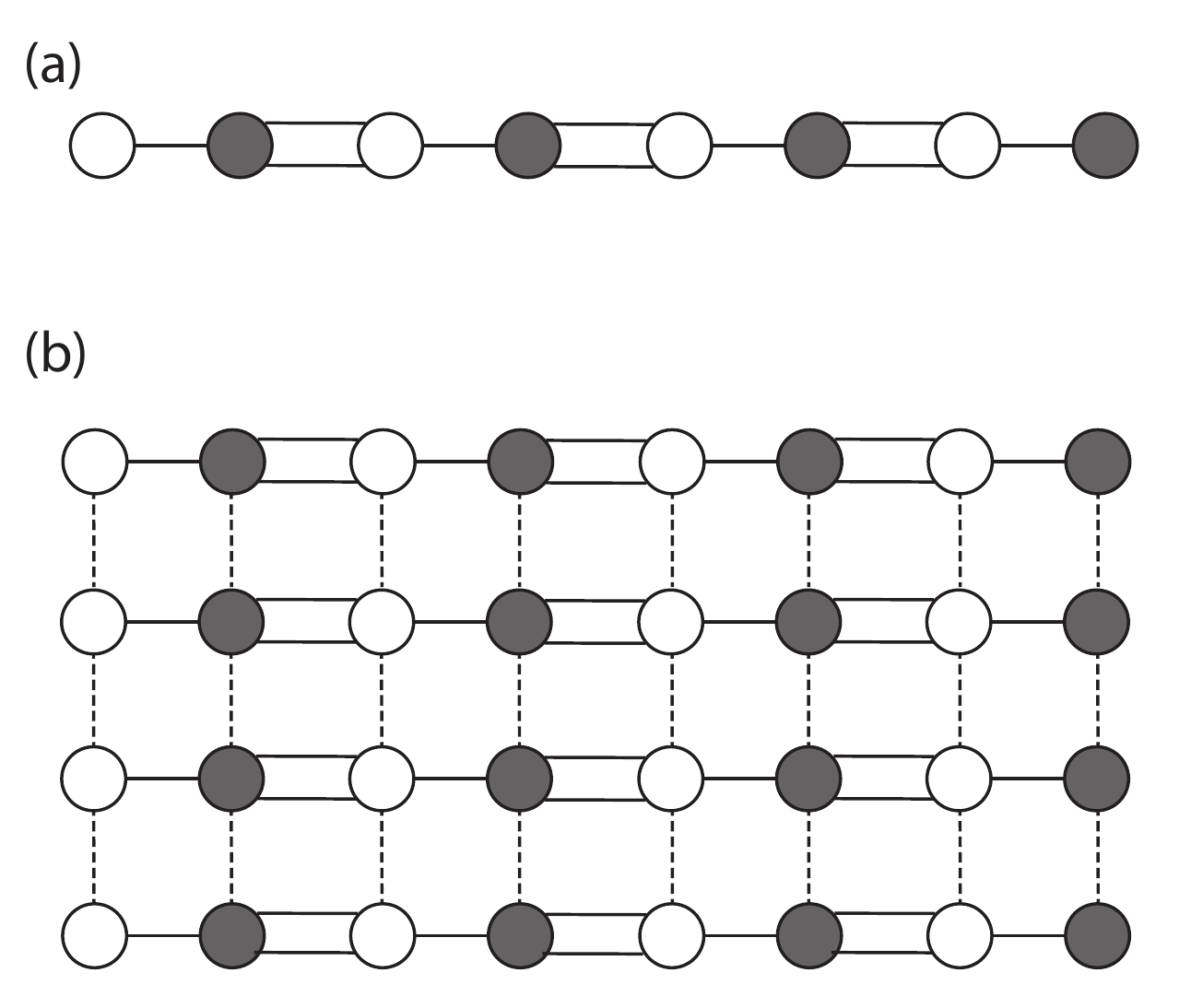}\\
  \caption{(a)1d Dimerized Chain (b)2d Dimerized square lattice model. Solid and dotted lines represent different hopping amplitudes. }
 \label{fig:DimerModels}
\end{figure}
\begin{figure}
  \includegraphics[width=7cm]{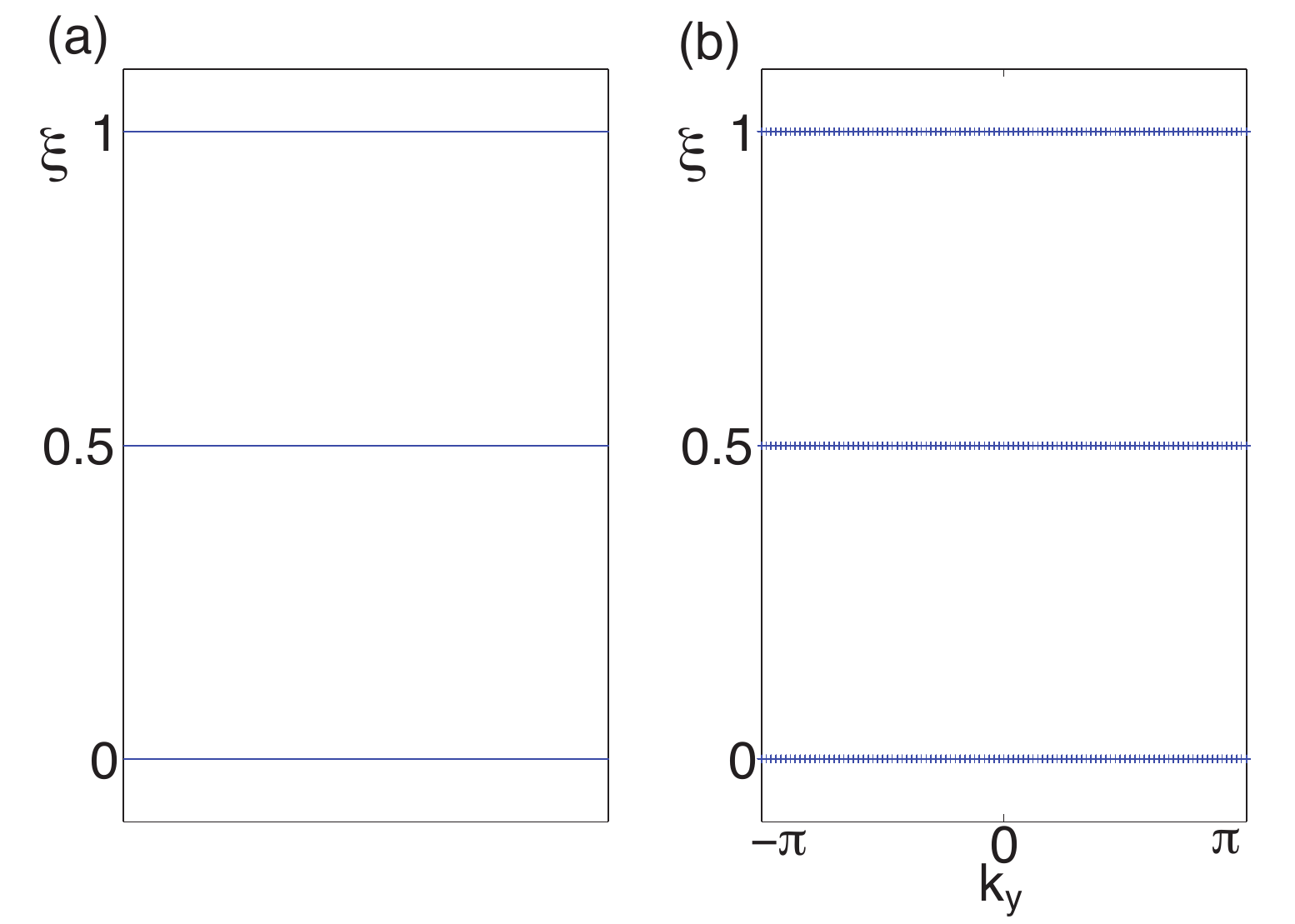}\\
  \caption{Entanglement Spectra for (a)Dimerized Chain(b)Dimerized Square Lattice}
 \label{fig:2dDimerEntanglement}
\end{figure}
As a final 1D test case we will look at spinless fermions hopping on a dimerized chain. This model is the familiar Su-Schrieffer-Heeger model for electrons in a polyacetylene chain.\cite{su1979} Later we will extend this model into 2D and 3D to illustrate anisotropic systems with non-trivial entanglement spectra. The layout of the chain is shown in Fig. \ref{fig:DimerModels}a along with the choice of two-atom unit cell. The Hamiltonian is given by
\begin{eqnarray}
H&=&\sum_{m}\Delta \left(c^{\dagger}_{mA}c_{mA}-c^{\dagger}_{mB}c_{mB}\right)\nonumber\\&+&\left((-t-\delta)c^{\dagger}_{mA}c_{mB}+(-t+\delta)c^{\dagger}_{mB}c_{m+1A}\right.\nonumber\\ &+&\left. \;{\textrm{h.c.}}\right)\label{eq:dimer1dH}
\end{eqnarray}\noindent where $A,B$ indicate sublattice $A$ or $B$ and $t>0.$ For our purposes we will set the onsite energy $\Delta=0.$ With this choice the Hamiltonian has an inversion symmetry with respect to the middle of a bond with ${\cal{P}}=\sigma^x$ \emph{i.e.} ${\cal{P}}$ exchanges sublattices $A$ and $B.$ 

The Hamiltonian can be Fourier transformed and becomes
\begin{eqnarray}
H&=&\sum_{k}\Psi^{\dagger}_{k}\left[\left(-(t+\delta)-(t-\delta)\cos k\right)\sigma^x\right. \nonumber\\&+&\left.(t-\delta)\sin k\sigma^y\right]\Psi_k\nonumber \\
\Psi_k&=&\left( c_{kA}\;\; c_{kB}\right)^{T}.\label{eq:1dDimer}
\end{eqnarray}\noindent  This model has two insulating phases (i)$\delta<0$ (ii) $\delta>0.$ At the two inversion invariant points we have the Bloch Hamiltonians $\hat{H}(0)=-2t\sigma^x$ and $\hat{H}(\pi)=-2\delta\sigma^x.$ As expected they both commute with ${\cal{P}}.$ For fixed $t=1$ we see that the inversion eigenvalue of the occupied band at $k=0$ is fixed to be $+.$ For $k=\pi$ the eigenvalue depends on the sign of $\delta.$ So for $\delta<0 (>0)$ we have $\zeta (\pi)= -1 (+1).$ We see that the non-trivial phase occurs when $\delta<0.$ In this case the Wannier center of the electrons is shifted to the mid-bond center \emph{between} unit cells. Thus if we cut the system between unit cells there will be a charge polarization. If $\delta>0$ the Wannier center is on the mid-bond center \emph{within} a unit cell. Thus our definition of unit cells is important and simply reflects the fact that the polarization is not well-defined absolutely but is gauge-variant. 

The entanglement spectra for this model for $\delta<0$ is shown in Fig. \ref{fig:2dDimerEntanglement}a. We expect that because the inversion eigenvalues change from positive to negative between $k=0$, and $\pi$ that there should be $1/2$ modes in the spectrum and this is confirmed numerically as seen in the figure. The topological invariant $\chi=0,\; 1$ for $\delta>0$ and $\delta<0$ respectively. The number of the entanglement modes in the spectrum matches the value of $2\chi$ as expected for a system with periodic boundary conditions and thus two entanglement cuts. 

There is a subtlety in the entanglement characterization of this model which we will now discuss. The real-space Hamiltonian as written in Eq. \ref{eq:dimer1dH} has broken translational symmetry.  As stated earlier, our classification method only applies to translationally invariant Hamiltonians since we need to evaluate inversion eigenvalues in momentum-space. However, by construction, the Hamiltonian for the dimerized chain only has a very mild breaking of translational symmetry. In fact, as implicitly assumed in the analysis, we can just define a unit cell encapsulating two-sites, and in terms of the larger unit cell the Hamiltonian is translationally invariant and our method applies. Our choice for a unit cell has already been implicitly assumed by the time we write the Bloch Hamiltonian in Eq. \ref{eq:1dDimer}. The choice made here is that sites connected by the hopping term $-(t+\delta)$ form a single unit cell and the hopping between unit cells is $(-t+\delta).$ After making this choice we are free to carry out the entanglement analysis by cutting the system \emph{between} unit cells. Cutting the system \emph{within} a unit cell is not a position-space cut of a translationally invariant Hamiltonian. 

Now, the important outstanding question is: is the analysis invariant under the choice of unit cell? It is, and we explicitly show it for this model.  Suppose that we take the unit cell to be sites connected by the hopping $(-t+\delta).$ Then the Bloch Hamiltonian becomes
\begin{eqnarray}
\bar{H}&=&\sum_{k}\Psi^{\dagger}_{k}\left[(-(t-\delta)-(t+\delta)\cos k)\sigma^x\right.\nonumber\\
&+&\left. (t+\delta)\sin k\;\sigma^y\right]\Psi_k\nonumber\\
\bar{H}(0)&=&-2t\sigma^x\;\;\; \bar{H}(\pi)=2\delta\sigma^x.
\end{eqnarray}\noindent If we fix $t>0$ then the inversion eigenvalue for the occupied band at $k=0$ is always  positive. Thus we see that for $\delta>0$ the system is in a non-trivial phase with opposite inversion eigenvalues at $0$ and $\pi,$  and for $\delta<0$ the system is trivial. The sign of $\delta$ that exhibits the non-trivial phase has changed when compared with the choice of the other unit cell. The physics, however, remains identical. Now when $\delta>0$ the Wannier centers are shifted to the bonds \emph{between} the new unit cells. This would have lied \emph{within} the unit cell for our previous choice and explains why the sign of $\delta$ has changed. For the choice of unit cell in $\bar{H}$ the entanglement spectrum will have mid-gap modes when $\delta>0$ \emph{i.e.} when the inversion eigenvalues are opposite at the two invariant momenta. Thus we see that for the new choice of unit cell the physics and entanglement analysis yields the same results. 
 
\subsection{2D Models}
\subsubsection{Dimerized Square Lattice Model}
The first 2D model we consider is a trivial extension of the dimerized chain as illustrated in Fig. \ref{fig:DimerModels}b. This extension has a Hamiltonian which is simply constructed from Eq. \ref{eq:1dDimer}:
\begin{eqnarray}
H&=&\sum_{k}\Psi^{\dagger}_{k}\left[\left(-(t+\delta)-(t-\delta)\cos k_x\right)\sigma^x\right. \nonumber\\&+&\left.(t-\delta)\sin k_x\sigma^y-2t_y\cos k_y\right]\Psi_k.
\end{eqnarray} This model has an inversion symmetry with ${\cal{P}}=\sigma^x.$ At the inversion invariant points we have
\begin{eqnarray}
\hat{H}(0,0)&=&-2t_y -2t\sigma^x\nonumber\\
\hat{H}(\pi,0)&=&-2t_y -2\delta\sigma^x\nonumber\\
\hat{H}(0,\pi)&=&2t_y-2t\sigma^x\nonumber\\
\hat{H}(\pi,\pi)&=&2t_y -2\delta\sigma^x.\nonumber
\end{eqnarray}\noindent We see that although this is more complicated than the 1D case everything still commutes with ${\cal{P}}.$ For simplicity we pick $t=2t_y=1$ and focus on the two gapped phases $\delta<0$ and $\delta>0.$ In these two phases we have the following set of eigenvalues
\begin{eqnarray}
\delta&>&0:\left\{\begin{array}{cc}\zeta(00)=& +\\ \zeta(\pi 0)= &+\\ \zeta(0\pi)=& +\\ \zeta(\pi\pi)= & +\end{array}\right.\nonumber\\
\delta&<&0:\left\{\begin{array}{cc}\zeta(00)=& +\\ \zeta(\pi 0)= &-\\ \zeta(0\pi)=& +\\ \zeta(\pi\pi)= & -\end{array}\right. .\nonumber
\end{eqnarray}\noindent The product of all the eigenvalues in each case is $+1$ so the parity of the first Chern number for these two cases is even. In fact, it is exactly zero for this model. Just as in the 1D case we see that the $\delta<0$ phase is interesting. Here the Wannier center for each electron is moved along the $x$-axis to the mid-bond center \emph{between} each unit cell. If we cut an edge which is perpendicular to the x-direction there will be a finite charge density on the boundary. Although the inversion symmetry is not enough to determine the polarization we see that this Hamiltonian also has a reflection symmetry ${\cal{M}}\hat{H}(k_x,k_y){\cal{M}}^{-1}=\hat{H}(-k_x,k_y)$ with ${\cal{M}}=\sigma^x={\cal{P}}.$ Thus, the charge polarization in the $x$-direction is quantized and equal to $P_1=e/2a$ where $a$ is the lattice constant in the $y$-direction.

Finally we can consider the entanglement spectra of these two phases. For $\delta>0$ the phase is adiabatically connected to the atomic limit and thus will not require the existence of $1/2$ modes. For $\delta<0$ we must first specify a cut direction to locate the $1/2$ modes. Let us first pick the cut to be parallel to the x-direction. Thus $k_x$ remains a good quantum number and we can ask whether or not there are $1/2$ modes at $k_x=0$ or $k_x=\pi.$ For $k_x=0$ we look at $\zeta (0,0)\zeta(0,\pi)=+1$ and for $k_x=\pi$ we consider $\zeta (\pi,0)\zeta(\pi,\pi)=+1.$ Thus for this cut there are no $1/2$ modes. Next we look at a cut parallel to the $y$-direction such that $k_y$ is a conserved quantum number. For $k_y=0$ we have $\zeta(0,0)\zeta (\pi,0)=-1$ and for $k_y=\pi$ we have $\zeta (0,\pi) \zeta(\pi,\pi)=-1$ which implies there will be $1/2$ modes at both $k_y=0$ and $\pi.$ The  entanglement spectrum for a cut parallel to the $y$-direction is shown in Fig. \ref{fig:2dDimerEntanglement}b. In this figure there are clear $1/2$ modes at $k_y=0,\pi.$ In fact, for this simple model there are $1/2$ modes for all values of $k_y$ though our criterion only constrains the modes at the inversion invariant points. 

\subsubsection{Chern Insulator}
\begin{figure*}
  \includegraphics[width=14cm]{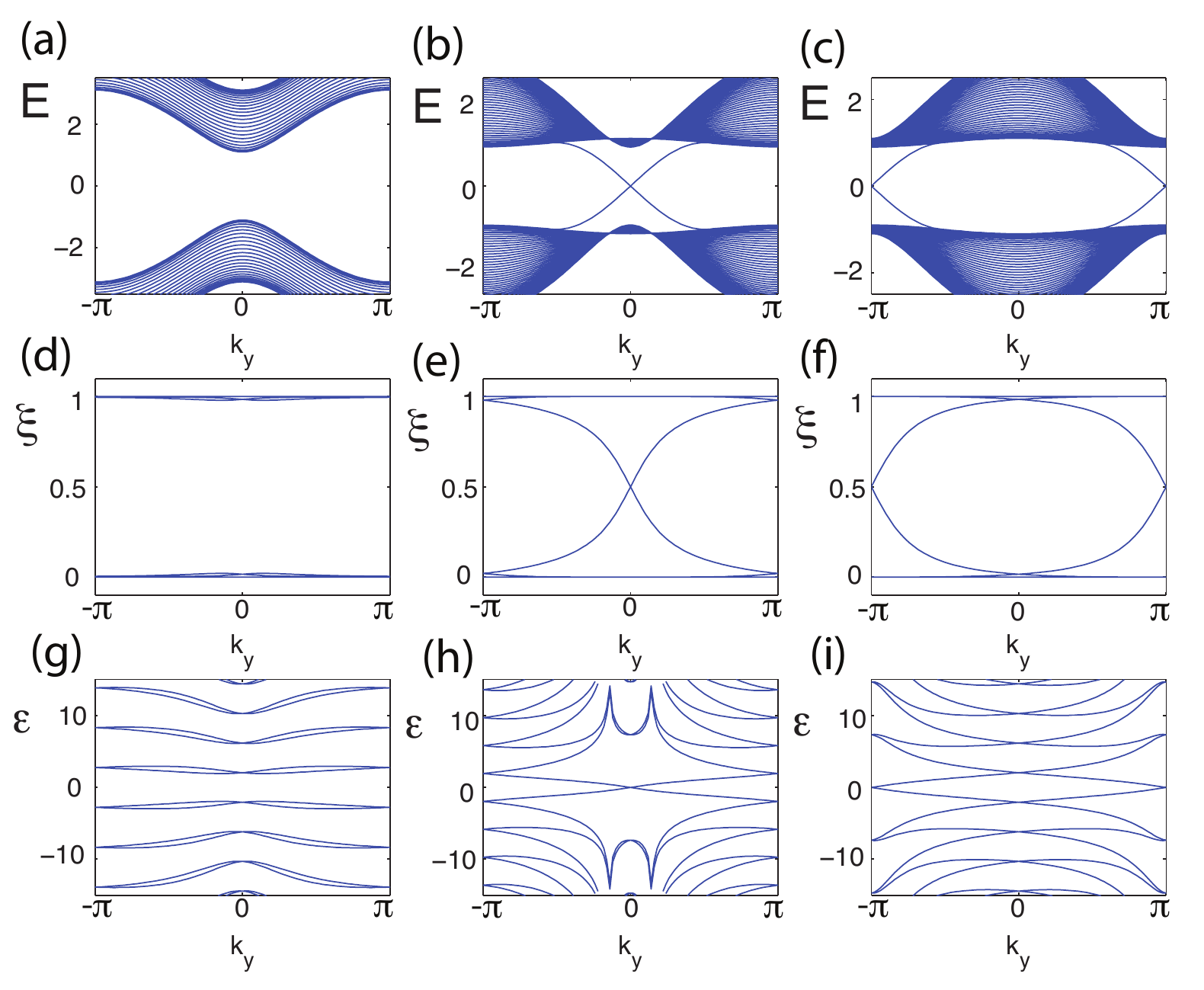}\\
  \caption{Energy spectrum with an open boundary, entanglement spectrum with a cut parallel to the y-direction, and entanglement energies for (Left panel a,d,g)A trivial insulator (Middle panel b,e,h)Non-trivial Chern insulator with negative inversion eigenvalue at $(k_x,k_y)=(0,0)$ implying entanglement mid-gap modes at $k_y=0$ (Right Panel c,f,i)Chern insulator with negative inversion eigenvalues at $(k_x,k_y)=(0,0), (\pi,0)$ and $(0,\pi).$ This implies entanglement mid-gap modes at $k_y=\pi.$}
 \label{fig:ChernInsulator}
\end{figure*}
Next we move on to study the well-known 2D topological insulators beginning with the Chern insulator (quantum anomalous Hall effect).\cite{Haldane1988} This is a 2D topological insulator which exhibits a quantum Hall effect and is classified by an integer invariant, the Chern number.\cite{thouless1982} Instead of studying the initially proposed honeycomb lattice model we will use the square lattice version which is nothing but massive Dirac fermions on a lattice. The Hamiltonian is
\begin{eqnarray}
H&=&\sum_{m,n}\left\{\frac{1}{2}\left[\Psi^{\dagger}_{m+1,n}(i\sigma^x-\sigma^z)\Psi_{m,n}\right.\right.\nonumber\\&+&\left.\left.\Psi^{\dagger}_{m,n+1}(i\sigma^y-\sigma^z)\Psi_{m,n}+{\textrm{h.c}}\right]\right.\nonumber\\&+&\left. (2-m)\Psi^{\dagger}_{m,n}\sigma^z\Psi_{m,n}\right\}.
\end{eqnarray} We can Fourier transform to get the Bloch Hamiltonian
\begin{eqnarray}
\hat{H}(k)&=&\sin k_x\sigma^x+\sin k_y\sigma^y+M(k)\sigma^z\\
M(k)&=&2-m-\cos k_x-\cos k_y.
\end{eqnarray}\noindent This model exhibits several different phases as a function of $m.$ For $m<0, m>4$ the system is in a trivial insulator phase and for $0<m<2, 2<m<4$ the system is in Chern insulator (quantum Hall) phases with Chern number $-1$ and $+1$ respectively. This Hamiltonian has an inversion symmetry with ${\cal{P}}=\sigma^z$ and at the four inversion invariant momenta we have
\begin{eqnarray}
\hat{H}(0,0)&=&-m\sigma^z \nonumber\\
\hat{H}(\pi,0)&=&2-m\sigma^z \nonumber\\
\hat{H}(0,\pi)&=&2-m\sigma^z \nonumber\\
\hat{H}(\pi,\pi)&=&4-m\sigma^z .\nonumber
\end{eqnarray}\noindent For $m<0,m>4$ the inversion eigenvalues are all positive/negative respectively. For $0<m<2$ $\zeta(0,0)=-1$ and $\zeta (\pi,0)=\zeta (0,\pi)=\zeta (\pi,\pi)=+1$ and the system must have a Chern number with odd parity. It does since $C_1=-1.$ For $2<m<4$ all the eigenvalues except $\zeta (\pi,\pi)$ are negative and again the Chern number must have odd parity $(C_1=+1)$.

The location of the $1/2$ modes in the entanglement spectrum is also clear. If we choose to cut along the $x$ or $y$ directions the picture remains the same. This indicates that we are not dealing with an (weak) anisotropic insulator as in the dimerized case but a fully 2D topological insulator state. This is similar to saying that for the quantum Hall effect, no matter what edge we cut in the system, there will be edge states. For definiteness assume that we cut parallel to $y$ so that $k_y$ is a good quantum number. For $0<m<2$ there will be a $1/2$ entanglement mode at $k_y=0$ and for $2<m<4$ there will be a $1/2$ mode at $k_y=\pi.$ In addition to these modes there is actually a dispersing set of modes that are localized on the cut. To clearly see the dispersing modes we look at the entanglement `energies' which are defined to be 
\begin{eqnarray}
\epsilon_m=\frac{1}{2}\log\left[\frac{1}{\xi_m}-1\right]
\end{eqnarray}\noindent where $\xi_m$ are the eigenvalues of the reduced correlation matrix $C_L.$ The entanglement energies show the full structure of the entanglement spectrum because they clearly separate the eigenvalues of $C_L$ which are clustered near zero and one. The energy, entanglement eigenvalues, and entanglement energies for the Chern insulator in the $m<0,0<m<2$ and $2<m<4$ phases are shown in Fig. \ref{fig:ChernInsulator}. In the two non-trivial phases the location of the $1/2$ mode is different. For $0<m<2$ it is at $k_y=0$ and for $2<m<4$ it is at $k_y=\pi.$ For the entanglement energies these become \emph{zero} modes. These entanglement spectra were cut from a torus geometry so that there are two entanglement cuts. This is the reason why there are entanglement modes dispersing in both directions in the $C_1\neq 0$ phases. 
\subsubsection{Quantum Spin Hall Insulator}
\begin{figure*}
  \includegraphics[width=14cm]{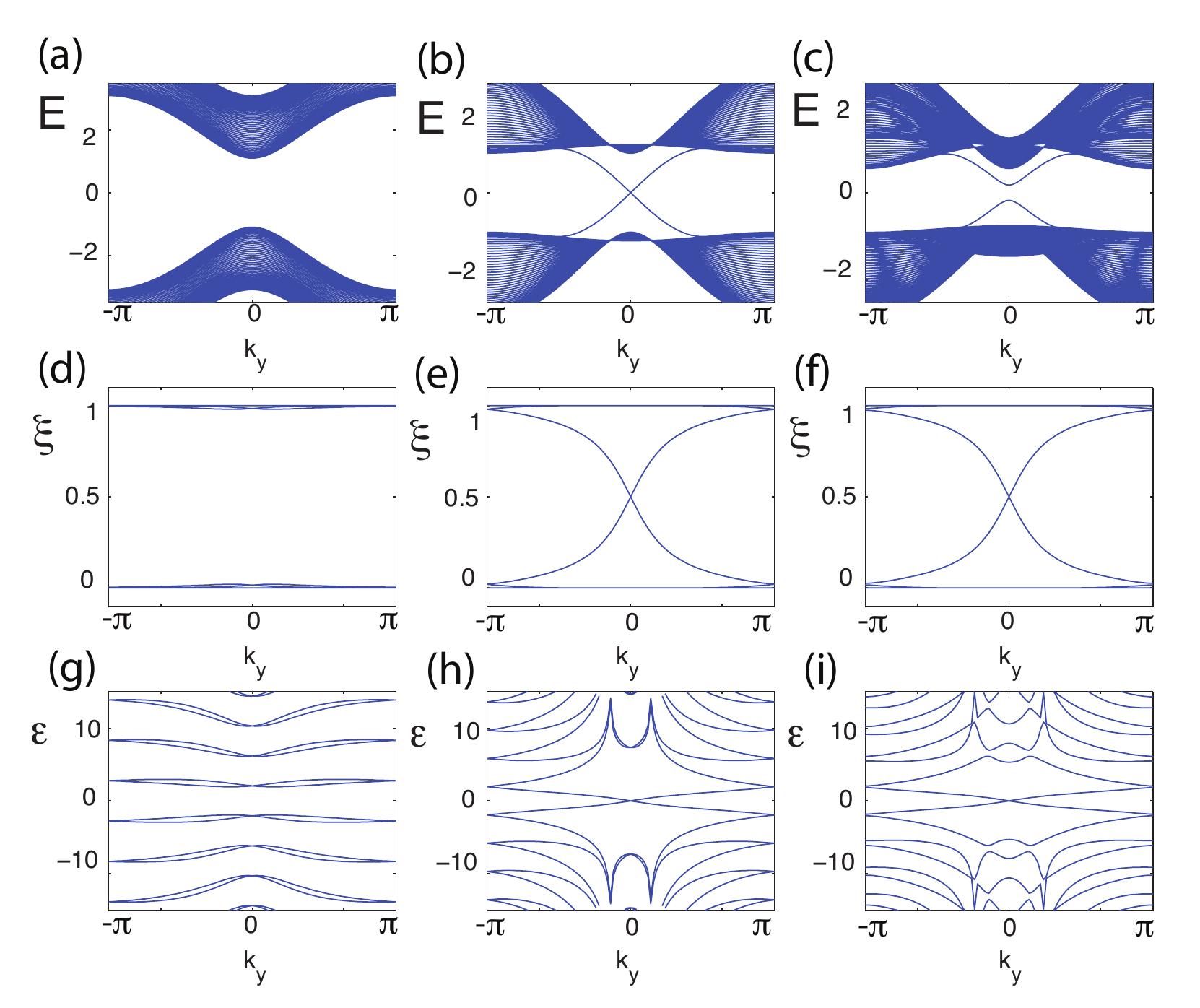}\\
  \caption{Energy spectrum with an open boundary, entanglement spectrum with a cut parallel to the y-direction, and entanglement energies for (Left panel a,d,g)A trivial insulator (Middle panel b,e,h)Non-trivial quantum spin Hall insulator with time-reversal symmetry (Right Panel c,f,i)Quantum spin Hall insulator with mildly broken time-reversal symmetry. Comparing h and i one can see that all Kramers' degeneracies are lifted when time-reversal is broken except the ones at $\epsilon=0.$ By just looking at e and f it is difficult to tell the difference in the two cases \emph{i.e.} that spectral flow is broken.}
 \label{fig:QSHInsulator}
\end{figure*}
The quantum spin Hall insulator (QSH) is a time-reversal invariant topological insulator\cite{Kane2005A,bernevig2006a,bernevig2006c} which is most easily thought of as two copies of the Chern insulator, one for each spin, with opposite chiralities. The realistic material in which this state is realized, HgTe/CdTe quantum wells, is best modeled by exactly this type of Hamiltonian. The effective HgTe Hamiltonian is a four-band model on the square lattice with a Hamiltonian given by
\begin{eqnarray}
\hat{H}(k)&=&\sin k_x\hat{\Gamma}_1+\sin k_y\hat{\Gamma}_2+M(k)\hat{\Gamma}_0\label{eq:QSH_H}\\
M(k)&=&2-m-\cos k_x-\cos k_y.\label{eq:MofK}
\end{eqnarray}\noindent where $\hat{\Gamma}_1=\sigma^z\otimes\tau^x, \hat{\Gamma}_2=1\otimes\tau^y, \hat{\Gamma}_0=1\otimes\tau^z$ where $\sigma^a$ is spin and $\tau^a$ is the orbital degree of freedom. For this system the time-reversal operator is $T= (i\sigma^y\otimes 1) K$ and the inversion operator is ${\cal{P}}=\hat{\Gamma}_0.$ This Hamiltonian is invariant under both symmetries. It exhibits phases directly analogous to the Chern insulator \emph{i.e.} it is a trivial insulator for $m<0,m>4$ and a topological quantum spin Hall insulator for $0<m<2$ and $2<m<4.$ The only difference with the Chern insulator is that now there are two occupied bands which are related by time-reversal symmetry. The total Chern number of the ground state vanishes but there is a  $Z_2$ invariant given (in the presence of time-reversal and inversion\cite{fu2007a}) by
\begin{eqnarray}
\chi_{Z_2}=\prod_{n\in occ./2}\zeta_n(0,0)\zeta_n(\pi,0)\zeta_n(0,\pi)\zeta_n(\pi,\pi)
\end{eqnarray}\noindent where the product runs over half the occupied bands. This invariant has the same formula as $\chi^{(2)}_{\cal{P}}$ defined above, but we distinguish it here to prevent confusion since this invariant only implies a non-trivial physical response when time-reversal is preserved. To specify which \emph{half} of the occupied bands you just take one from every Kramers' pair of bands. If we focus on the transition when $m\sim 0$ we see that for $m<0$ the inversion eigenvalues of both occupied bands are all positive. For $2>m>0$  the inversion eigenvalues at $(k_x,k_y)=(0,0)$ for \emph{both} occupied bands are negative while all others are positive. The product over inversion eigenvalues of \emph{all} the occupied bands is trivial, but if we only multiply over half the Kramers' pairs we find that $\chi_{Z_2}=-1$ and is non-trivial. Since the product over all the bands is trivial this means that the parity of the first Chern number is even, in fact it is zero for this case. The energies, entanglement eigenvalues, and entanglement energies for these two phases are shown in the left and middle panels of Fig. \ref{fig:QSHInsulator}. Without time-reversal symmetry this type of inversion invariant (product over half the occupied states) does not uniquely specify a topological response in 2D since a $C_1=2$ quantum anomalous Hall state could have the same inversion eigenvalue structure.  

\subsubsection{Quantum Spin Hall Insulator without time-reversal symmetry}
So far the studies of the Chern Insulator and QSH insulator have just been reconfirmed by recognizing the importance of inversion eigenvalues when there is an inversion symmetry. The most interesting prospect is when we take the QSH effect and \emph{break} time-reversal but keep inversion. The importance of inversion symmetry for this type of case in 3D was emphasized in Ref. \onlinecite{turner2009}. To break time-reversal symmetry we will consider an additional Zeeman term in the QSH Hamiltonian\cite{koenig2008}:
\begin{eqnarray}
\hat{H}(k)&=&\sin k_x\hat{\Gamma}_1+\sin k_y\hat{\Gamma}_2+M(k)\hat{\Gamma}_0\nonumber\\
&+&B_x\hat{\Gamma}_B\\
\hat{\Gamma}_{B}&=&\left(\begin{array}{cccc}0 &0&1&0\\0 &0&0&0\\1&0&0&0\\0&0&0&0\end{array}\right).\label{eq:QSHTRB}\nonumber
\end{eqnarray}\noindent For small values of $B_x$ this term lifts the Kramers' degeneracy of the occupied bands but does not cause any crossings at the Fermi-level. Although time-reversal is broken, inversion is still preserved and we can still see that $\chi_{Z_2}=-1.$ This is still well defined because the product of all inversion eigenvalues at a particular $k_{inv}$ is still trivial for all $k_{inv}.$ Thus, this system is an \emph{inversion invariant} topological insulator. It was first noted that such states could exist in Ref. \onlinecite{turner2009} where it was suggested that as long as inversion symmetry was not broken the entanglement spectra for the time-reversal preserved and broken cases were the same.  However, there is an important difference between the two. For the time-reversal broken QSH state we show the entanglement eigenvalues and entanglement energies in the right panel of Fig. \ref{fig:QSHInsulator}. Comparing with the middle panel we see that the entanglement eigenvalues seemingly show very little difference due to the fact that most are exponentially close to zero or one, but the entanglement energies are quite different. The mode at $1/2$ \emph{is} protected by inversion symmetry but all of the other ``Kramers'" degeneracies are lifted \emph{e.g.} all the crossings at $k_y=-\pi$ and $k_y=\pi$ are lifted. This occurs because the spectral flow between the bulk valence and conduction bands is cut-off when time-reversal is broken. The edge states no longer tie together both bands and there is no ``anomaly"-type structure. Thus states on the left and right half of the system are no longer tied together through the bulk in a topological way. 

We ask now if there is anything interesting in this system once time-reversal is broken? As long as inversion is preserved we cannot connect this state to a trivial atomic limit while preserving inversion symmetry and there must always be a finite entanglement entropy.  The finite entanglement entropy is due to the fact that the mode at $1/2$ is protected and cannot be removed. Thus the system is not trivial in the sense that it cannot ever be continuously deformed to a trivial atomic insulator; the entanglement spectrum clearly shows this. Now we can also ask if there is any non-trivial physical response? The robust electromagnetic response discussed in Refs. \onlinecite{qi2008,qi2008A,qi2008B} comes from coupling the quantum spin Hall state to varying adiabatic parameters. For example, applying a magnetic domain wall to the edge of a QSH system induces a mid-gap state and fractional charge localized on the domain wall. This response requires time-reversal symmetry to be broken and thus could still exist in the presence of a Zeeman term. The Zeeman term is an external time-reversal breaking field and it opens a gap in the edge states. If we are able to create a magnetic domain wall on the edge, \emph{i.e.} a field strong enough to reverse the direction of the Zeeman field in some region of the edge then there will be trapped domain wall states. Thus the state is a topological insulator in a very physical sense as well.  Unfortunately the  inversion invariant  ($\chi^{(2)}_{\cal{P}}$) in 2D does not imply that this \emph{must} be the physical response. As mentioned above a $C_1=2$ quantum anomalous Hall effect can have the same value of the $Z_2$ invariant. 
However, for this model Hamiltonian we are saved because there is an additional mirror symmetry ${\cal{M}}\hat{H}(k_x,k_y){\cal{M}}^{-1}=\hat{H}(k_x,-k_y)$ with ${\cal{M}}=\sigma^x\otimes\tau^{z}.$ Note that $[{\cal{M}},{\cal{P}}]=[{\cal{M}},\hat{\Gamma}_{B}]=0.$ Thus, the inversion eigenvalues are still valid labels and the Zeeman term does not break the mirror symmetry. This symmetry forbids a non-zero $C_1$ and thus the QSH response is the unique result. 

\subsubsection{2D 8-band model}
\begin{figure}
  \includegraphics[width=7cm]{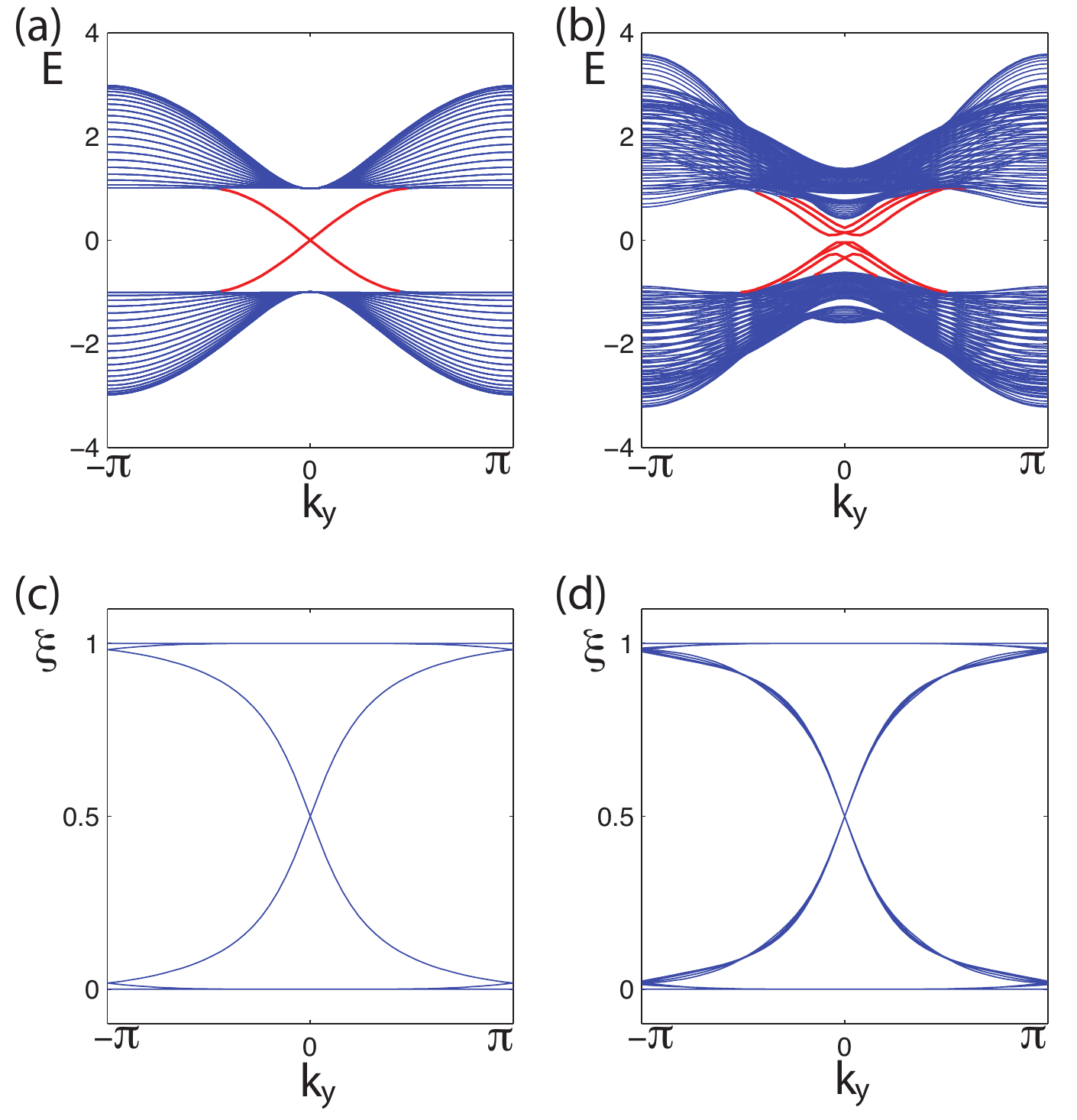}\\
  \caption{Energy spectrum with open boundary conditions for (a)8-band model with $T$ symmetry (b) 8-band model with random inversion preserving perturbation. Entanglement spectrum for (c) 8-band with $T$ (d) 8-band with random inversion preserving perturbation.}
 \label{fig:EightBandInsulator}
\end{figure}
\begin{figure}
  \includegraphics[width=7cm]{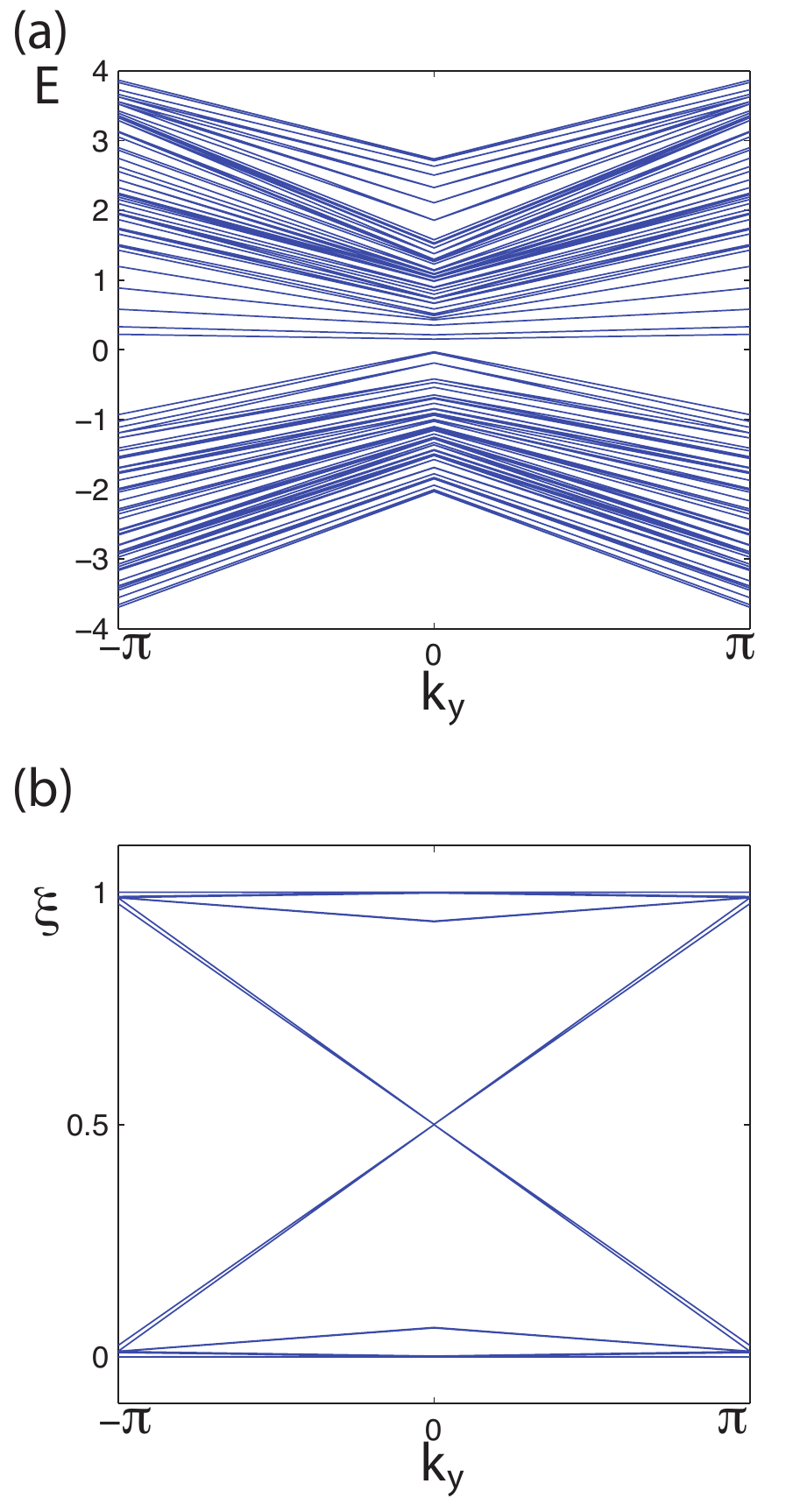}\\
  \caption{(a)Energy spectrum for the small anisotropic 8-band Hamiltonian with periodic boundary conditions. Difference in negative parity eigenvalues at $k_y=0,\pi$ is three. (b) Entanglement spectrum for the same. At $k_y=0$ there are six modes at $\xi=1/2.$ The lines are only filled in as guides to the eye. The only allowed k-values are $k_y=0$ and $k_y=\pi.$ This is why the spectra appear to have kinks. }
 \label{fig:EightBandQ1D}
\end{figure}

Here we consider a more complicated case of a model with $8$-bands total and four occupied bands. The model we choose is simply two-copies of the QSH model. The Bloch Hamiltonian is given by
\begin{eqnarray}
\hat{H}(k)=\sin k_x\gamma^x+\sin k_y \gamma^y+M(k)\gamma^z\label{eq:eightband}
\end{eqnarray}\noindent with $M(k)$ given in Eq. \ref{eq:MofK} and $\gamma^a=1_{2\times 2}\otimes \Gamma^a.$ This model preserves time-reversal symmetry with $T=(1\otimes i\sigma^y\otimes 1 )K$ and inversion symmetry with ${\cal P}=\gamma^z.$ In the presence of time-reversal symmetry this model does not yield any non-trivial topological insulators since you always get an even number of pairs of edge states. If one adds perturbations which break time-reversal symmetry then it is possible to generate non-trivial states such as Chern insulator states. At half-filling this model will have four occupied bands, and will exhibit edge states for the same range of parameters as the quantum spin Hall model above in Eq. \ref{eq:QSH_H}. These edge states are not protected generically \emph{i.e.} one can add a time-reversal and inversion \emph{preserving} perturbation to the model which will open a gap in the edge states. We use this model to illustrate three points (i) even though there is no topological invariant in the system associated with time-reversal symmetry, the presence of inversion symmetry predicts that there will be non-trivial mid-gap states in the entanglement spectrum (ii) these mid-gap states are completely stable as long as you do not break inversion symmetry (iii) the number of mid-gap states is proportional to the difference in negative parity eigenvalues at the invariant momenta. 

In Fig. \ref{fig:EightBandInsulator} we show the energy and entanglement spectra for two different cases. In Fig. \ref{fig:EightBandInsulator}a,c we simply diagonalize Eq. \ref{eq:eightband} with $m=1.0$ on an open boundary. One can clearly see the (unstable) edge states lying in the mid-gap region. The entanglement spectrum is shown for a system with periodic boundary conditions cut parallel to the $y$-axis. The difference between the number of negative inversion eigenvalues at $(k_x,k_y)=(0,0)$ and $(\pi,0)$ is four giving a total number of $2\times 4=8$ modes at $1/2.$ If we leave out the identity matrix, and ${\cal{P}}$ itself, there are $30$ matrices which commute with ${\cal{P}}.$ To illustrate the stability of the $1/2$ modes in the entanglement spectrum we add a perturbation which includes all $30$ matrices with random couplings chosen from a uniform distribution $[-\delta/2,\delta/2]$ where $\delta>0$ is chosen small enough not to close the bulk gap. This perturbation will break all of the `accidental' symmetries in the problem but preserves inversion symmetry. We see in Fig. \ref{fig:EightBandInsulator} where $\delta=0.20$ that the energy spectrum still has states in the gap but the crossing-points have been lifted. The entanglement spectrum, however, still has eight exact $1/2$ modes.

Finally we take a very small anisotropic sized system with $L_x=20$ and $L_y=2.$ We cut the system in the middle between $x=10$ and $x=11$ and plot the entanglement spectrum vs $k_y.$ There are only two allowed values for $k_y=0,\pi.$ The energy spectrum for such a system (with the random perturbation included but chosen from a uniform distribution $[0,\delta]$ which is no longer symmetric about zero) is gapped and has an entanglement spectrum with six mid-gap $1/2$ modes when $\delta=0.19.$ Counting the occupied states we find numerically that there is a difference of three negative inversion eigenvalues which agrees with the entanglement spectrum. The energy and entanglement spectra are shown in Fig. \ref{fig:EightBandQ1D}.

\subsection{3D Models}

\subsubsection{Dimerized Cubic Lattice}
The 3D dimerized model on a cubic lattice is a trivial extension of the 2D dimerized case into 3D. The Hamiltonian is given by
\begin{eqnarray}
H&=&\sum_{k}\Psi^{\dagger}_{k}\left[\left(-(t+\delta)-(t-\delta)\cos k_x\right)\sigma^x\right. \nonumber\\&+&\left.(t-\delta)\sin k_x\sigma^y-2t_y\cos k_y-2t_z\cos k_z\right]\Psi_k.\nonumber \\
\end{eqnarray} \noindent For $t=2t_y=2t_z=1$ there are two different phases $\delta<0$ and $\delta>0.$ As before, for $\delta>0$ the Wannier center of the electron is located within a unit cell and all inversion eigenvalues are $+1.$ For $\delta<0$ the Wannier center is shifted along the x-axis to the mid-bond site between unit cells. The inversion eigenvalues are $\zeta(000)=\zeta(00\pi)=\zeta(0\pi 0)=\zeta(0\pi\pi)=+1$ and $\zeta(\pi 00)=\zeta(\pi 0\pi)=\zeta (\pi \pi 0)=\zeta(\pi\pi\pi)=-1.$ Again this system has a non-trivial charge polarization on a surface with $\hat{x}$ as a normal vector. However, this is protected by the reflection symmetry about the $yz$-plane \emph{i.e.} ${\cal{M}}\hat{H}(k_x,k_y,k_z){\cal{M}}^{-1}=\hat{H}(-k_x.k_y.k_z)$ with ${\cal{M}}=\sigma^x={\cal{P}}.$ Thus, since the product of the reflection eigenvalues is $-1$ for each reflection invariant line in the Brillouin zone,  the polarization on a surface perpendicular to the x-axis is $P_{1}=e/2a^2$ where $a^2$ is the area of a plaquette in the $yz$ plane.  The product of all the parity eigenvalues is trivial as it must be in 3D, and additionally the product of the eigenvalues in every plane is trivial. For an entanglement cut such that $k_y$ and $k_z$ are good quantum numbers it is clear that there will be $1/2$ modes at $(k_y,k_z)=(0,0),(\pi,0),(0,\pi),$ and $(\pi,\pi).$ On the other translationally invariant cuts parallel to the $xz$ or $xy$ planes there will be no $1/2$ modes. 

\subsubsection{3D Quantum Hall Effect}
\begin{figure}
  \includegraphics[width=8cm]{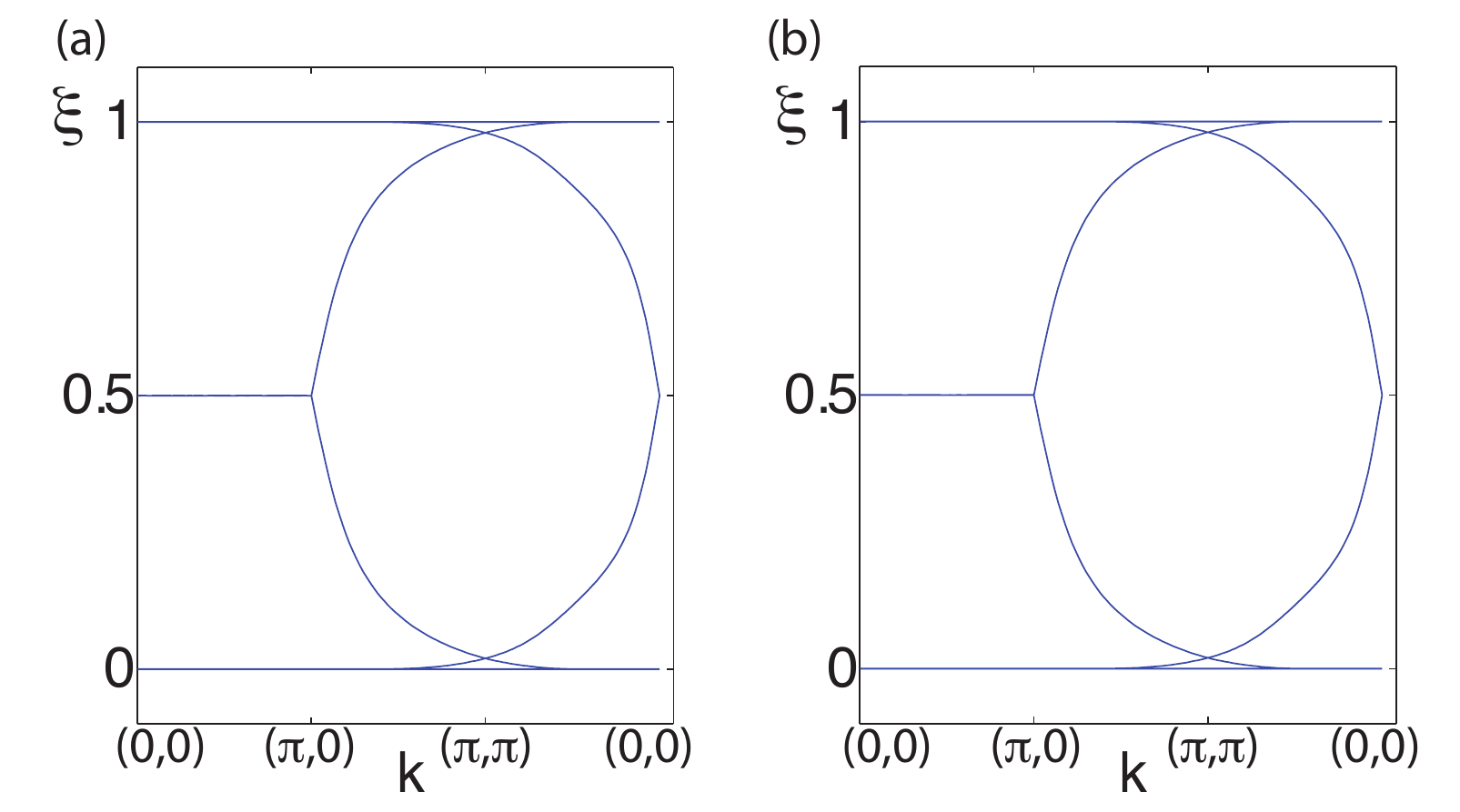}\\
  \caption{Entanglement Spectra for (a)3DQHE (b)3d WTI with a cut parallel to the x-y plane plotted along a line in the Brillouin zone. The states in (b) are all doubly degenerate compared to figure (a). }
 \label{fig:3dqhe3dwti}
\end{figure}
The 3D quantum Hall effect state can be thought of as stacks of 2d quantum Hall states which are connected together. We will use a very simple model for the 3D quantum (anomalous) Hall effect which is a trivial extension of the 2D Chern insulator. The Hamiltonian is 
\begin{eqnarray}
\hat{H}(k)&=&\sin k_x\sigma^x+\sin k_y\sigma^y+M(k)\sigma^z\\
M(k)&=&2-m-\cos k_x-\cos k_y-t_{\perp}\cos k_z.\label{eq:3dQHEHam}
\end{eqnarray}\noindent This system has an inversion symmetry with ${\cal{P}}=\sigma^z.$ This model exhibits several different phase transitions but we will only focus on one, namely the phase transition that occurs with a gapless point at $(k_x,k_y,k_z)=(0,0,0).$  For $m<-t_{\perp}$ the system is in a trivial insulating phase with all inversion eigenvalues positive. At $m=-t_{\perp}$ the system becomes gapless and stays gapless until $m>t_{\perp}.$ For $m>t_{\perp}$ the system is in a 3D quantum Hall effect phase\cite{halperin1987} with effectively 2D quantum Hall states stacked up in the $z$-direction. The Bloch Hamiltonians at the inversion invariant points which switch eigenvalues are
\begin{eqnarray}
\hat{H}(0,0,0)=-(m+t_{\perp})\sigma^z\nonumber\\
\hat{H}(0,0,\pi)=(-m+t_{\perp})\sigma^z\nonumber
\end{eqnarray} At $m=-t_{\perp}$ the eigenvalue around the $\Gamma$-point switches from positive to negative but the system is still gapless. Then at $m=+t_{\perp}$ the eigenvalue at $(0,0,\pi)$ switches and the system becomes a gapped insulator. The product over all the eigenvalues is trivial as expected but if we restrict the product to the $k_z=0$ or $k_{z}=\pi$ planes the product is negative. We proved earlier that this indicates a non-trivial 3D QHE response.

For this arrangement of eigenvalues we can calculate the location of the $1/2$ modes in the entanglement spectrum. For a cut parallel to the $xy$-plane there will be no $1/2$ modes at the inversion invariant points. If we take the cut \emph{e.g.} parallel to the $zx$-plane then there will be $1/2$ modes at $(k_z,k_x)=(0,0)$ and $(\pi,0).$ These nodes are shown in Fig. \ref{fig:3dqhe3dwti}a. The figure shows, in addition to the $1/2$ modes at the two invariant momenta, a line of $1/2$ modes between $(k_z,k_z)=(0,0)$ and $(\pi,0).$

\subsubsection{3D Weak Topological Insulator}
There are several different classes of the recently proposed 3D time-reversal invariant topological insulators. The anisotropic classes, the so-called weak topological insulators, are effectively 2D quantum spin Hall states stacked into 3D. This is similar to the 3D quantum Hall effect and is essentially just two copies of that system, one for each spin. We use the following model: 
\begin{eqnarray}
\hat{H}(k)&=&\sin k_x\hat{\Gamma}_1+\sin k_y\hat{\Gamma}_2+M(k)\hat{\Gamma}_0\\
M(k)&=&2-m-\cos k_x-\cos k_y-t_{\perp}\cos k_z.
\end{eqnarray}\noindent where the $\hat{\Gamma}_a$ matrices are the same as in the quantum spin Hall state. This system is time-reversal and inversion invariant with an inversion operator ${\cal{P}}=\hat{\Gamma}_0.$ It exhibits phase transitions at the same values of $m$ as the 3D quantum Hall effect and the only difference is that there are two occupied bands instead of one. For $2-t_{\perp}>m>t_{\perp}$ The system has two pairs of negative inversion eigenvalues, one pair at $(0,0,0)$ and one at $(0,0,\pi).$ The rest of the eigenvalues are all positive. The total product of inversion eigenvalues is trivial, and unlike the 3D quantum Hall case the product of the eigenvalues when restricted to the $k_z=0,\pi$ planes is also trivial. However, there is still something non-trivial here which arises from taking the eigenvalues from only one of the Kramers' pairs at each invariant momentum. We see that this product is non-trivial and indicates an anisotropic inversion invariant topological insulator. In this case, since time-reversal symmetry is preserved, it is a weak topological insulator state.\cite{fu2007a} However, without time-reversal symmetry the inversion invariant being non-trivial does not require it is  a weak topological insulator. As a counter example it could be a 3D quantum Hall effect with a Chern number ``per layer'' that is an odd multiple of two (unless there is a reflection symmetry which requires the Chern number to vanish in each plane). The interesting thing about this model is that even when time-reversal is softly broken (\emph{i.e.} broken without causing a phase transition) the system still is not in a trivial topological state and even though the surface states are no longer protected it can exhibit non-trivial behavior in the entanglement. The entanglement spectrum for the T-invariant case is shown in Fig. \ref{fig:3dqhe3dwti}b and exhibits the same 1/2 mode structure as the 3D quantum Hall effect model but with twice as many modes. 
\subsubsection{3D Strong Topological Insulator}
\begin{figure}
  \includegraphics[width=7.5cm]{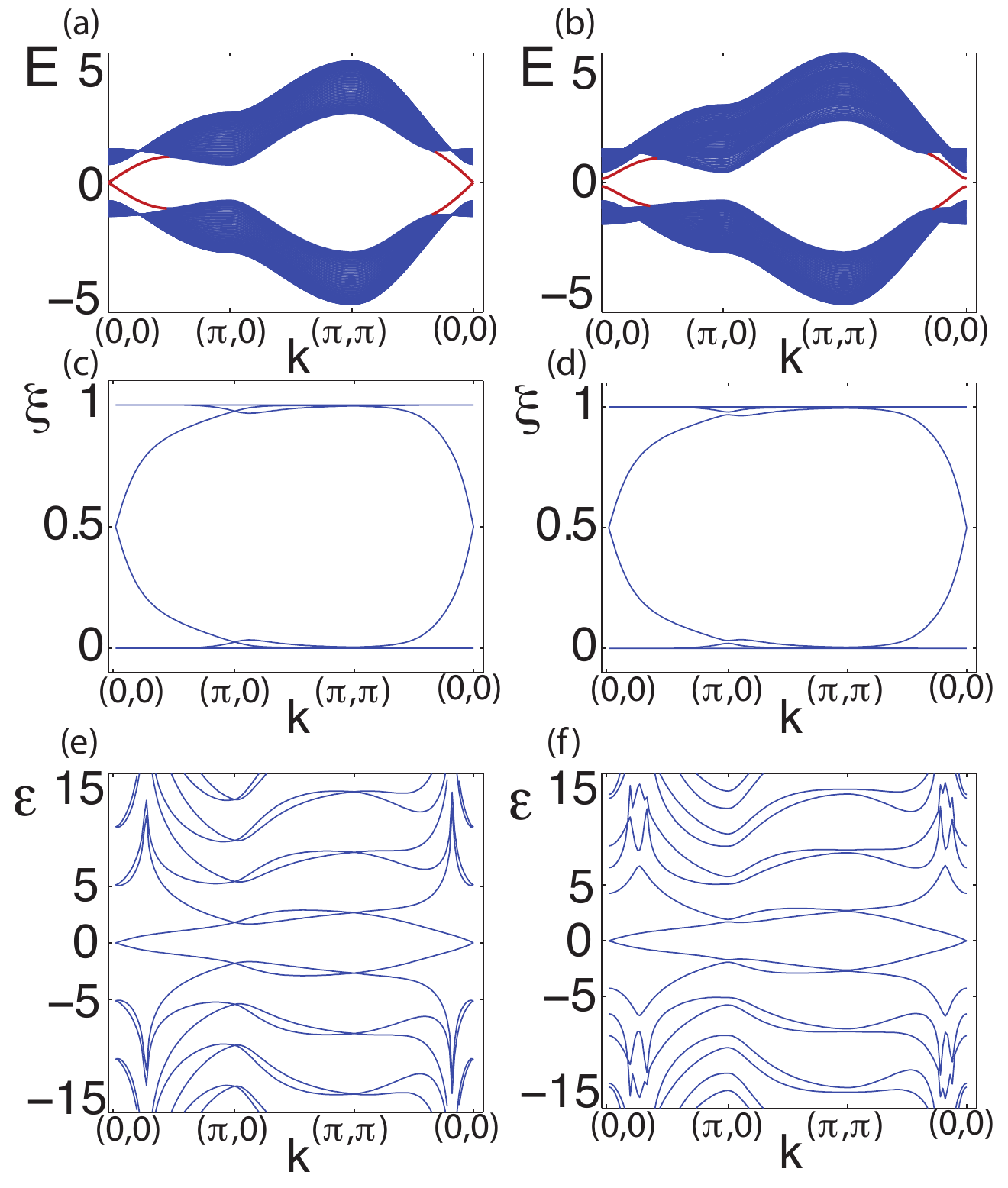}\\
  \caption{ (Left column)Strong topological insulator (a) Energy spectrum plotted along a line in the Brillouin zone for open boundary conditions along the z-direction (c)entanglement spectrum with a cut parallel to the x-y plane with periodic boundary conditions (e)entanglement energies. (Right column)Strong topological insulator with time reversal breaking (b)energy spectrum (d)entanglement spectrum (f) entanglement energies. In (a) and (b) the mid-gap states are localized on the surfaces. In (b) there is a gap in the surface states due to time-reversal symmetry breaking. In (f) all Kramers' degeneracies are lifted except the one at $\epsilon =0.$  }
 \label{fig:3dSTI}
\end{figure}
The last class of models we will consider is the 3D lattice Dirac model which is the minimal model for time-reversal invariant strong topological insulators in 3D. The Bloch Hamiltonian is given by
\begin{eqnarray}
\hat{H}(k)&=&\sin k_x \hat{\Gamma}_1 + \sin k_y\hat{\Gamma}_2+\sin k_z\hat{\Gamma}_3 +M(k)\hat{\Gamma}_0\nonumber\\
M(k)&=&3-m-\cos k_x-\cos k_y-\cos k_z \label{eq:STI}
\end{eqnarray}\noindent where $\hat{\Gamma}_3=\sigma^y\otimes \tau^x.$ As a function of $m$ this model exhibits many phase transitions. We will focus on the range $m<0$ and $0<m<3.$ There is a phase transition at $m=0$ with a band crossing at the $\Gamma$-point in k-space. Four bands meet at this point and a pair of inversion eigenvalues is exchanged. For $m<0$ the inversion eigenvalues are positive for both occupied bands at all the invariant momenta. The Bloch Hamiltonian at $k=0$ is $\hat{H}(0)=-m\hat{\Gamma}_0$ and thus when $m$ switches sign the inversion eigenvalues at $k=0$ are exchanged. For $0<m<3$ both inversion eigenvalues at $k=0$ are negative. In this phase the product over all inversion eigenvalues is trivial, but if we only keep one of the Kramers' pairs, then the product of all the eigenvalues of half the occupied bands is non-trivial. In the presence of inversion and time-reversal, which is the case here, this invariant is the strong topological $Z_2$ index.\cite{fu2007a} Physically this index has two implications: (1) the presence of an odd-number of massless Dirac cones on \emph{any} surface (2) a topological magneto-electric effect. To see the topological response one must apply a time-reversal breaking field on the surface to open a gap in the gapless Dirac fermions. This induces a quantum Hall effect confined to the surface which leads to the magneto-electric response. For the topological phase we picked $(0<m<3),$ the surface states are located around the surface $\Gamma$-point and the (pair of) entanglement modes will be located at the $\Gamma$-point of the conserved momenta parallel to the entanglement cut. The energy spectrum, entanglement eigenvalues, and entanglement energies are shown in Fig. \ref{fig:3dSTI}a,c,e respectively. The energy spectrum is shown with periodic boundary conditions along $x$ and $y$ and open boundary conditions along $z.$ The surface states are shown in red and the spectrum is plotted along a 1D path in the Brillouin zone. 

\subsubsection{3D Strong Topological Insulator without time-reversal symmetry}
Now we can consider the more interesting question of the properties of the system when we break time-reversal but keep inversion. We add the same Zeeman term shown in Eq. \ref{eq:QSHTRB} to the bulk of the insulator. We only break time-reversal softly which opens a gap in the surface states but does not close the bulk gap. Thus, while we can no longer consider the $Z_2$ invariant protected by time-reversal, this system will still exhibit a magneto-electric effect since the Zeeman field simply establishes a quantum Hall effect on the surface. This distinction was first considered in Ref. \onlinecite{turner2009}. Additionally, although we can add surface potentials to push the surface states into the bulk bands we still cannot adiabatically connect this insulator with an atomic limit. This is clearly shown in the entanglement spectrum where there are still modes protected at $1/2$ which cannot be removed without breaking inversion or passing through a phase transition. Thus, it is impossible to continuously connect this to a system with vanishing entanglement entropy. Fig. \ref{fig:3dSTI}b,d,f show the energy, entanglement eigenvalues, and entanglement energies respectively. For the energies the red surface states are now clearly gapped, but the entanglement eigenvalues are hard to distinguish between  the time-reversal invariant and breaking cases. The distinction, we see, comes when we look at the entanglement energies which show that while there is still a pair of zero modes, all of the other degeneracies at the inversion invariant momenta (which arose from Kramers' degeneracies) are lifted generically. This shows that the spectral flow between the valence and conduction bands of the entanglement spectrum has been cut off. 

Now we will prove that $P_3\neq 0$ for this model. To be explicit we take
\begin{equation}
\begin{array}{c}
\hat{H}(k)=\sin k_1 \hat{\Gamma}_1+\sin k_2 \hat{\Gamma}_2+\sin k_3 \hat{\Gamma}_3 \medskip \\
+(5/2-\cos k_1 - \cos k_2 - \cos k_3 )\hat{\Gamma}_0+\frac{1}{4}  \hat{\Gamma}_{B}
\end{array}
\end{equation}
This Hamiltonian can be connected through a gapped interpolation with the trivial Hamiltonian:
\begin{equation}
\begin{array}{c}
\hat{H}_0(k)=\sin k_1 \hat{\Gamma}_1+\sin k_2 \hat{\Gamma}_2+\sin k_3 \hat{\Gamma}_3 +5/2 \hat{\Gamma}_0
\end{array}
\end{equation}
using the inversion symmetric homotopy
\begin{equation}\label{interpol}
\begin{array}{c}
\hat{H}(k,\theta)=\frac{1}{2}(1+\cos \theta)\hat{H}(k) \medskip \\
+\frac{1}{2}(1-\cos \theta)\hat{H}_0(k)+\sin \theta  \hat{\Gamma}_5
\end{array}
\end{equation}\noindent where $\hat{\Gamma}_5=\hat{\Gamma}_0 \hat{\Gamma}_1 \hat{\Gamma}_2 \hat{\Gamma}_3.$
The second Chern number generated by $\hat{H}(k,\theta)$ is $C_2$=1 (odd) and consequently $P_3=\frac{1}{2}\ \mbox{mod}(Z)$ for $\hat{H}(k)$ since we specifically chose the form of $\hat{H}_0(k)$ to be trivial. Note, that as with the cases shown in Appendix \ref{app:invExamples} we calculated $C_2$ numerically using the standard gauge invariant formula in terms of ground-state projection operators.

\section{Conclusions}
The question of what makes an insulator ``topological" has many answers. In this article we presented an answer which encompasses all of the known topological insulators. The fundamental distinction between an ordinary band insulator and a topological insulator is the inability to adiabatically connect a topological insulator to the atomic limit. This distinction can have many manifestations including non-trivial topological responses to external fields and robust boundary states, however these properties are not necessary conditions for a topological insulator. In fact, we have seen examples in this paper without protected boundary states, and  examples with no topological response. At first sight these insulators seem to have no characteristics which distinguish them from  trivial band insulators. Admittedly these are not the most interesting systems to consider experimentally, but they still show a striking signature in the entanglement spectrum. In fact, all known topological insulators show a signature in the entanglement spectrum when the bi-partition is a position-space cut. Instead of taking the whole spectrum, one can calculate just the entanglement entropy, which for all topological insulators cannot be adiabatically deformed to zero. This fact is what serves as the basis for our definition and unifies the inversion symmetric insulators with the ones that are invariant under other discrete symmetries.  The experimental relevance of inversion symmetric topological insulators is unclear, but not out of the question.\cite{wan2010} Although in principle disorder immediately destroys any stability of the topological state (unlike the typical topological insulators\cite{schnyder2008}), the robustness of the insulator state is ultimately a question to be answered in practice. Many of the well-known topological insulators simplify when inversion symmetry is required along with the discrete symmetry that stabilizes the topological state. Thus it seems like the most interesting inversion symmetric insulators are ones which are derived from parent topological insulator states with weakly broken $T$ or $C$ symmetries. These types of materials would be the first place to search for signatures of topological protection due to inversion symmetry. 

{\it Note:}  During the last stages of preparation of this manuscript, we became  aware of a paper by  A. Turner, Yi Zhang, R. Mong and A. Vishwanath dealing with similar issues.

{\it Acknowledgements}
We acknowledge useful conversations with C.-K. Chiu, E. Fradkin, and F.D.M. Haldane. TLH was supported in part by the NSF DMR 0758462 at the University of Illinois, and by the ICMT.  EP acknowledges a support from the Research Corporation for Science Advancement. BAB was supported by Princeton Startup Funds, Alfred P. Sloan Foundation, NSF DMR-095242, and NSF China  11050110420, and MRSEC grant at Princeton University, NSF DMR-0819860. TLH and BAB  thank the Institute of Physics in Beijing, China for generous hosting. BAB thanks  Ecole Normale Superieure and Microsoft Station Q for generous hosting. 
\appendix 

\section{Proof that the one-body correlation function is a projection operator}\label{app:Gprojector}
Here we prove that the one-body correlation function over the full system (not only over part of the system) is a projector. This can easily be proved: $C_{ij}$ (which is a matrix at each $i,j$  i.e. $C_{ij}^{\alpha \beta}$) has the property:
\begin{widetext}
\begin{eqnarray}
&\sum_j C_{ij}(k_y)C_{jk}(k_y)    =  \sum_j \frac{1}{N} \sum_{k_x} e^{i k_x (i-j)} \hat{P}(k_x,k_y) \frac{1}{N} \sum_{k'_x} e^{i k'_x (j-k)} \hat{P}(k'_x,k_y) = \nonumber \\ &  =   \sum_{k_x, k'_x} \sum_j \frac{1}{N^2}  e^{i (k'_x - k_x) j}   e^{i k_x  i}\hat{P}(k_x,k_y)  e^{-i k'_x k} \hat{P}(k'_x,k_y) =   \sum_{k_x, k'_x} \frac{1}{N}  \delta_{k'_x - k_x }  e^{i k_x  i} \hat{P}(k_x,k_y)  e^{-i k'_x k} \hat{P}(k'_x,k_y)   = \nonumber \\ &  =  \frac{1}{N}   \sum_{k_x}  e^{i k_x ( i-k)} \hat{P}(k_x,k_y)  \hat{P}(k_x,k_y)=  \frac{1}{N}   \sum_{k_x}  e^{i k_x ( i-k)} \hat{P}(k_x,k_y) =C_{ik}(k_y).
\end{eqnarray} 
\end{widetext}\noindent Note that we have used the notation $\hat{P}(k_x,k_y)$ to represent the k-dependent projector onto the occupied states instead of $\hat{P}_k$ to make the momentum dependence easier to see. 

\section{Entanglement eigenvalues for two occupied bands  and four sites}\label{app:foursite}

We will analyze only one case of inversion eigenvalues, \emph{i.e.} when both inversion eigenvalues at $k=0$ differ from the inversion eigenvalues at $k=\pi$.  For simplicity and without loss of generality, we particularize to $\zeta_{1}(0)= \zeta_{2}(0)=1$, $\zeta_{1}(\pi)= \zeta_{2}(\pi)=-1 $, where the orthogonality relations between the wavefunctions at the same $k$ and at different inversion symmetric $k$'s hold due to the opposite inversion eigenvalues. Let the wavefunctions of the two occupied bands be $\psi_1(k)$ and $\psi_2 (k).$
Similar to the above the arguments in Sec \ref{sec:1band4site}, the entanglement wavefunctions take the form $(\psi_A, m {\cal{P}} \psi_A)$ where $\psi_A$ diagonalizes the operator $\hat{C}_0 + m \hat{C}_1 {\cal{P}} =
\frac{1}{4}[(1+ m)( \hat{P}_0 + \hat{P}_\pi) + \hat{P}_{\frac{\pi}{2}} {\cal{P}}({\cal{P}} + i m) + {\cal{P}} \hat{P}_{\frac{\pi}{2}} ({\cal{P}}- im)] $.  We expand the wavefunction $\psi_A$:
\beq 
\psi_A= a_1 \psi_1(0) +a_2 \psi_2(0) + b_1 \psi_1(\pi) + b_2 \psi_2 (\pi).
\eneq With this expansion, we have to look at the solutions $\psi_A$ which are in the nullspace of $ \hat{P}_{\frac{\pi}{2}} {\cal{P}}({\cal{P}} + i m) + {\cal{P}} \hat{P}_{\frac{\pi}{2}} ({\cal{P}}- im).$ Once we have found such solutions, we know they have $1/2$ eigenvalues for $m=1$ since $(\hat{C}_0 + \hat{C}_1 {\cal{P}})\psi_A = \frac{1}{2} (\hat{P}_0 + \hat{P}_\pi) \psi_A= \frac{1}{2} \psi_A$ due to the fact that wavefunctions at different inversion symmetric momenta are orthogonal if their inversion eigenvalues are different. We denote the overlaps:
\begin{eqnarray}
&\langle \psi_1 (\frac{\pi}{2}) | \psi_1(0) \rangle = \alpha_1 ,\;\;\;  \langle \psi_1 (\frac{\pi}{2}) | \psi_2(0) \rangle = \alpha_2 \nonumber \\  &\langle \psi_1 (\frac{\pi}{2}) | \psi_1(\pi) \rangle = \alpha_3 ,\;\;\;  \langle \psi_1 (\frac{\pi}{2}) | \psi_2(\pi) \rangle = \alpha_4 \nonumber \\  &\langle \psi_2 (\frac{\pi}{2}) | \psi_1(0) \rangle = \beta_1 ,\;\;\;  \langle \psi_2 (\frac{\pi}{2}) | \psi_2(0) \rangle = \beta_2 \nonumber \\  &\langle \psi_2 (\frac{\pi}{2}) | \psi_1(\pi) \rangle = \beta_3 ,\;\;\;  \langle \psi_2 (\frac{\pi}{2}) | \psi_2(\pi) \rangle = \beta_4 
\end{eqnarray} These are the only unknown overlaps, as  the wavefunction at $3\pi/2$ is related to the one at $\pi/2$ by inversion symmetry. Thus its overlaps with eigenstates of ${\cal{P}}$ are, up to a sign, identical to the ones above. We find the following two exact half modes of the entanglement spectrum:
\begin{eqnarray} 
&(a_1, a_2, a_3, a_4) = \nonumber \\ & (i(\beta_3 \alpha_4-\beta_4 \alpha_3), 0, \beta_4 \alpha_1-\beta_1 \alpha_4,\beta_3 \alpha_1-\beta_1 \alpha_3) ;\nonumber \\ & (\beta_2 \alpha_3-\beta_3 \alpha_2, \beta_3 \alpha_1 - \beta_1 \alpha_3, I(\beta_2 \alpha_1-\beta_1 \alpha_2),0)
\end{eqnarray}
\noindent  The other two mid-gap entanglement eigenvalues  are obtained when $m=-1$, in which case we have to diagonalize the operator $\frac{1}{4}[\hat{P}_{\frac{\pi}{2}} {\cal{P}}({\cal{P}} - i ) + {\cal{P}} \hat{P}_{\frac{\pi}{2}} ({\cal{P}}+ i)] $. In this case, we expand the eigenstate:
\beq
\ket{\psi_A} =( c_1 + c_3 {\cal{P}}) \ket{\psi_1(\frac{\pi}{2})} +( c_2  + c_4 {\cal{P}}) \ket{\psi_2(\frac{\pi}{2})}   
\eneq we find $1/2$ modes if, just like in the previous section:
\begin{eqnarray} 
&(c_1, c_2, c_3, c_4) =  (0,1,0,i)\;, \;\;\;\;\; (1,0,i,0)
\end{eqnarray} We then see that we have $4$ robust modes at exactly $1/2$ in the entanglement spectrum, or exactly twice the difference of negative inversion eigenvalues  at the two inversion symmetric momenta. 

\section{Explicit Proof for the generic two-site problem with $N$ occupied bands}\label{app:manybands}
We now show that the two-site  problem with $n_1$ negative inversion eigenvalues at $k=0$ and $n_2$ negative inversion eigenvalues at $k=\pi$  out of a number $N$ of occupied bands contains $2|n_1- n_2|$ zero modes in the entanglement spectrum. Without loss of generality denote the eigenstates of the original Hamiltonian $\ket{\psi_1(0)} \ldots \ket{\psi_{n_1} (0)}$ as the ones with negative inversion eigenvalue at $k=0$,  $\ket{\psi_{n_1+1}(0)} \ldots \ket{\psi_{N }(0)}$ as the ones with positive inversion eigenvalue at $k=0$,  $\ket{\psi_1(\pi)} \ldots \ket{\psi_{n_2} (\pi)}$ as the ones with negative inversion eigenvalue at $k=\pi$,  $\ket{\psi_{n_2+1}(\pi)} \ldots \ket{\psi_{N }(\pi)}$ as the ones with positive inversion eigenvalue at $k=\pi$. Bands at the same momentum are \emph{all} orthogonal, while bands at different momenta are orthogonal if they have opposite inversion eigenvalues. The simplest case of the above, which should be obvious from our previous examples, is that of all the  $N$ inversion eigenvalues at $k=0$ being identical and negative of the $N$ eigenvalues at $k=\pi$. In this case, the projector at  one of the inversion symmetric $k$'s annihilates all the eigenstates at the other inversion symmetric $k$, and the $2N$ occupied eigenstates of the original two-site Hamiltonian are  also the eigenstates of the entanglement spectrum at fixed eigenvalue $1/2$. Due to their orthogonality, they are linearly independent. From here it is clear that our formula is physically correct: adding the same eigenvalues to both  $k=0, \pi$ cannot change the result.

We again expand the eigenstates $\ket{\psi_A}$ of the correlation function $C_L= (\hat{P}_0+ \hat{P}_\pi)/2$ as a sum over all the occupied eigenstates, even though these might not be (and in general are not) orthogonal:
\begin{eqnarray}
&\psi_A= \sum_{m=1}^{n_1} a_m \psi_m(0) + \sum_{m=n_1+1}^{N} a_m \psi_{m} (0) + \nonumber \\ &+  \sum_{m=1}^{n_2} b_m \psi_m(\psi) + \sum_{m=n_2+1}^{N} b_m \psi_{m} (\pi)
\end{eqnarray} In the generic case, we assume that the norms that are not fixed to vanish by symmetry (such as different inversion eigenvalues) are all nonzero.  In general, it might be the case that not all the eigenstates in the expansion above are linearly independent i.e. $\psi_{m}(0)$ might not be linearly independent from a sum of the eigenvalues $\psi_m (\pi)$ which have identical inversion eigenvalues. In building the  matrix to be diagonalized, we take this into consideration, but when writing the eigenvalue equation, we \emph{assume} they are linearly independent - generically, they will be, because there are $2N$ wavevectors of $2N$ components. The vector $(a_1,\ldots, a_{n_1}, a_{n_1+1}, \ldots, a_{N}, b_1,\ldots, b_{n_2}, b_{n_2+1}, \ldots, b_{N})$ has to diagonalize the matrix:

 \beq
 \left( {\begin{array}{cccc}
 \frac{1}{2}_{n_1\times n_1} &  0  &  B_{n_1 \times n_2} & 0  \\
 0 & \frac{1}{2}_{N- n_1 \times N-n_1} & 0 & A_{N - n_1\times N- n_2}\\
  B^\dagger_{n_2 \times n_1}&0 & \frac{1}{2}_{n_2\times n_2} & 0 \\
  0 & A^\dagger_{N- n_2 \times N - n_1} & 0 & \frac{1}{2}_{N- n_2 \times N-n_2} 
 \end{array} } \right)
\eneq where $B_{ij} = \langle \psi_i (0) |\psi_j (\pi)\rangle$ with $i=1,\ldots, n_1$, $j=1, \ldots, n_2$ and $A_{ij} = \langle \psi_i (0) |\psi_j (\pi)\rangle$ with $i=n_1+1,\ldots,N $, $j=n_2+1, \ldots, N $. It is easy to see that this matrix has $2|n_1- n_2|$ eigenvalues at exactly $\frac{1}{2}$ irrespective of the $A, B$ matrices. We show it for $n_2=0$, the generalization to $n_2 \ne 0$ being straightforward. For $n_2=0$, the matrix reads:
 \beq
 \left( {\begin{array}{cccc}
 \frac{1}{2}_{n_1\times n_1} &  0   & 0  \\
 0 & \frac{1}{2}_{N- n_1 \times N-n_1}  & A_{N - n_1\times N}\\
  0 & A^\dagger_{N\times N - n_1} & \frac{1}{2}_{N\times N} 
 \end{array} } \right)
\eneq Half of the entanglement eigenvalues at $1/2$ are obvious -- they are the eigenvalues of the $\psi_i(0)$, $i=1,\ldots, n_1$ eigenstates. The remaining eigenvalues must then be part of the eigenvalues of the matrix:
\beq
R^{N- n_1}_{N} = \left( {\begin{array}{cc}
  \frac{1}{2}_{N- n_1 \times N-n_1}  & A_{N - n_1\times N}\\
  A^\dagger_{N\times N - n_1} & \frac{1}{2}_{N\times N} 
 \end{array} } \right)
\eneq where we have indexed the matrix by the  dimension $N- n_1$ of the upper block-diagonal square matrix and by the dimension $N$ of the lower block diagonal square matrix.  We need to compute the determinant of:
\beq
R^{N- n_1}_{N} = \left( {\begin{array}{cc}
  (\frac{1}{2}- \lambda)_{N- n_1 \times N-n_1}  & A_{N - n_1\times N}\\
  A^\dagger_{N\times N - n_1} &( \frac{1}{2}- \lambda) _{N\times N} 
 \end{array} } \right)
\eneq
\noindent
We can prove that this matrix has $n_1$ eigenvalues independent of what the matrices $A$ are: as such, we denote by $M^\alpha_\beta$ a matrix of the form above, but with any random numbers instead of the matrix made out of norms matrix $A_{ij}=\langle \psi_i (0) |\psi_j (\pi)\rangle $. We want to compute the determinant of $M_{N}^{N- n_1}$. By expanding first on the last column of the matrix,  then immediately after, expanding on the  last row of all the matrices obtained, we find the recurrence relation:
\beq
\det(M_{N}^{N-n_1} ) = (\frac{1}{2} - \lambda) \det{M^{N-n_1}_{N-1}} + x\cdot \det(M_{N-1}^{N-n_1-1} )
\eneq which, applied successively, leads to:
\begin{eqnarray}
&\det(M_{N}^{N-n_1} ) = (\frac{1}{2} - \lambda)^r \det{M^{N-n_1}_{N-r}} + \nonumber \\  &+  \sum_{i=1}^r x_i \cdot  (\frac{1}{2} - \lambda)^{i-1} \det(M_{N-i}^{N-n_1-1} ) \label{firstrecursion}
\end{eqnarray}  where $r\le N-1$ and $x_r$ are numbers.  We choose $r= N-1$ Notice we have so far applied the recursion relation only to the lower index of the matrix. We now apply it to the $\det(M_{N-i}^{N-n_1-1} )$ to obtain:
\begin{eqnarray}
&\det(M_{N-i}^{N-n_1-1} ) = (\frac{1}{2} - \lambda)^{p-i} \det{M^{N-n_1-1}_{N-p}} + \nonumber \\  &+  \sum_{q=1}^{p-i} x_q \cdot  (\frac{1}{2} - \lambda)^{q-1} \det(M_{N-i-q}^{N-n_1-2} ) \label{secondrecursion}
\end{eqnarray} where $p>i$ is an integer and the $x_q$ are complex numbers not necessarily equal to the $x_i$'s in the previous recursion. We again choose $p=N-1$, which means $i\le N-2$ in the above formula. By plugging in $\det(M_{N-i}^{N-n_1-1} )$ into Eq[\ref{firstrecursion}], and separating the sum to take into account the restriction $i\le N-2$ for which Eq[\ref{secondrecursion}] is valid, we find:
\begin{eqnarray}
&\det(M_{N}^{N-n_1} ) = O((\frac{1}{2} - \lambda)^{N-2}) + \nonumber \\ &+\sum_{q_0=1}^{N-1} \sum_{q_1= 1}^{N-1-  q_0} x_{q_0}x_{q_1} (1- \delta_{q_0, N-1}) \cdot \nonumber \\ & \cdot (\frac{1}{2}- \lambda)^{q_0+q_1-2} \det(M_{N - (q_0+q_1)}^{N-n_1-2} ) \nonumber 
\end{eqnarray} where $ O((\frac{1}{2} - \lambda)^{N-2}) $ means terms proportional to at least a the $(\frac{1}{2} - \lambda)^{N-2}$. By applying the recursion relation successively, we obtain:
\begin{eqnarray}
&\det(M_{N}^{N-n_1} )= O((\frac{1}{2} - \lambda)^{N-(l+1) }) +  \nonumber \\ &  \sum_{q_0=1}^{N-1} \sum_{q_1= 1}^{N-1-  q_0}   \sum_{q_2= 1}^{N-1-  (q_0+q_1) }...  \sum_{q_l= 1}^{N-1- (q_0 +... + q_{l-1})} \nonumber \\ &   (1- \delta_{q_0, N-1})(1- \delta_{q_0+q_1, N-1}) ... (1- \delta_{q_0+ ... + q_{l-1}, N-1})\cdot \nonumber \\ & \cdot (\frac{1}{2} - \lambda)^{(q_0+ \ldots +q_l - (l+1)}\det( M^{N- n_1-(l+1)}_{N - (q_0+ \ldots+ q_l)})
\end{eqnarray} when $N - n_1 - (l+1)=0$, the matrix $M^{N- n_1-(l+1)}_{N - (q_0+ \ldots+ q_l)}$ is fully diagonal and has determinant $(\frac{1}{2} - \lambda)^{N - (q_0+ \ldots+ q_l)}$. Hence the term in the sum is:
\beq
(\frac{1}{2} - \lambda)^{q_0+ \ldots +q_l - l-1} (\frac{1}{2} - \lambda)^{N - q_0- \ldots- q_l} = (\frac{1}{2} - \lambda)^{n_1}
\eneq while the lowest order term in $ O((\frac{1}{2} - \lambda)^{N-(l+1) })$ is again $ (\frac{1}{2} - \lambda)^{n_1}$.  We hence see that the determinant of $\det(M_{N}^{N-n_1} ) \sim (\frac{1}{2} - \lambda)^{n_1}$, and these are the remaining $n_1$ eigenvalues. Combined with the first $n_1$ eigenvalues, we see we have a total of $2 n_1$ eigenvalues in the entanglement spectrum at $1/2$, just as the formula predicts.

\section{Properties of the 1D adiabatic connection under inversion symmetry}\label{app:PolOdd}

We begin by recalling some basic facts of inversion symmetric Hamiltonians. 
Assume $\ket{u_i(k) } $ is an eigenstate of the Hamiltonian  at energy
\beq
\hat{H}(k) \ket{u_i(k)} = E_i(k) \ket{u_i(k)}
\eneq Then ${\cal P}\ket{u_i(k)} $ is necessarily an eigenstate at $-k$ of the same energy:
\begin{eqnarray}
{\cal P} \hat{H}(k) {\cal P}^{-1} {\cal P}\ket{u_i(k)}&=& \hat{H}(-k) {\cal P}\ket{u_i(k)}= {\cal P} \hat{H}(k) \ket{u_i(k)}\nonumber\\
 &=& {\cal P}  E_i(k) \ket{u_i(k)}=  E_i(k) {\cal P} \ket{u_i(k)}.\nonumber
\end{eqnarray} We assumed no degeneracies in the spectrum which means
\beq
\ket{u_i(-k)} = e^{i \alpha_k} {\cal P} \ket{u_i(k)}.
\eneq Thus we have:
\beq
\ket{u_i(k)} = e^{-i \alpha_k} {\cal P} \ket{u_{m}(-k)}\label{eq:appInvWf}
\eneq with $E_i(k)= E_i(-k).$

We assume we are in an insulating state where the charge polarization is  \begin{widetext} 
  \beq
P_{1}  = - ie \int_{-\pi}^\pi \frac{ dk}{2\pi}  \sum_{E_i(k) <0}  \bra{u_{k, i}} \partial_k \ket{u_{k,i}}.
\eneq  Using Eq. \ref{eq:appInvWf} we get
\begin{eqnarray}
&P_{1}  = -i e\int_{-\pi}^\pi \frac{dk}{2\pi} \sum_{i\in {\rm{occ.}}} \bra{{\cal P} u_i(-k)} e^{i \alpha_k} \partial_k e^{-i \alpha_k} {\cal P}  \ket{u_{i}(-k)} = \nonumber \\ &=   -ie \int_{-\pi}^\pi \frac{dk}{2\pi} \sum_{i\in {\rm{occ.}}} \bra{{\cal P}  u_i(-k)} \partial_k {\cal P}  \ket{u_{i}(-k)}    -i e\int_{-\pi}^\pi \frac{dk}{2\pi} \sum_{i\in {\rm{occ.}}} (-i) \partial_k \alpha_k .
\end{eqnarray} The last term is an integer because it depends only on $\alpha_\pi - \alpha_{-\pi}$ which can at most be $2 \pi j$, so we will drop it and be left with:
\begin{eqnarray}
&P_{1}  = -ie \int_{-\pi}^\pi \frac{dk}{2\pi} \sum_{i\in {\rm{occ.}}} \bra{ u_i(-k)} \partial_k  \ket{u_{i}(-k)} = i e\int_{\pi}^{-\pi} \frac{dk}{2 \pi}   \sum_{i\in {\rm{occ.}}} \bra{ u_i(k)} \partial_{-k}  \ket{u_{i}(k)} = \nonumber \\ &=- ie \int_{-\pi}^{\pi} \frac{dk}{2 \pi}   \sum_{i\in {\rm{occ.}}} \bra{ u_i(k)} \partial_{-k}  \ket{u_{i}(k)} = ie \int_{-\pi}^{\pi} \frac{dk}{2 \pi}   \sum_{i\in {\rm{occ.}}} \bra{ u_i(k)} \partial_{k}  \ket{u_{i}(k)} = - P_{1} .
\end{eqnarray} Since $P_{1} $ is defined only mod $m$ the two values consistent with inversion symmetry are:
\beq
P_1 = 0, \frac{e}{2}.
\eneq  We define the trivial insulator as $P_1=0$, while the topological one is $P_1 = \frac{e}{2}$. The $e/2$ value means that when the system is cut in two, there exists \emph{half} a charge on each end. 
\end{widetext}

\section{Relation between adiabatic connection and $B(k)$ in 1D}\label{app:PolZ2}
We now prove that the matrix
\begin{equation}
B_{ij}=\langle u_{i,-k}\vert {\cal P}  \vert u_{j,k}\rangle.\nonumber
\end{equation} is unitary and that
\begin{eqnarray}
&\hat{A}(-k) = -\hat{A}(k)+ i Tr[B(k)  \nabla_k B^\dagger(k) ]   \nonumber
\end{eqnarray}

We define the matrix $B_{ij}$ as the matrix connecting the bands at $k$ with the ones at $-k$:
\beq
\ket{u_i (-k) } = B_{ij}^*(k)  {\cal P}  \ket{u_j(k)} \label{inversionmatrix}
\eneq where $i,j$ run over the occupied bands $1,...,N$. From this formula we see that

That $B_{ij}$ has to be unitary can be easily seen by extending the matrix to belong to \emph{all} bands, occupied \emph{and} unoccupied. Since we have a full gap in the system none of the occupied bands at $k$ can transform to unoccupied bands at $-k$ and vice-versa (otherwise we would not have an insulator).  This means that the full matrix $B^{\text{full}} = diag(B^{\text{occupied}}, B^{\text{unoccupied}})$ is block diagonal.  One can prove the full $B$ is unitary by using completeness:
\beq
\sum_{i \in \text{all bands}} \ket{u_i(k)} \bra{u_i(k)} = I
\eneq we have (double index means summation)
\begin{eqnarray}
(B^\dagger B)_{ij}&=& (B^{\dagger})_{im} B_{mj} =  \bra{u_{i,k}}{\cal P} ^\dagger \ket{u_{m,-k}}  \bra{u_{m,-k}}{\cal P} \ket{u_{j,k}}\nonumber\\ &=&  \bra{u_{i,k}}{\cal P} ^\dagger {\cal P} \ket{u_{j,k}}=  \bra{u_{i,k}}\ket{u_{j,k}} = \delta_{ij}\nonumber
\end{eqnarray} Hence, since  the full matrix $B$ is unitary, so are the $B^{\text{(un)occupied}}$.

We now want to express a connection between $A(-k)$ and $A(k).$ We have
\begin{widetext}
\begin{eqnarray}
&\hat{A}(-k) = -i \bra{u_{i, -k}} \nabla_{-k} \ket{u_{i, -k}} = i (u_i(-k))^*_\alpha  \nabla_k (u_i(-k))_\alpha =  i B_{il}(k) {\cal P}^*_{\alpha \beta} (u_{l,k})^*_\beta \nabla_k (B_{ij}^*(k) {\cal P} _{\alpha \theta} (u_{j , k})_\theta)= \nonumber \\ &=   i B_{il}(k)  [\nabla_k B_{ij}^*(k) ]  {\cal P}^*_{\alpha \beta} (u_{l,k})^*_\beta {\cal P} _{\alpha \theta} (u_{j , k})_\theta + i B_{il}(k) B_{ij}^*(k) {\cal P}^*_{\alpha \beta} (u_{l,k})^*_\beta \nabla_k  {\cal P} _{\alpha \theta} (u_{j , k})_\theta = \nonumber \\ & =   i B_{il}(k)  [\nabla_k B_{ij}^*(k) ]   (u_{l,k})^*_\beta \delta_{\beta \theta} (u_{j , k})_\theta + i B_{il}(k) B_{ij}^*(k) \delta_{\beta \theta} (u_{l,k})^*_\beta \nabla_k   (u_{j , k})_\theta = \nonumber \\ & =   i B_{il}(k)  [\nabla_k B_{ij}^*(k) ]  \delta_{jl} + i B_{il}(k) B_{ij}^*(k) \ (u_{l,k})^*_\beta \nabla_k   (u_{j , k})_\beta = \nonumber \\ & =   i B_{il}(k)  [\nabla_k B_{il}^*(k) ]  + i\delta_{jl} \ (u_{l,k})^*_\beta \nabla_k   (u_{j , k})_\beta = i B_{il}(k)  [\nabla_k B_{il}^*(k) ]  + i \ (u_{j,k})^*_\beta \nabla_k   (u_{j , k})_\beta= \nonumber \\ & =   i Tr[B(k)  \nabla_k B^\dagger(k) ]   - \hat{A}(k) 
\end{eqnarray}\noindent where repeated indices are summed over. 
\end{widetext}

\section{Magneto Electric Polarization as Winding Number of the $B(k)$ matrix}\label{app:windNumb}

The Abelian $\hat{A}(k)$ in the previous section obeys a special case of the more general non-Abelian transformation
\beq {\hat{\textbf{A}}}(-k) = - B {\hat{\textbf{A}}}(k) B^\dagger + i B(k) \vec{\nabla} B^\dagger(k)
\eneq where the non-Abelian adiabatic connection is is $\hat{A}^{\alpha \beta}_i(k)= -i \bra{u_{\alpha, k}} \nabla_{k_i} \ket{u_{\beta, k}}$. The above implies that the Berry gauge fields at $k$ and $-k$ are non-abelian gauge transformed of each other. The field strength gauge transformation is:
\begin{eqnarray}
 \hat{F}_{ij}(-k) &=&   B(k)  \hat{F}_{ij}(k) B^\dagger(k)
\end{eqnarray} 
  From here the magneto-electric polarizability is easy but tedious to obtain:

\begin{widetext}
\begin{eqnarray}
& P_3 =\frac{1}{16 \pi^2}\int d^3k \epsilon_{ijk} Tr[(\hat{F}_{ij} (k) - \frac{2}{3} i \hat{A}_i(k) \hat{A}_j(k) ) \hat{A}_k(k)]=\nonumber \\ &=  \frac{1}{16 \pi^2}\int d^3k \epsilon_{ijk} Tr[(\hat{F}_{ij} (-k) - \frac{2}{3} i \hat{A}_i(-k) \hat{A}_j(-k) ) \hat{A}_k(-k)]= \nonumber \\ &=  \frac{1}{16 \pi^2}\int d^3k \epsilon_{ijk} Tr[ (  B(k) \hat{F}_{ij}(k) B^\dagger - \frac{2}{3} i (B(k) \hat{A}_i (k) \hat{A}_j(k) B^\dagger(k)  - i B(k)  \hat{A}_i(k)  \partial_j B^\dagger(k) -\nonumber \\ &+  i  (\partial_i B(k) ) \hat{A}_j(k) B^\dagger(k)   + \partial_i B \partial_j B^\dagger)   (- B(k) \hat{A}_k (k) B^\dagger(k) + i B(k) \partial_k B^\dagger(k))] = \nonumber \\ &= - \frac{1}{16 \pi^2}\int d^3k \epsilon_{ijk} Tr[(\hat{F}_{ij} (k) - \frac{2}{3} i \hat{A}_i(k) \hat{A}_j(k) ) \hat{A}_k(k)] - \frac{1}{24 \pi^2} \int d^3 k \epsilon_{ijk} Tr[(B(k) \partial_i B^\dagger)  (B(k) \partial_j B^\dagger)  (B(k) \partial_k B^\dagger)  ] + \nonumber \\\ & + \frac{i}{8 \pi^2} \int d^3 k \epsilon_{ijk} \partial_i (B(k) \hat{A}_j(k) \partial_k B^\dagger) = - P_3- \frac{1}{24 \pi^2} \int d^3 k \epsilon_{ijk} Tr[(B(k) \partial_i B^\dagger)  (B(k) \partial_j B^\dagger)  (B(k) \partial_k B^\dagger)  ]
\end{eqnarray} 
which proves the formula in the text.
\end{widetext}

\section{Proof of Fu-Kane Formula\cite{fu2007a}}\label{app:kaneFuFormula}
In this appendix we provide an alternative proof for the Fu-Kane formula for the $Z_2$ invariant of 3D $T$ and ${\cal{P}}$ invariant insulators. 
Consider the Bloch Hamiltonian $\hat{H}(k)$ of an Êinsulator with both inversion ${\cal{P}}$ and time-reversal symmetry $T$. We have
\begin{eqnarray}
{\cal{P}}^2=1, \;\; T^2&=&-1, \;\; [{\cal{P}},T] =0\nonumber\\
T\hat{H}(k) T^{-1} &=& \hat{H}(-k)\nonumber\\
 Ê{\cal{P}}\hat{H}(k) {\cal{P}}^{-1} &=& \hat{H}(-k)
\end{eqnarray} and hence:
\beq
{\cal{P}}T \hat{H}(k) ({\cal{P}}T)^{-1} = \hat{H}(k), \;\;\;\;\; ({\cal{P}}T)^2=-1
\eneq which proves that the insulator has doubly degenerate bands at each momentum $k.$ This does not depend on the dimensionality of the space. The two ingredients we will use to prove  Eq. \ref{eq:kaneInvariant} are Ê(i) band crossing arguments between a 3D insulator in the trivial atomic limit  and  a topologically non-trivial insulator, and (ii) the fact that a $4+1$-d Dirac Hamiltonian changes its 4D `Hall conductance' by $1$ Êif there is a band crossing of \emph{four} bands - these are actually two doubly degenerate bands. Without loss of generality we consider a topological insulator with four bands - two occupied bands and two unoccupied. As we take a generic $T$ and ${\cal P}$ invariant insulator through a phase transition, four ( \emph{i.e.} two doubly degenerate) bands are generically needed. 

Assume we have two $T$ and ${\cal P}$ symmetric Hamiltonians in $3D$  $\hat{h}_1(k)$ and $\hat{h}_2(k)$ with:
\begin{eqnarray}
T\hat{h}_{1,2}(k) T^{-1} &=& \hat{h}_{1,2}(-k) \nonumber\\ {\cal P}\hat{h}_{1,2}(k) {\cal P}^{-1} &=& \hat{h}_{1,2}(-k).\nonumber
\end{eqnarray} We choose $\hat{h}_1(k)$ to be trivial (in the atomic limit with all hoppings taken to vanish). We now construct a time-reversal and inversion invariant interpolation between these two Hamiltonians. ÊWe first prove that such a gapped interpolation exists. To see this, it is easiest to remain with our reduced Hamiltonian with $4$-bands, which represents the generic effective Hamiltonian of the two doubly-degenerate bands immediately above and below the Fermi level, out of the total $N$ bands in the insulator. In this basis, the $3D$ effective insulating Hamiltonian  with inversion and time-reversal symmetry has co-dimension 2 (there are $3$ momenta $k_x,k_y,k_z$ and $5$ Clifford generator $\Gamma^a$ matrices in which a $4$-band insulating Hamiltonian with doubly degenerate bands can be expanded). In $4D$, a topological insulator with inversion and time-reversal symmetry has co-dimension 1(there are $4$ momenta and $5$ $\Gamma^a$ matrices), so there is still Êalways a way to make it gapped. This shows that generically, a $3D$ or $4D$  insulator with inversion and time-reversal is always gapped. It can be made gapless by tuning $2$ and $1$ parameter(s) respectively. ÊNow,  let the gapped interpolation between $\hat{h}_1(k)$ and $\hat{h}_2(k)$ be $\hat{h}(k, \theta)$ which satisfies the properties:
\begin{eqnarray}
\hat{h}(k, 0) = \hat{h}_1(k) &;&\;\;\;\ \hat{h}(k,\pi)= \hat{h}_2(k)\nonumber\\ Ê
 T\hat{h}(k, \theta) T^{-1} &=& \hat{h}(-k, -\theta)\nonumber\\ {\cal{P}} \hat{h}(k, \theta) {\cal{P}}^{-1} &=& \hat{h}(-k, -\theta).
\end{eqnarray} The interpolation between the 3D Hamiltonians is chosen this way so that if one interprets $\theta$ as a fourth momentum then the resulting $4D$ Hamiltonian would  respect inversion and time-reversal. ÊIt was shown in Ref. \onlinecite{qi2008B} that $2 (P_3(\hat{h}_2(k))-P_3 (\hat{h}_1(k)))=C_2(\hat{h}(k,\theta))$ where $C_2$ is the second Chern number of the $4D$ Hamiltonian $\hat{h}(k,\theta).$  Since we chose $\hat{h}_1(k)$ to be trivial 
\beq
P_3(\hat{h}_1) = 0 \mod n\in Z.
\eneq Hence, if the second Chern number of the $4$-D $\hat{h}(k, \theta)$ is \emph{ odd}, we have
\beq
P_3(\hat{h}_2) = 1/2 Ê\mod n \in Z
\eneq giving rise to the result that if $C_2$ of the $4D$ Hamiltonian is odd, then either $\hat{h}_1(k)$ or $\hat{h}_2(k)$ is a non-trivial topological insulator. Since we pick $\hat{h}_1(k)$ to be our reference trivial Hamiltonian then this would imply that $\hat{h}_2(k)$ is non-trivial.

To understand how to classify the $3D$ insulators we first need to understand how to get a $4D$ insulator with an odd second Chern number. The $4D$ trivial Hamiltonian is simply a momentum-independent interpolation between $\hat{h}_1(k)$ and itself. This clearly has vanishing $C_2.$  Since $C_2$ is a topological invariant, we must have a gap-closing phase transition to change it.  As the system is inversion and TR invariant, we Êhave to analyze the crossings between two doubly-degenerate bands. This is the generic case, even for insulators with an arbitrary number of bands $N,$ because we can build an ``effective" Hamiltonian close to the transition which will be a $4$-band model. With time-reversal and inversion, the Bloch Hamiltonian has to be of the form $\hat{H}(k)=d_a(k) \Gamma^a.$ We first consider transitions which occur away from the invariant momenta. Because of inversion (or time-reversal) a gap closing at $k$ must be matched by one at $-k.$  Such transitions can be tuned by a single parameter. Since the gap closing and re-opening happens away from an invariant momentum, the inversion eigenvalues of the occupied bands remain unmodified. 
The non-abelian adiabatic field strengths at the two k-points are equal (up to a gauge transformation, and a minus sign in case of time-reversal). Thus,  a gap closing at two points, $k$ and $-k$ makes the total change in second Chern number which is  \emph{even}.  Thus a 4D  Chern insulator with inversion and time reversal has even second Chern number if the inversion eigenvalues of the occupied bands are the same as in the atomic limit.

We now look at the case where the gap closing and re-opening happens at an inversion symmetric point. In this case, the Hamiltonian is still $4\times 4$ but the co-dimension of any two doubly degenerate bands is $5$ because we have $5$ Clifford matrices and no tunable momenta -- the momenta are fixed at the inversion symmetric points. ÊIf the bands involved in the phase transition at the inversion symmetric points have the same inversion eigenvalues (\emph{e.g.} all positive)  then, up to a gauge choice, the inversion matrix is the identity operator. This means that $[{\cal{P}},\hat{H}(0)]$ is trivially satisfied and does not provide an additional constraint and so the crossing is always avoided with such a large co-dimension. However, if the bands have different inversion eigenvalues,(\emph{e.g.} bonding and anti-bonding bands), Êso that the inversion matrix in a specific choice of basis is ${\cal{P}}=1_{2\times 2}\otimes\tau^z$, then we find that the effective ÊHamiltonian matrix at the inversion symmetric points Ê$G_i/2$ is 
\beq
\hat{H}(G_i/2) = M_i {\cal{P}}
\eneq where $M_i$ is the mass at the inversion symmetric point $G_i/2$.\cite{Murakami2007A,Murakami2007B, Murakami2008} As such, a gap closing transition whereby the $M_i$ changes sign is accompanied by two effects: (i) the second Chern number will change by $\pm 1$ if only one mass goes through zero, or more generally, it will change by the number of masses that go through zero (multiplied by a sign) and (ii) the inversion eigenvalues of a \emph{pair} of  occupied bands change sign as a result of every $M_i$ that switches sign. This is the crucial difference between phase transitions which change the first and second Chern numbers. In two dimensions a change in the first Chern number is accompanied by a single inversion eigenvalue switch since the minimal crossing is between two-bands. In 4D, changes in the second Chern number require a four-band crossing and thus two inversion eigenvalues are exchanged. This carries over to 3D where we are interested in the $Z_2$ valued parity of the second Chern number. A change of the parity is effected when a 4-band crossing switches two inversion eigenvalues. 
We hence proved that
a 4D Chern insulator with inversion and time-reversal symmetry has odd second Chern number if the product of \emph{half} of the inversion eigenvalues (half meaning each Kramers' pair is only counted once) Êat all TR invariant points is $-1.$ By keeping $\hat{h}_1(k)$ constant and changing only $\hat{h}_2(k)$  through a gap closing and reopening we can make $\hat{h}(k,\theta)$ a 4D insulator with an odd second Chern number. Thus $\hat{h}_2(k)$ is a non-trivial $Z_2$ insulator when Eq. \ref{eq:kaneInvariant} is negative and we have proved the Fu and Kane formula.

\section{Proof that the winding number is independent of a smooth $U(1)$ phase}\label{app:U1ind}
We begin with the generic form of a $U(2)$ block of the entire $B(k)$ matrix
\beq
B(k) = e^{i \phi(k) } (f(k) I + i g_a(k) \sigma_a)
\eneq where
\beq
(f(k))^2 + g_a(k) g_a(k) =1.
\eneq
We drop the argument $(k)$ which is implied from now on. By using:
\beq
\sigma_a \sigma_b = \delta_{ab} + i \epsilon_{abc} \sigma_c
\eneq  we find:
\beq
B^\dagger \partial_i B = i \left[ \partial_i \phi + (f\partial_i g_c - g_c \partial_i f + \epsilon_{bac} g_b \partial_i g_a )\sigma_c \right] 
\eneq Define   
\beq
G_{ic} =  (f\partial_i g_c - g_c \partial_i f + \epsilon_{bac} g_b \partial_i g_a )
\eneq so that:
\beq
B^\dagger \partial_i B = i \left[ \partial_i \phi  + G_{ic} \sigma_c \right] .
\eneq

The winding number density is:
\begin{eqnarray}
 n_{W}&=&\epsilon_{ijk} Tr[B^\dagger \partial_i B B^\dagger \partial_j B B^\dagger \partial_k B ]\nonumber\\ &=&   -i \epsilon_{ijk}  Tr\left[ ( \partial_i \phi  + G_{ia} \sigma_a  )   ( \partial_j \phi  + G_{jb} \sigma_b  )   ( \partial_k \phi  + G_{kc} \sigma_c )  \right] = \nonumber \\ & =& 2 \epsilon_{ijk}\epsilon_{abc} G_{ia} G_{jb} G_{kc}.
\end{eqnarray}
Thus the $U(1)$ phase does not contribute if it is smooth.

\section{Proof that the $U(1)$ phase is smooth in 3D }\label{app:smoothPhase}
We have seen that the $\hat{B}(k)$ matrix decomposes into trivial one dimensional and non-trivial 2 dimensional diagonal blocks. Each 2 dimensional block can be seen as map from $T^3$ into $U(2)$. Here we show that the $U(1)$ phase of these maps is smooth and for this reason it can be taken out from winding number calculation. For this, we will show that a discontinuity in the phase implies the appearance of the $(^+_-)$ eigenvalue pattern of inversion eigenvalues at one $k_{\mbox{\tiny{inv}}}$ point. Since this pattern is excluded for the $U(2)$ blocks that cannot be reduced to $U(1)\times U(1)$, it follows that the phase is actually smooth.

Now, a simple example of a map with a non-smooth phase, to keep in mind for intuition is:
\begin{equation}\label{Exa}
\begin{array}{c}
B(k)=e^{\frac{1}{2}ik}\big ( \cos \frac{k}{2}+i \sin \frac{k}{2}  \ \sigma_z \big)\equiv e^{i\theta(k)}U(k)
\end{array}
\end{equation}
where $U(k)\in SU(2).$
Note that this $B(k)$ satisfies all the required constraints and we can see that in Eq.~\ref{Exa} everything is smooth as $k$ is continuously varied, except when we are close to the end of the $2\pi$ cycle, that is when we close the loop around the Brillouin torus. If we start for example from $k$=$-\pi$, one can see that the phase and the $SU(2)$ factor take different values when $k$ approaches $\pi$, but as a whole $B(k)$ reaches the same value as the one we started with at $k$=$\pi$. It is important to note that the discontinuity in the phase cannot be removed by any continuous deformation of $B(k)$.

 In the following we argue that such a discontinuity in the phase immediately implies the existence of a $(^+_-)$ pattern of inversion eigenvalues at some $k_{\mbox{\tiny{inv}}}$ point. As we already pointed out, the non-removable discontinuities in the phase are entirely due to the nontrivial topology of the Brillouin torus. The question is what happens when $k$ is taken over a nontrivial loop of the torus? So let us start from a $k_{\mbox{\tiny{inv}}}^1$ point and continuously vary $k$ until it advances by $2\pi$ along one of the $k_1$, $k_2$ or $k_3$ directions, that is, we are taking $k$ around non-trivial loop of the torus.  As before, we will work with the following expression of the $U(2)$ matrix:
 \begin{equation}
 B(k)=e^{i\theta_k}(f_k+i\vec{g}_k \vec{\sigma}).
 \end{equation}
 At the $k_{\mbox{\tiny{inv}}}$ points, $\theta$ can take only integer or half-integer values (in units of $\pi$), and an important observation is that, if $\theta$ is half-integer at a $k_{\mbox{\tiny{inv}}}$ point, then $B(k_{\mbox{\tiny{inv}}})$ necessarily takes the form $\hat{n}\vec{\sigma}$ ($\hat{n}$= unit vector), which has $\pm 1$ inversion eigenvalues.  So let us assume that $\theta$=0 at $k_{\mbox{\tiny{inv}}}^1$ (otherwise the $(^+_-)$ pattern will show up). After we complete the loop around the torus, in general, we can have:
\begin{equation}
\theta_{k_{\mbox{\tiny{inv}}}^1+2\pi}=\theta_{k_{\mbox{\tiny{inv}}}^1}+\left \{
\begin{array}{cc}
2n\pi & \mbox{(a)}\\
(2n+1)\pi & \mbox{(b)}
\end{array} \right .
\end{equation}
In case (a) the phase factor returns to its original value so it is smooth, in which case there is no problem removing it from the winding number calculation. In case (b), the phase factor changes sign. Now let us consider the inversion properties of $B(k)$ relative to the $k_{\mbox{\tiny{inv}}}^2$=$k_{\mbox{\tiny{inv}}}^1$+$\pi$ point, which also belongs to the loop. We have:
\begin{equation}
\begin{array}{c}
e^{i\theta_{k_{\mbox{\tiny{inv}}}^2+k}}\big (f_{k_{\mbox{\tiny{inv}}}^2+k}+i \vec{g}_{k_{\mbox{\tiny{inv}}}^2+k} \vec{\sigma}\big ) \medskip \\
=e^{-i\theta_{k_{\mbox{\tiny{inv}}}^2-k}}\big (f_{k_{\mbox{\tiny{inv}}}^2-k}-i \vec{g}_{k_{\mbox{\tiny{inv}}}^2-k} \vec{\sigma}\big ).
\end{array}
\end{equation}
Since $f_k$ and $\vec{g}_k$ take only real values, we can see that if $
\theta_{k_{\mbox{\tiny{inv}}}^2+k}$ is half-integer then necessarily $\theta_{k_{\mbox{\tiny{inv}}}^2-k}$ is also half-integer. In other words, the half-integer values of $\theta_k$ come in pairs, unless $k$ is the $k_{\mbox{\tiny{inv}}}^2$ invariant point itself. Since $\theta$ varies from 0 to $(2n$+$1)\pi$ as we vary $k$ by $2\pi$, there will be an odd number of times when $\theta_k$ assumes a half-integer value. It is then evident that one such half-integer value must occur at $k_{\mbox{\tiny{inv}}}^2$ and consequently the $(-+)$ inversion eigenvalue pattern will show up at $k_{\mbox{\tiny{inv}}}^2$.

\section{Inversion eigenvalue patterns and the corresponding magneto-electric polarizability}\label{app:invExamples}

In this appendix we provide some explicit examples of our proof of the connection between inversion eigenvalues and the magneto-electric polarization for inversion invariant insulators. We focus on 4-band, gapped Hamiltonians $\hat{H}(k)$ in 3-dimensions displaying various patterns of inversion eigenvalues. Assuming two occupied bands, we will write these inversion eigenvalue patterns as 
\begin{equation}
\begin{array}{c}
++++++++ \\
----++++,
\end{array}
\end{equation}
for example, where one should understand that the two non-zero eigenvalues of $P_{k_{\mbox{\tiny{inv}}}}{\cal P}P_{k_{\mbox{\tiny{inv}}}}$ are $+1$ and $-1$ at four $k_{\mbox{\tiny{inv}}}$ points and $+1$ and $+1$ at the remaining $k_{\mbox{\tiny{inv}}}$ points. The order of the $k_{inv}$ is not important for the magneto-electric polarizability, but to be precise we will order them as \begin{eqnarray}\{ &(0,0,0),&(\pi,0,0),(0,\pi,0), (0,0,\pi),(\pi,\pi,0),(\pi, 0,\pi),\nonumber\\ &(0,\pi,\pi), & (\pi,\pi,\pi)\}.\nonumber\end{eqnarray}

For each case, o compute the magneto-electric polarizability $P_3$, we will begin with a trivial reference Hamiltonian $(P_3=0)$ and find a gapped and inversion symmetric interpolation $\hat{H}(k,\theta)$  between this and our example Hamiltonians. We can compute $P_3$ as half the second Chern number $C_2$ generated by the interpolation $\hat{H}(k,\theta)$ which is considered as a 4+1-d inversion symmetric Hamiltonian. An odd $C_2$ corresponds to a non-trivial insulator with half-integer magneto-electric polarizability, while an even $C_2$ indicates a trivial insulator. This method of calculation is convenient because $C_2$ can be computed using a gauge-invariant  projector method , so we can bypass the task of finding a smooth gauge. All of our numerical observations are in agreement with our mathematical proofs above.

1.) The pattern 
\begin{equation}
\begin{array}{c}
++++++++ \\
--++++++
\end{array}
\end{equation}
is seen with the following gapped Hamiltonian:
\begin{equation}
\begin{array}{c}
\hat{H}(k)=\sin k_1 \hat{\Gamma}_1+\sin k_2 \hat{\Gamma}_2+\sin k_3 \hat{\Gamma}_3 \medskip \\
+(-2.5+\cos k_1 + \cos k_2 + \cos k_3 )\hat{\Gamma}_0 \medskip \\
+0.75(\cos k_1 + \cos k_2 ) (\hat{\Gamma}_{15}+\hat{\Gamma}_{25}-\hat{\Gamma}_{35}-\hat{\Gamma}_{12}). 
\end{array}
\end{equation}
This Hamiltonian can be connected through a gapped interpolation with the Hamiltonian:
\begin{equation}
\begin{array}{c}
\hat{H}_0(k)=\sin k_1 \hat{\Gamma}_1+\sin k_2 \hat{\Gamma}_2\medskip \\
+(-2.5+\cos k_1 + \cos k_2 )\hat{\Gamma}_0 \medskip \\
+0.75(\cos k_1 + \cos k_2 ) (\hat{\Gamma}_{15}+\hat{\Gamma}_{25}-\hat{\Gamma}_{35}-\hat{\Gamma}_{12}). 
\end{array}
\end{equation}
using the inversion symmetric homotopy
\begin{equation}
\begin{array}{c}
\hat{H}(k,\theta)=\frac{1}{2}(1+\cos \theta)\hat{H}(k)
+\frac{1}{2}(1-\cos \theta)\hat{H}_0(k).
\end{array}
\end{equation}
The second Chern number generated by $\hat{H}(k,\theta)$ is $C_2$=0 (even) and the $P_3$ of $\hat{H}_0(k)$ is zero because it depends only on $k_1$ and $k_2$. Consequently, the $P_3$ of $\hat{H}(k)$ is $0 \ \mbox{mod}(Z)$.

2.) The pattern 
\begin{equation}
\begin{array}{c}
++++++++ \\
----++++
\end{array}
\end{equation}
is seen using the following gapped Hamiltonian:
\begin{equation}
\begin{array}{c}
\hat{H}(k)=\sin k_1 \hat{\Gamma}_1+\sin k_2 \hat{\Gamma}_2+\sin k_3 \hat{\Gamma}_3 \medskip \\
+(-2.5+\cos k_1 + \cos k_2 + \cos k_3 )\hat{\Gamma}_0  \medskip \\
-(2-\cos k_3)(\cos k_1 + \cos k_2 ) (\hat{\Gamma}_{15}+\hat{\Gamma}_{25}-\hat{\Gamma}_{35}-\hat{\Gamma}_{12}). 
\end{array}
\end{equation}
This Hamiltonian can be connected through a gapped interpolation with the Hamiltonian:
\begin{equation}
\begin{array}{c}
\hat{H}_0(k)=\sin k_1 \hat{\Gamma}_1+\sin k_2 \hat{\Gamma}_2  \medskip \\
+(-2.5+\cos k_1 + \cos k_2) \hat{\Gamma}_0 \medskip \\
-2(\cos k_1 + \cos k_2 ) (\hat{\Gamma}_{15}+\hat{\Gamma}_{25}-\hat{\Gamma}_{35}-\hat{\Gamma}_{12}). 
\end{array}
\end{equation}
using the inversion symmetric homotopy
\begin{equation}
\begin{array}{c}
\hat{H}(k,\theta)=\frac{1}{2}(1+\cos \theta)\hat{H}(k)
+\frac{1}{2}(1-\cos \theta)\hat{H}_0(k).
\end{array}
\end{equation}
The second Chern number generated by $\hat{H}(k,\theta)$ is $C_2$=0 (even) and the $P_3$ of $\hat{H}_0(k)$ is zero because it depends only on $k_1$ and $k_2$. Consequently, the $P_3$ of $\hat{H}(k)$ is $0 \ \mbox{mod}(Z)$.

3.) The pattern 
\begin{equation}
\begin{array}{c}
----++++ \\
----++++
\end{array}
\end{equation}
is seen with the following gapped Hamiltonian:
\begin{equation}
\begin{array}{c}
\hat{H}(k)=\sin k_1 \hat{\Gamma}_1+\sin k_2 \hat{\Gamma}_2+\sin k_3 \hat{\Gamma}_3 \medskip \\
+(-\frac{1}{2}+\cos k_1 + \cos k_2 + \cos k_3 )\hat{\Gamma}_0+\frac{1}{2}\hat{\Gamma}_{25}.
\end{array}
\end{equation}
This Hamiltonian can be connected through a gapped interpolation with the trivial Hamiltonian:
\begin{equation}
\begin{array}{c}
\hat{H}_0(k)=\sin k_1 \hat{\Gamma}_1+\sin k_2 \hat{\Gamma}_2+\sin k_3 \hat{\Gamma}_3 -\frac{1}{2}\hat{\Gamma}_0,
\end{array}
\end{equation}
using the inversion symmetric homotopy of Eq.~\ref{interpol}. The second Chern number generated by $\hat{H}(k,\theta)$ is $C_2$=2 (even) and consequently $P_3=0 \ \mbox{mod}(Z)$.

4.) The pattern 
\begin{equation}
\begin{array}{c}
++++++++ \\
--------
\end{array}
\end{equation}
is seen using the following gapped Hamiltonian:
\begin{equation}
\begin{array}{c}
\hat{H}(k)=\sin k_1 \hat{\Gamma}_1+\sin k_2 \hat{\Gamma}_2+\sin k_3 \hat{\Gamma}_3 - \frac{1}{2} \hat{\Gamma}_0 \medskip \\
+ (2+ \cos k_1 + \cos k_2 + \cos k_3) \hat{\Gamma}_{15} \medskip \\
+ (-2+ \cos k_1 + \cos k_2 + \cos k_3) \hat{\Gamma}_{25}
\end{array}
\end{equation}
This Hamiltonian can be connected through a gapped interpolation with the trivial Hamiltonian:
\begin{equation}
\begin{array}{c}
\hat{H}_0(k)=\sin k_1 \hat{\Gamma}_1+\sin k_2 \hat{\Gamma}_2+\sin k_3 \hat{\Gamma}_3 -\frac{1}{2}\hat{\Gamma}_0
\end{array}
\end{equation}
using the inversion symmetric homotopy
\begin{equation}
\begin{array}{c}
\hat{H}(k,\theta)=\sin k_1 \hat{\Gamma}_1+\sin k_2 \hat{\Gamma}_2+\sin k_3 \hat{\Gamma}_3 - \frac{1}{2} \hat{\Gamma}_0 \medskip \\
+ \frac{1}{2}(1+\cos \theta)(2+ \cos k_1 + \cos k_2 + \cos k_3) \hat{\Gamma}_{15} \medskip \\
+ \frac{1}{2}(1+\cos \theta) (-2+ \cos k_1 + \cos k_2 + \cos k_3) \hat{\Gamma}_{25} \medskip \\
+ \sin \theta \ \hat{\Gamma}_5.
\end{array}
\end{equation}
The second Chern number generated by $\hat{H}(k,\theta)$ is $C_2$=0 (even) and consequently $P_3=0 \ \mbox{mod}(Z)$.

5.) The pattern 
\begin{equation}
\begin{array}{c}
--++++++ \\
--++++++
\end{array}
\end{equation}
is seen using the following gapped Hamiltonian:
\begin{equation}
\begin{array}{c}
\hat{H}(k)=\sin k_1 \hat{\Gamma}_1+\sin k_2 \hat{\Gamma}_2 \medskip \\
+ (-\frac{3}{2}+ \cos k_1 + \cos k_2) \hat{\Gamma}_0 \medskip \\
+ (-\frac{1}{4}+ 0.1 (\cos k_1 - \cos k_2 )) \hat{\Gamma}_{25}.
\end{array}
\end{equation}
This Hamiltonian can be connected through a gapped interpolation with the trivial Hamiltonian:
\begin{equation}
\begin{array}{c}
\hat{H}_0(k)=\sin k_1 \hat{\Gamma}_1+\sin k_2 \hat{\Gamma}_2+\sin k_3 \hat{\Gamma}_3 -\frac{3}{2}\hat{\Gamma}_0
\end{array}
\end{equation}
using the inversion symmetric homotopy
\begin{equation}
\begin{array}{c}
\hat{H}(k)=\sin k_1 \hat{\Gamma}_1+\sin k_2 \hat{\Gamma}_2 + \frac{1}{2}(1-\cos \theta)\sin k_3 \hat{\Gamma}_3 \medskip \\ 
+(-\frac{3}{2}+\frac{1}{2}(1+\cos \theta)(\cos k_1 + \cos k_2 ))\hat{\Gamma}_0  \medskip \\
+\frac{1}{2}(1+\cos \theta)(-\frac{1}{4}+ 0.1 (\cos k_1 - \cos k_2 )) \hat{\Gamma}_{25} \medskip \\
+ \sin \theta \ \hat{\Gamma}_5.
\end{array}
\end{equation}
The second Chern number generated by $\hat{H}(k,\theta)$ is $C_2$=0 (even) and consequently $P_3=0 \ \mbox{mod}(Z)$.

\section{Response Theory Argument in Higher Dimensions}\label{app:higherD}
Although we do not rigorously prove anything in higher dimensions we note that the physical response arguments presented in this work and Refs. \onlinecite{qi2008B,turner2009} continue to apply in higher dimensions.  This gives us a hint that there are interesting inversion symmetric topological  insulators in higher dimensions. For \emph{even} spacetime dimensions the topological response actions all take a standard form
\begin{eqnarray}
S_{eff}[A_\mu]=\int d^{2n}x P_n\epsilon^{a_1 a_2\ldots a_{2n-1}a_{2n}}F_{a_1a_2}\ldots F_{a_{2n-1}a_{2n-2}}\nonumber
\end{eqnarray}\noindent where $P_n$ is a response coefficient and $F_{ab}$ is the electromagnetic field-strength tensor. If these were all dynamical fields then the entire action must transform like a scalar which means that the intrinsic response coefficients $P_n$ must transform the same way as the product of the electromagnetic fields. Thus,  $P_n$ must be odd under inversion symmetry for all $n.$ Additionally for even (odd) $n,$ $P_n$ is odd under T (C) symmetry. Since we are talking about external electromagnetic fields we do not transform them under the symmetry operation and hence only the $P_n$ are changed.  Thus in alternating even spacetime dimensions topological insulators are protected by either T or C symmetry. However, in \emph{every} even spacetime dimension there is a topological insulator protected by inversion symmetry which has a topological response. In all of these cases $P_n$ is not gauge invariant under transformations of the occupied wavefunctions and the oddness under the different symmetries quantizes $P_n$ to take only two independent values. This yields a $Z_2$ classification for the \emph{response coefficient} in line with the arguments presented in Ref. \onlinecite{qi2008B}. 

In \emph{odd} space-time dimensions the generalized Chern-Simons terms describing the electro-magnetic responses in topological insulators are \emph{compatible} with inversion symmetry. The general action is
\begin{eqnarray}
S^{(odd)}_{eff}[A_\mu]&=&\frac{C_{n}}{(n+1)!(2\pi)^n}\nonumber\\ &\times &\int d^{2n+1} \epsilon^{a_1 a_2\ldots a_{2n+1}}A_{a_1}F_{a_2 a_3}\ldots F_{a_{2n} a_{2n+1}}.\nonumber
\end{eqnarray}\noindent The quantity $C_{n}$ is the $n$-th Chern number and is even under inversion symmetry and thus not restricted by the requirement of preserving inversion symmetry.  If $n$ is even (odd) then the action is also compatible with $T$ ($C$) symmetry.  We have not proven it, but our intuition suggests that the parity of $C_n$ can be characterized by the inversion topological invariants given by  the $\chi^{(n)}_{\cal{P}}$ defined above when calculated in $(2n+1)-d.$

\end{document}